\documentclass[conference]{IEEEtran}
\IEEEoverridecommandlockouts
\newcommand{\sys}{\texttt{UTune}\xspace}

\usepackage{amsthm}
\usepackage{amsmath}
\usepackage{tabularx}
\usepackage{booktabs}
\usepackage[linesnumbered,ruled,vlined]{algorithm2e}
\usepackage{graphicx}
\usepackage{subcaption}
\usepackage{xspace}
\usepackage{textcomp}
\usepackage{hyperref}
\usepackage{multirow}
\usepackage{enumitem}
\usepackage{makecell}

\usepackage{xcolor}
\usepackage[most]{tcolorbox}

\definecolor{myblue}{HTML}{1c61a1}
\definecolor{mygreen}{HTML}{588b42}
\definecolor{myorange}{HTML}{c68302}
\definecolor{myred}{HTML}{ca5b65}
\newcommand{\rwone}[1]{{#1}}    
\newcommand{\rwthree}[1]{{#1}}   
\newcommand{\rwfour}[1]{{#1}}
\newcommand{\rwfive}[1]{{#1}}

\newtheorem{example}{Example}
\newtheorem{definition}{Definition}

\def\BibTeX{{\rm B\kern-.05em{\sc i\kern-.025em b}\kern-.08em
    T\kern-.1667em\lower.7ex\hbox{E}\kern-.125emX}}
\usepackage[numbers,sort&compress]{natbib}

\makeatletter
\newcommand{\MidHuge}{%
  \@setfontsize\MidHuge{22pt}{26pt}%
}
\makeatother

\begin{document}

\title{
\MidHuge
\spaceskip=0.2em
UTune: Towards Uncertainty-Aware Online Index Tuning
}
\author{
\IEEEauthorblockN{Chenning Wu*, Sifan Chen*, Wentao Wu$\dagger$, Yinan Jing*, Zhenying He*, Kai Zhang*, X. Sean Wang*}
\IEEEauthorblockA{\textsuperscript{*}Fudan University, Shanghai, China}
\IEEEauthorblockA{\textsuperscript{\dag}Microsoft Research, Washington, USA}
\IEEEauthorblockA{%
\{wucn23,sfchen23\}@m.fudan.edu.cn,\;
wentao.wu@microsoft.com,\;
\{jingyn,zhenying,zhangk,xywangCS\}@fudan.edu.cn}
}
\maketitle

\begin{abstract}
There have been a flurry of recent proposals on learned benefit estimators for index tuning.
Although these learned estimators show promising improvement over what-if query optimizer calls in terms of the accuracy of estimated index benefit, 
they face significant limitations when applied to online index tuning, an arguably more common and more challenging scenario in real-world applications.
There are two major challenges for learned index benefit estimators in online tuning: (1) limited amount of query execution feedback that can be used to train the models, and (2) constant coming of new unseen queries due to workload drifts.
The combination of the two hinders the generalization capability of existing learned index benefit estimators.
To overcome these challenges, we present \sys, an uncertainty-aware online index tuning framework that employs operator-level learned models with improved generalization over unseen queries.
At the core of \sys is an \emph{uncertainty quantification} mechanism that characterizes the inherent uncertainty of the operator-level learned models given limited online execution feedback.
We further integrate uncertainty information into index selection and configuration enumeration, the key component of any index tuner, by developing a new variant of the classic $\epsilon$-greedy search strategy with uncertainty-weighted index benefits.
Experimental evaluation shows that \sys not only significantly improves the workload execution time compared to state-of-the-art online index tuners but also reduces the index exploration overhead, resulting in faster convergence when the workload is relatively stable.
\end{abstract}

\setlength{\textfloatsep}{2pt plus 1pt minus 1pt}   
\setlength{\floatsep}{4pt plus 1pt minus 1pt}       
\setlength{\intextsep}{4pt plus 1pt minus 1pt}      

\SetAlFnt{\small}                 
\setlength{\algomargin}{0.8em}    
\IncMargin{-0.5em}                
\SetAlCapSkip{0.3ex}              

\setlength{\abovedisplayskip}{5pt plus 1pt minus 1pt}
\setlength{\belowdisplayskip}{5pt plus 1pt minus 1pt}
\setlength{\abovedisplayshortskip}{0pt plus 1pt}
\setlength{\belowdisplayshortskip}{4pt plus 1pt minus 1pt}
\everydisplay\expandafter{\the\everydisplay\small}

\setcounter{page}{1}
\pagestyle{plain}
\section{Introduction}


Building appropriate indexes inside a database system is crucial for accelerating query performance.
The problem of index tuning/recommendation therefore has been studied intensively and extensively in the past decades (see~\cite{siddiqui2024ml,WuZZL24} for recent surveys).
A classic scenario is \emph{offline} index tuning, where a fixed workload of queries is given and the goal is to recommend a set (a.k.a. \emph{configuration}) of indexes that can minimize the workload execution cost with respect to certain constraints such as the maximum number of indexes or the maximum amount of storage that can be taken by the indexes~\cite{chaudhuri1997efficient}. 
To evaluate the quality of candidate indexes, a common approach is to use the ``\emph{what-if}'' API provided by the query optimizer, which allows for cost estimation of a query \emph{without} materializing the candidate indexes~\cite{chaud1998autoadmin}.
The downside of what-if cost estimation is its potential inaccuracy, as it is built on top of query optimizer's cost models that are known to be error-prone~\cite{WuCZTHN13,ding2019ai}.

To improve the accuracy of cost estimation in the context of index tuning, recent work has introduced learned index benefit estimators~\cite{ding2019ai, shi2022learned, yu2024refactoring}, which leverage historical execution data to predict the potential benefit of various index configurations. These learned models, typically trained offline using large amounts of query execution telemetry data, have shown promise in improving the quality of recommended indexes. 
However, they face significant limitations when it comes to \emph{online} index tuning, which is arguably a more common and more challenging scenario in real-world applications~\cite{das2019automatically,YadavVZ23}. 
In online scenarios, workload drifts can occur from time to time with incoming new queries that may not be observed in the past workload execution history. 
Without appropriate and timely adjustments, pre-trained index benefit estimators can hardly generalize well for the unseen queries.

We propose \sys, an online index tuning framework equipped with a learned index benefit estimator and an index selector specifically customized for online scenarios.
Instead of having a holistic query-level index benefit estimator, as was proposed by recent work~\cite{shi2022learned}, in \sys we maintain a learned cost model for each individual type of relational operator.
Although operator-level learned cost models have been studied in the literature~\cite{AkdereCRUZ12-brown-icde,LiKNC12,Wu25}, their application to online index tuning has not yet been explored.
Our motivation of using operator-level instead of query-level model is driven by the observation that the amount of query execution time feedback collected in online index tuning is often limited.
Compared to query-level model, 
with limited training data, the operator-level model often exhibits better generalization capability for unseen queries~\cite{AkdereCRUZ12-brown-icde,LiKNC12}, a fundamental and inevitable challenge faced by online index tuners/advisors.

\begin{figure}
    \centering
    \includegraphics[width=\linewidth]{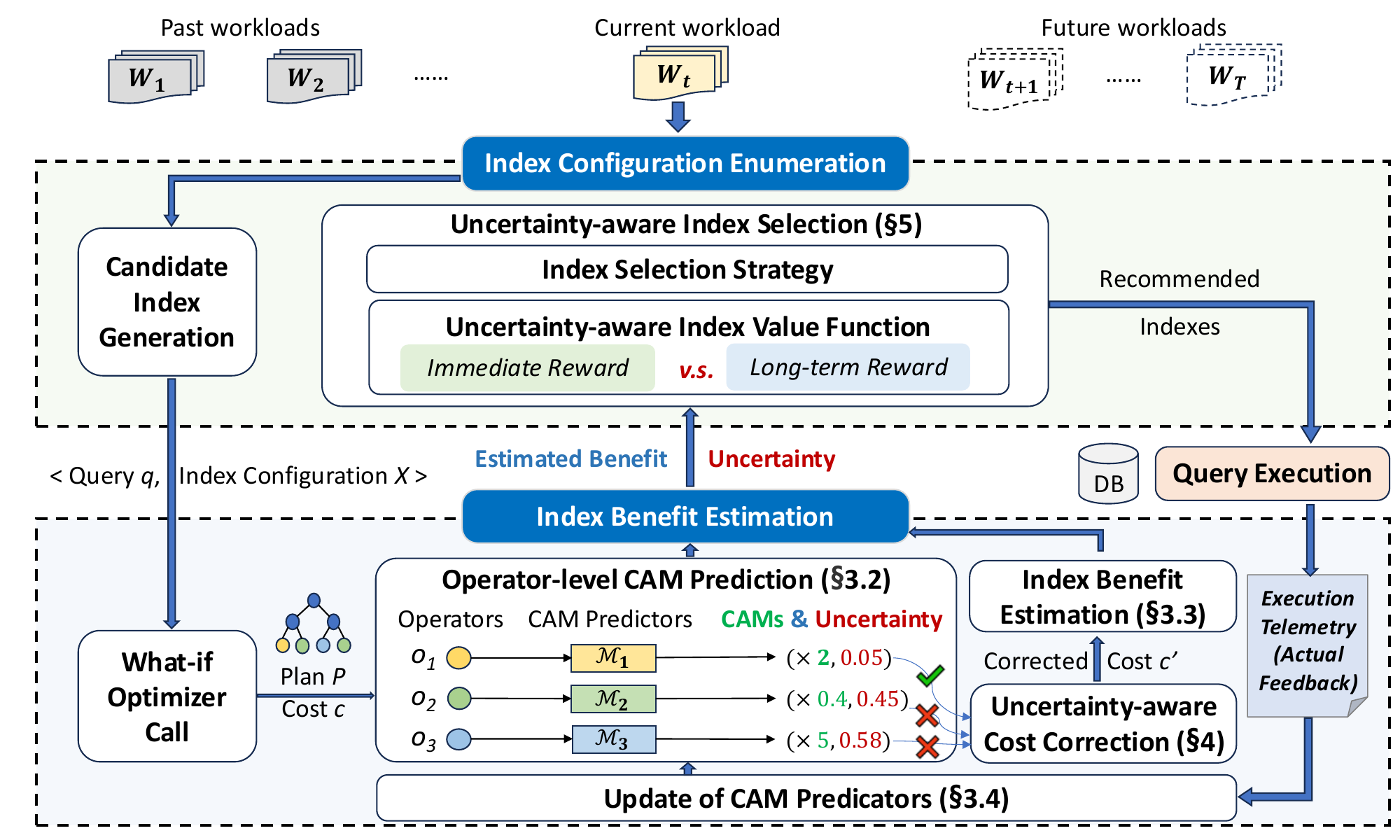}
    \vspace{-1.5em}
    \caption{Overview of \sys.}
    \vspace{-0.5em}
 \label{fig:system-overview}
\end{figure}

At the core of \sys is an \emph{uncertainty quantification} mechanism that measures the reliability of the operator-level cost models.
Considering model uncertainty plays an important role in the effectiveness of \sys, as it both improves the cost estimates themselves and reduces the exploration overhead of RL-based online index selection, a paradigm that has been adopted by a stream of recent work on online index tuning~\cite{perera2021dba, zhou2022autoindex, perera2023no, sharma2022indexer++,sadri2020online}, which \sys also follows (see Figure~\ref{fig:system-overview}).
Below, we provide a brief overview of this uncertainty-driven design underlying \sys.

\paragraph*{Uncertainty-aware Cost Estimation Correction}
Existing work on index benefit estimation often assumes the availability of abundant historical execution data~\cite{shi2022learned}, which is often impractical in real-world online settings where proactive data collection can be prohibitively expensive~\cite{ding2019ai,Wu25},
and future workloads cannot be observed in advance.
As a result, in reality, one may have to start online index tuning from scratch.
This implies a \emph{cold-start} phase during which the ML-based index benefit estimator is not reliable due to the lack of execution feedback.
This issue can be further compounded by frequent workload drifts, which can damage the fidelity of past execution data, and thus the usability of the trained ML models.
There has been recent work on modeling index tuning as contextual bandits~\cite{oetomo2024warm,oetomo2023cutting,zhang2019warm} that attempted to address this ``cold-start'' challenge by incorporating historical knowledge and transfer learning to better initialize the RL models. 
However, this line of work remains dependent on the simplified assumptions of a linear function between index features and their benefits, which can fail to capture more complex index benefit relationships present in real-world workloads.

Unlike previous work, \sys mitigates the ``cold-start'' challenge with an uncertainty-aware cost estimation correction mechanism.
Specifically, we train and maintain a model for each (type of) relational operator to predict its desired ``cost adjustment multiplier'' (CAM), which is the correction factor applied to the operator's what-if cost estimate.
That is, instead of asking for the absolute execution time prediction of an operator, we ask for the \emph{degree of adjustment/correction} that it requires.
In this way, we are not completely \emph{clueless} in the ``cold-start'' phase; rather, we can start by relying on the what-if cost estimate and gradually transit to relying on learned cost estimates via CAM-based cost correction.
However, it is important to determine the \emph{reliability} of the learned CAM during this transition.
For example, if the learned CAM is based on a model with very little execution feedback, then it is difficult to trust.
Therefore, we develop a metric to quantify the \emph{uncertainty} of each CAM model and only update the operator costs when the corresponding models have sufficiently \emph{low uncertainty} (i.e., high confidence).
\paragraph*{Uncertainty-aware Index Selection}
Most state-of-the-art online index tuners adopt RL-based technologies that essentially make trade-offs between \emph{exploration} and \emph{exploitation}~\cite{perera2021dba, zhou2022autoindex, perera2023no, sharma2022indexer++,sadri2020online}.
This ``trial-and-error'' behavior often incurs significant computational overhead by creating and dropping indexes only for exploration purposes.
The uncertainty metric developed for measuring the reliability of an operator-level CAM model offers new opportunities to reduce the exploration overhead of existing RL-based online index tuners.
Specifically, we propose a new \emph{index value function} that considers both the estimated index benefit and its uncertainty.
Given that RL-based online index tuners aim for an optimal policy that decides the next index to be selected for exploration, this new index value function 
can be easily integrated to revise existing policies.
In particular, in \sys we develop a new variant of the classic $\epsilon$-greedy policy~\cite{sutton2018reinforcement} that explores candidate indexes with probabilities proportional to their new uncertainty-weighted benefits.
Factoring model uncertainty into index selection reduces the chance of exploring indexes for which models already have high confidence (i.e., low uncertainty), saving the computational overhead of creating and dropping such indexes that would bring little extra information beyond what the models have already learned.
This is in the same spirit of \emph{active learning}.
\vspace{0.5em}
To summarize, this paper makes the following contributions:
\begin{itemize}[leftmargin=*]
    \item We propose \sys, an uncertainty-aware online index tuning framework that employs operator-level cost models for more effective utilization of limited query execution feedback and better generalization capability over unseen queries observed during workload drifts (Section~\ref{sec:overview}).
    \item We propose measuring the uncertainty of operator-level cost models and incorporating model uncertainty into both cost estimation (Section~\ref{sec:uncertainty:cost-correction}) and index selection (Section~\ref{sect:enumeration}), which not only improves the reliability of the estimated costs but also reduces the exploration overhead of RL-based online index selection.
    \item We evaluate \sys on top of three benchmarks \textbf{TPC-H}, \textbf{TPC-DS}, and \textbf{JOB}, with a variety of static and dynamic workloads, and our evaluation results show that \sys can significantly outperform existing state-of-the-art online index tuners in terms of the overall workload execution time improvement (Section~\ref{sec:evaluation}).
\end{itemize}

From a more broad viewpoint, given the inherent dynamism of online index tuning, we believe that uncertainty is an ubiquitous notion that can be factored into other components of an index tuner beyond the scope that has been studied in this paper.
For instance, while in \sys we reuse the super-arm selection method from \emph{DBA bandits}~\cite{perera2021dba} for configuration enumeration, it is perhaps worthwhile to further develop an uncertainty-aware configuration enumeration method itself.
As another example, there has been work on applying ML-based \emph{workload forecasting} to online index tuning~\cite{ma2018query}.
One may want to further consider the uncertainty in such workload forecasting models and use that to influence the index selection strategies of RL-based index tuners.
Furthermore, uncertainty is not restricted to the operator-level cost models used by \sys---existing and/or future learned index benefit estimators may also be improved by taking model uncertainty into consideration.
In this regard, we view this work as just the start of a new direction for fertile ground of future work.

\section{An Overview of \sys}
\label{sec:overview}

Online index tuning aims at minimizing the execution time of incoming workload queries by deploying a set of indexes subject to certain constraints, such as the maximum number of indexes allowed or the maximum storage space that can be taken by the deployed indexes.
We start by giving a formal problem definition.
We then present an overview of \sys.

\vspace{-0.5em}
\subsection{Problem Formulation}

\begin{table}[t]
    \centering

    \caption{Summary of notations} 
    \vspace{-0.5em}
    \label{tab:symbols}
    \begin{tabularx}{\linewidth}{lXc}
    \toprule
       \textbf{Symbol}  &  \textbf{Description} & \textbf{Section} \\ 
       \midrule
       $X_t$  & Index configuration at time $t$  & 2.1\\
       $W_t$ & Mini-workload at time $t$  & 2.1\\
       $K$ & The maximum number of indexes allowed & 2.1\\
       $C_{\text{exe}}(W_t,X_t)$ & Execution time of $W_t$ under $X_t$ & 2.1\\
       \midrule
       $\mathcal{M}$ & Operator-level CAM predictor & 3.2\\ 
       $c(q, X)$ & Estimated cost of the query $q$ with the index configuration $X$ & 3.3\\
       $b_c(X,q)$ & Estimated benefit of the index configuration $X$ for the query $q$ & 3.3\\
       $t(q, X)$ & Actual execution time of the query $q$ with the index configuration $X$ & 3.4\\
       $b_t(X,q)$ & Actual benefit of the index configuration $X$ for the query $q$ & 3.4\\ 
       \midrule
       $U(o,\mathcal{M})$ & Uncertainty of $\mathcal{M}$ for the operator $o$  & 4.1\\
       $\alpha$ & Combination weight of data uncertainty and model uncertainty & 4.1 \\
       $\rho$ & Uncertainty threshold  & 4.2\\
       \midrule
       $EB(x,W)$ & Estimated benefit of the index $x$ for the mini-workload $W$ & 5.1\\
       $EV(x,W,\mathcal{M})$ & Exploratory value of the index $x$ for the mini-workload $W$ w.r.t. $\mathcal{M}$ & 5.1\\
       $V(x,W)$ & Total value of the index $x$ for the mini-workload $W$ & 5.1\\
       $\lambda$ & Exploration weight in index selection & 5.1\\
       $\gamma$ & Decay factor for index exploration & 5.2\\
       \bottomrule
       
    \end{tabularx}
    
\end{table}

Following existing work~\cite{BrunoC07,perera2021dba}, we model the input workload as a \emph{sequence} of \emph{mini-workloads} $\mathcal{W}=(W_1, W_2,...,W_T)$ up to some time $T$, where each mini-workload $W_t$ represents a set of queries and their corresponding frequencies for $1\leq t\leq T$.
We define $\mathcal{X}$ as the set of all possible indexes that could be created.
An index \emph{configuration} $X$ is a subset of $\mathcal{X}$.
Table~\ref{tab:symbols} summarizes the notation.

Our goal is to recommend an index configuration $X_t$ for each $W_t$ to minimize the execution time $C_{\text{exe}}(W_t, X_t)$. 
The recommended $X_t$ is subject to the following constraints: (1) $|X_t|\leq K$, where $K$ is the maximum number of indexes allowed; and (2) $X_t$ is chosen based on only observing the history, i.e., the \emph{execution telemetry} of the mini-workloads $\{W_i\}_{i=1}^{t-1}$ w.r.t. the corresponding index recommendations $\{X_i\}_{i=1}^{t-1}$, and the \emph{characteristics} of the present mini-workload $W_t$.
We can now define the optimization problem associated with online index tuning as to select $X_1$, ..., $X_T$ to minimize
\begin{displaymath}
    \sum\nolimits_{t=1}^{T}C_{\text{exe}}(W_t,X_t), \quad s.t.\ |X_t| \leq K \ \ \forall t\in\{1,2,...,T\}.
\end{displaymath}





\subsection{Workflow of \sys}


As shown in Figure~\ref{fig:system-overview}, 
\sys contains two main components: (1) index benefit estimation and (2) index configuration enumeration.
Below is an overview of the two components.

\paragraph*{Index Benefit Estimation}
\sys employs operator-level cost models for better generalization over unseen queries.
For a given pair of query and index configuration, we first make a what-if call to obtain the corresponding query plan and its estimated cost.
We then traverse the query plan to refine the estimated costs of individual operators, based on the corresponding operator-level learned cost models.
Specifically, each operator-level learned cost model will output a ``cost adjustment multiplier'' (CAM), and we multiply the CAM with the what-if cost of the operator as its \emph{corrected cost}.
Given that the operator-level learned cost models are often trained with limited amounts of query execution feedback, we develop an uncertainty quantification mechanism to measure the model uncertainty.
We will only correct the what-if cost of an operator if its corresponding model has low uncertainty.
We present the details of the operator-level CAM predictors and the uncertainty-aware cost correction mechanism in Sections~\ref{sec:operator-level:cam-pred} and~\ref{sec:uncertainty:cost-correction}, respectively.

\paragraph*{Index Configuration Enumeration}
\sys follows a typical online index tuning workflow that involves candidate index generation, exploration, and selection.
For the current mini-workload $W_t$, \sys first generates candidate indexes based on well-known technologies to find \emph{indexable columns}~\cite{chaudhuri1997efficient}, e.g., columns that appear in filter or join predicates.
It then selects a subset (i.e., a configuration) from the candidate indexes that aims to minimize the execution time of $W_t$, subject to the given constraint $K$ on the maximum number of indexes allowed.
Unlike existing RL-based online index tuners that only rely on estimated index benefit, \sys employs a novel variant of the $\epsilon$-greedy index exploration algorithm that considers both the estimated index benefit \emph{and} model uncertainty when evaluating the value of a candidate index.
This improves the effectiveness of exploration by focusing more on indexes that could further reduce model uncertainty, therefore resulting in faster convergence and lower overhead of index selection during the stable period of $W_t$.
\sys selects the top-valued indexes as the final configuration, deploys them in the database, and collects query execution telemetry data to update the operator-level CAM predictors. 

\section{Index Benefit Estimation}
\label{sec:operator-level:cam-pred}
\begin{figure*}[t]
    \centering
     \begin{subfigure}[b]{0.5\textwidth}
     \centering
        \includegraphics[height = 4.5cm]{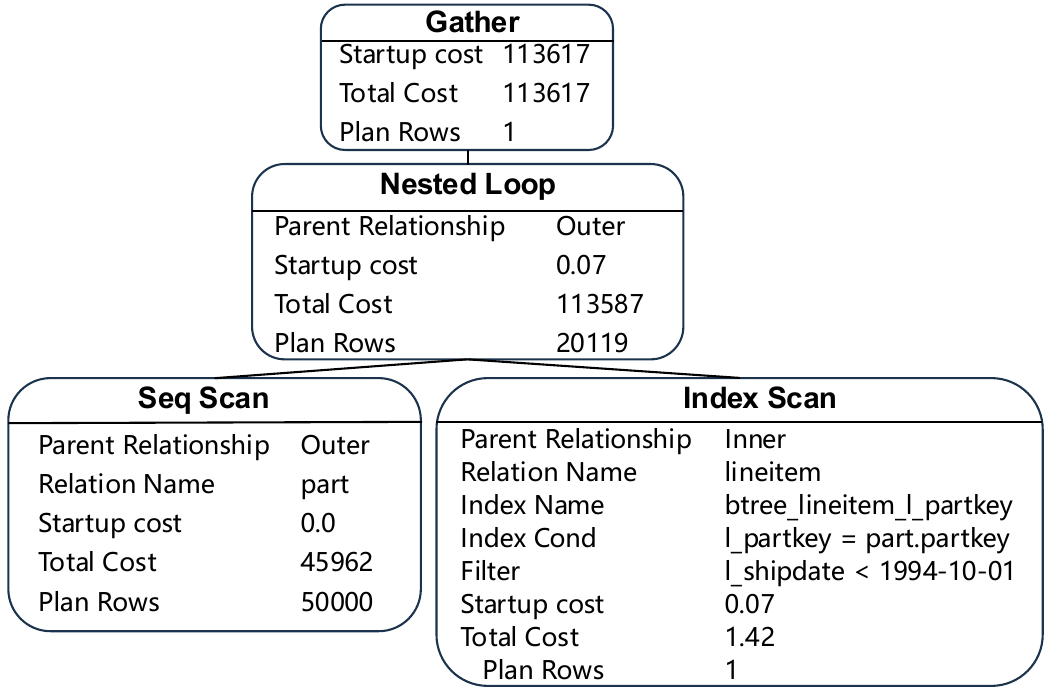}
          \caption{An example query plan}
  \label{fig:query-plan-structure}
    \end{subfigure}
    \begin{subfigure}[b]{0.45\textwidth}
    \centering
        \includegraphics[height = 4.5cm]{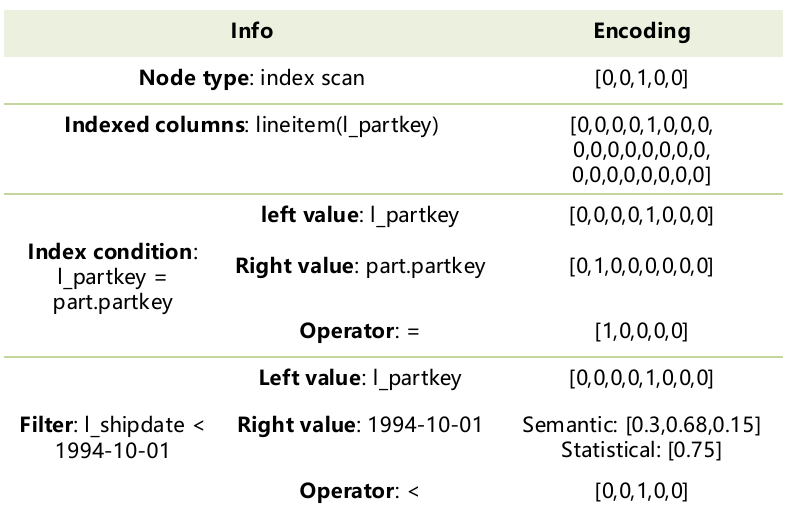}
        \caption{Encoding of the \texttt{Index Scan} operator}
  \label{fig:encoding}
    \end{subfigure}
    \vspace{-0.5em}
    \caption{Feature representation of relational operators for operator-level cost models in \sys}
    \vspace{-2em}
    \label{fig:query-plan}
\end{figure*}

Recent work on learned index benefit estimators~\cite{ding2019ai, shi2022learned, yu2024refactoring} has been done to address the inaccuracy of what-if cost estimates. 
However, such learned index benefit estimators require a considerable amount of training data that is typically unavailable in online index tuning.
Moreover, they were designed for offline index tuning where workload is presumed to be fixed and they do not offer agile model update mechanisms to deal with workload drifts.

To address the above ``data starvation'' and workload drift challenges in online index tuning, unlike existing learned index benefit estimators that take the entire query plan as input (referred to as query-level models in this paper), we propose having one learned model for each different type of (relational) operator.
Although the idea of learning operator-level cost models has been studied in the literature~\cite{AkdereCRUZ12-brown-icde,LiKNC12,Wu25}, it has not yet been applied to online index tuning to the best of our knowledge.
Moreover, the specifics of our learned operator-level cost modeling techniques are different from existing work, as we will detail in this section.

\rwfour{Operator-level models are structurally more sample-efficient than query-level models. 
Query-level models map query plan features 
directly to query execution time.
They suffer from the credit assignment problem~\cite{marcus2019Neo}, as it is difficult for query-level models to discern which part of the plan contributes to the query-level cost/latency estimation error.
As a result, query-level models often require a considerable number of training samples to converge. 
In contrast, operator-level models decompose the estimation error and execution feedback, 
and apply them directly to the corresponding operators. 
This allows for finer-grained error correction, enabling operator-level models to learn efficiently even from the sparse training samples available in online tuning (Section~\ref{sect:estimation:CAM}).
Moreover, 
query-level models often struggle to deal with workload drift because they perceive new query templates as out-of-distribution samples.
Operator-level models are known to generalize better for unseen queries from new templates~\cite{AkdereCRUZ12-brown-icde,WuCZTHN13,Wu25}.}

\vspace{-0.5em}
\subsection{Featurization of Relational Operators}\label{section:operator-level:encoding}

The raw feature representation of a relational operator consists of two parts: \emph{structural information} (SI) and \emph{predicate information} (PI).
Structural information includes the operator type and the table columns that the operator processes. 
For example, the structural information for the \texttt{Index Scan} in Figure~\ref{fig:query-plan-structure} is represented as 
$$\{\textbf{node\_type}: \text{index\_scan}, \textbf{indexed\_column}: \text{l\_partkey}\}.$$

\paragraph*{Structural information}
We use \emph{one-hot encoding} to encode the node type.
For indexed columns, we apply one-hot encoding to each column and concatenate the resulting vectors in sequence to preserve their \emph{positional information}. To ensure a fixed input length, we set a maximum index width (which is 3 in our experimental evaluation) and zero-pad shorter indexes.
\vspace{-0.5em}
\paragraph*{Predicate information}
This includes both filter and join conditions, represented as $\langle \text{column}, \text{operator}, \text{value/column} \rangle$ triplets. 
\emph{`Table columns'} and \emph{`comparison operators'} (e.g., $=$, $>$, $<$, $>=$, $<=$, $!=$,\texttt{LIKE}, and \texttt{SIMILAR TO}) are one-hot encoded.
\emph{`Values'} are min-max normalized for numbers, while strings are encoded using both semantic and statistical information.
We encode its semantic information with a \texttt{word2vec}~\cite{grohe2020word2vec} model pre-trained on strings appearing in the initial workload $W_0$.
For unseen predicates with string value $s$ during workload drifts, we search the vocabulary for the entry $v^*$ that shares the longest prefix with $s$, and then use the embedding \texttt{word2vec}($v^*$) to represent $s$, thereby allowing the model to generalize to previously unseen predicate values based on their lexical similarity to known strings.
Statistical information is captured via normalized cardinality rank. Figure~\ref{fig:encoding} illustrates the encoding of the operator \texttt{Index Scan} in the query plan of Figure~\ref{fig:query-plan-structure}.

\vspace{-0.5em}
\subsection{Cost Adjustment Multipliers}
\label{sect:estimation:CAM}
Rather than having an operator-level model to directly predict the execution cost/time of the operator as in previous work~\cite{AkdereCRUZ12-brown-icde,LiKNC12,Wu25}, an operator-level model in \sys instead predicts a \emph{cost adjustment multiplier} (CAM) to correct the query optimizer's original cost estimate of the operator. 
This is motivated by the observation that the query optimizer often exhibits similar degrees of cost estimation errors for similar operators (e.g., \texttt{Index Scan}s on similar indexes with similar predicates), even though their true execution costs may differ across queries. 
For example, an \texttt{Index Scan} used as the inner child of a \texttt{Nested-Loop Join} can incur much higher cost than standalone execution, due to repeated loops. However, in both standalone and nested scenarios, the query optimizer may underestimate the execution cost by a similar factor.
As a result, learning CAMs that capture these systematic estimation biases helps the model generalize better across queries.

We frame the operator-level CAM prediction problem as a multi-class classification task. 
Specifically, given a relational operator encoding $o$ based on the featurization in Section~\ref{section:operator-level:encoding}, the CAM predictor $\mathcal{M}$ selects one CAM from a given set $\Omega$.
Formally, $\mathcal{M}$ represents a function
$\mathcal{M}(o)=\omega$, for some $\omega\in \Omega$.
The distribution of $\omega\in\Omega$ needs to cover both underestimation and overestimation errors up to certain degrees.
As a concrete implementation, we consider the following buckets of estimation errors: (1) $\Omega_1=\{0.01, 0.02, ..., 0.09\}$, (2) $\Omega_2=\{0.1, 0.2, ..., 0.9\}$, (3) $\Omega_3=\{1,2,...,9\}$ and (4) $\Omega_4=\{10, 20, ..., 100\}$.
We set $\Omega=\cup_{i=1}^{4}\Omega_i$, and we use a simple multi-layer perceptron (MLP) classifier as $\mathcal{M}$. 
\rwfour{
This design of $\Omega$ follows a \emph{non-uniform quantization} strategy. We use buckets with denser incremental steps for small estimation errors ($\Omega_2$ for underestimation and $\Omega_3$ for overestimation) and buckets with wider incremental steps for large estimation errors ($\Omega_1$ for underestimation and $\Omega_4$ for overestimation).
This scheme provides both precise adjustments for relatively accurate query optimizer estimates (e.g., for \textbf{TPC-H}) and drastic correction of gross estimation errors (e.g., for \textbf{JOB}).
}

\paragraph*{Remark}
An alternative approach is to model CAM prediction as a \emph{regression} problem instead of a multi-class classification problem.
\rwfour{
The core reason for the classifier's superiority lies in the simplification of the learning objective. Instead of strictly minimizing error for every individual data point, the classifier focuses solely on learning the boundaries of the decision. 
This formulation aligns better with index recommendation, where identifying the relative magnitude of index benefit is sufficient for optimal decision-making~\cite{Wu25}. 
By relaxing the requirement from precise numerical estimation to coarse-grained quantization, the classification-based CAM model achieves better learning efficiency and is more robust to noise in online index tuning.
We have further implemented a regression-based CAM predictor and compared it with our classification-based approach (Section~\ref{sect:exp:model-validation}).
}

\vspace{-0.5em}
\subsection{Index Benefit Estimation}
\label{sect:estimation-infer}
To estimate the index benefit, we aggregate the corrected costs of all individual operators in the query plan.
For a query $q$ with a proposed (hypothetical) index configuration $X$, the optimizer's what-if call yields a plan $P(q,X)$ with cost $c(q,X)$.
Let $o_1, \ldots, o_p$ be the different types of relational operators in $P(q, X)$, with $\mathcal{M}_j$ as the CAM predictor for $o_j$ ($1 \leq j \leq p$).
After applying operator-level cost correction for each $o_j$ based on its corresponding $\mathcal{M}_j$ (details are presented in Section~\ref{sec:uncertainty:cost-correction}), we get the corrected plan cost $c'(q, X)$. The estimated index benefit is then
$$b_c(X,q) = \frac{c(q,\emptyset)-c'(q,X)}{c(q,\emptyset)}=1-\frac{c'(q,X)}{c(q,\emptyset)},$$
where $c(q, \emptyset)$ is the estimated cost of $q$ without indexes.

\vspace{-0.5em}
\subsection{Update of CAM Predictors}
\label{section:benefit-estimation:update}
Unlike offline index benefit estimators that only need one-time training, the operator-level CAM predictors that we proposed for online index tuning require continuous updates as more query execution feedback is available.
Although we can obtain operator-level execution telemetry (e.g., by utilizing the \texttt{EXPLAIN ANALYZE} command of PostgreSQL), it is difficult to use this execution feedback directly since the what-if cost and execution time are not based on the same \emph{unit of measurement}.
One solution could be to perform a calibration of the query optimizer's cost modeling system~\cite{WuCZTHN13}, which would incur non-trivial computation overhead. We instead propose a low-overhead approach to effectively use the query-level execution feedback without extra calibration. 

We choose to focus on \emph{leaf} (table access) operators on \emph{indexed} columns, such as \texttt{Sequential Scan}, \texttt{Index Scan}, \texttt{Index Only Scan}, or \texttt{Bitmap Index scan} in PostgreSQL, due to their dominant impact on query performance in the context of index tuning~\cite{Wu25}.
For each $X$-related operator in the query plan $q$, we independently correct its cost estimation to make the estimated index benefit $b_c(X, q)$ close to the \emph{actual index benefit}, which is naturally defined as
$$b_t(X,q) = \frac{t(q,\emptyset)-t(q,X)}{t(q,\emptyset)}=1-\frac{t(q,X)}{t(q,\emptyset)}.$$
Here, $t(q, X)$ is the actual execution time of $q$ with $X$.

Algorithm~\ref{Algo:feedback-analysis} illustrates this execution feedback utilization in detail.
\rwfive{It aims to identify an appropriate CAM that aligns estimated and actual index benefits after cost correction~(line 7). This is realized through a bottom-up propagation mechanism \texttt{UpdateCost} of cost corrections~(line 5).}
The details of \texttt{UpdateCost} can be found in Algorithm~\ref{Algo:update-cost}.

\begin{algorithm}
\caption{Execution Telemetry Utilization}
\label{Algo:feedback-analysis}
\small
\KwIn{$P$, query plan with execution time feedback; $X$, index configuration; $\Omega$, candidate CAMs.}
\KwOut{$L=[(o_1,\omega_1),(o_2,\omega_2),...,((o_{|\mathcal{O|}},\omega_{|\mathcal{O|}})]$, a list of operator-CAM pairs for model updating.}
$\mathcal{O}\leftarrow X\text{-related}  \ \text{leaves on P}, \quad L \gets \emptyset$\;
\ForEach{$o \in \mathcal{O}$}{
$\delta\leftarrow|b_t(X,q) - b_c(X,q)|$, $\omega\leftarrow 1$\;
\ForEach{$\omega'\in \Omega$}{
    $\texttt{UpdateCost}(P,o,\omega')$\;
    $b_{c'}(X, q)\leftarrow 1-\frac{c'(q, X)}{c(q,\emptyset)}$\;
    \If{$|b_t(X,q) - b_{c'}(X,q)| < \delta$}{
        $\omega\leftarrow \omega'$\;
        $\delta\leftarrow|b_t(X,q) - b_{c'}(X,q)|$\;
    }
}
$L \gets L \cup (o,\omega)$
}
\Return $L$\;
\end{algorithm}

\vspace{-0.5em}
\paragraph*{Discussion}
\rwone{
Algorithm~\ref{Algo:feedback-analysis} does not explicitly model \emph{cross-operator} index interactions.
Although this could have been done by training a query-level CAM model that captures cross-operator effects, it requires much more training data to cover the interaction space, which would severely delay convergence in ``data-starved'' online tuning scenarios.
Instead, \sys captures index interactions \emph{implicitly} based on its \emph{uncertainty-aware} index benefit estimation and index selection strategy. 
As detailed in Appendix~\ref{sect:exp:index-interaction}, when an index interacts with others (i.e., the index benefit on the same query diverges significantly depending on the presence or absence of another index), it manifests performance volatility in the operator-level feedback, which increases the model-level \emph{aleatoric uncertainty} (see Section~\ref{sec:uncertainty-quantify}).
The uncertainty-aware index selection strategy of \sys is then prompted to further explore such indexes in different configurations.
Eventually, when the beneficial index combination is sampled, the query performance stabilizes, leading \sys to converge.
}

\vspace{-0.5em}
\section{Uncertainty-aware Cost Correction}
\label{sec:uncertainty:cost-correction}

We now discuss how to correct the operator-level what-if costs to obtain the adjusted plan cost $c'(q, X)$, by considering the learned operator-level CAMs.
To this end, we develop an uncertainty-aware cost correction framework.

\vspace{-0.5em}
\subsection{Uncertainty Quantification}
\label{sec:uncertainty-quantify}

Given the limited amount of query execution feedback and the constant change of workload queries, uncertainty is ubiquitous in online index tuning.
To quantify the degree of uncertainty with respect to the operator-level CAM predictors, we consider two major sources of uncertainty:

\begin{itemize}[leftmargin=*]
    \item \textit{Data uncertainty} (a.k.a. \emph{aleatoric uncertainty}~\cite{deep2020comprehensive}), which stems from the inherent noise and variability in online workload execution, due to environmental factors such as caching and resource contention. 
    As a result, the same operator may observe different execution times, introducing irreducible variance in the training data.
    \item \textit{Model uncertainty} (a.k.a. \emph{epistemic uncertainty}~\cite{deep2020comprehensive}), which arises from the limitations of a model's ``knowledge.''
    For example, some operators may appear less often than others. As a result, they have fewer training data points and the corresponding operator-level models are less confident when making predictions.
\end{itemize}

\paragraph*{Quantification of Data Uncertainty}
We propose using \textbf{entropy} to measure data uncertainty, defined as follows:
\begin{definition}[Entropy]
Given a \emph{softmax} output of an operator-level CAM predictor $\mathcal{M}$ for the operator $o$, $\mathcal{M}(o) = [p_1,p_2,...,p_{|\Omega|}]$, where $p_j$ is \emph{probability} of the CAM $\omega_j\in\Omega$,
$$\text{Entropy}(o, \mathcal{M}) = -\sum\nolimits_{\omega_j\in\Omega} p_j \log(p_j).$$ 
\end{definition}
Here, we have abused the notation $\mathcal{M}(o)$ to also represent the softmax output of $\mathcal{M}$.
Intuitively, entropy measures the inherent variance observed in the execution time of an operator.
When an operator-level CAM predictor sees noisy or ambiguous data during training, its entropy increases.
In fact, the entropy is maximized when all CAM outputs are equally likely, implying a highly uncertain situation where the predicted CAM degenerates to a random guess.

\paragraph*{Quantification of Model Uncertainty}
We propose using \textbf{Monte Carlo dropout (MCD)}~\cite{gal2016dropout} to measure model uncertainty.
It allows us to assess how confident a model is about its predictions across different regions of the input space. 
Dropout is a standard regularization technique used in the \emph{training} of deep neural networks to prevent overfitting by randomly deactivating neurons.
MCD extends this dropout idea to model \emph{inference}, with a very different purpose of measuring the variance of model outputs with random probing of the structure of the neural network.


Specifically, we pass the same encoded relational operator $o$ through the CAM predictor $\mathcal{M}$ for $m$ times.
In this way, we obtain slightly different predictions: $\hat{y}_1,\hat{y}_2,...,\hat{y}_m$, where each $\hat{y}_i$ is a softmax output vector $\hat{y}_i = [p_{i,1},p_{i,2},...,p_{i,|\Omega|}]$. 
We then compute the variance for each possible CAM output $\omega_j\in\Omega$ as 
$\text{Var}(\omega_j) = \frac{1}{m} \sum\nolimits_{i=1}^m (p_{i,j} - \bar{p}_j)^2,$
where $\bar{p}_j$ represents the mean predicted probability of $\omega_j$. 
We take the maximum variance across the CAM outputs to obtain the MCD:
\begin{displaymath}
    MCD(o,\mathcal{M}) =\max_{\omega_j\in\Omega} \text{Var}(\omega_j) = \max_{\omega_j\in\Omega} \frac{1}{m} \sum\nolimits_{i=1}^m (p_{i,j} - \bar{p}_j)^2. 
\end{displaymath}
Intuitively, for regions in the input space where the model has not found strong functional mappings to the outputs (e.g., due to lack of training data), MCD introduces randomness in model inference where each ``reduced network'' with randomly deactivated neurons predicts differently, resulting in higher prediction variability with the same input.

\paragraph*{Combined Uncertainty}
We combine data and model uncertainty with the following weighing mechanism:
\begin{equation}
\label{equation:uncertainty}
    U(o,\mathcal{M}) = \alpha \cdot \text{MCD}(o,\mathcal{M}) + (1-\alpha) \cdot \text{Entropy}(o,\mathcal{M}),
\end{equation}
where $0<\alpha<1$ is a user-specified weight parameter. 
We set $\alpha=0.5$ by default, meaning that we treat data and model uncertainty equally.
However, one can adjust $\alpha$ based on different situations.
For example, one may want to set $\alpha< 0.5$ if query execution exhibits high variance, implying higher data uncertainty on average.

\begin{table*}[htbp]
    \centering

    \caption{Cost formulas of relational operators supported by PostgreSQL} 
    \label{table:cost-functions}
    \vspace{-0.5em}
 
    \begin{tabularx}{0.9\linewidth}{lll}
      \toprule
      Operator $o$ & Startup Cost Variation $\Delta c_s(o)$ & Execution Cost Variation $\Delta c_e(o)$ \\
      \midrule
       \texttt{Nested-Loop Join} & $\sum \Delta c_s(\texttt{child})$ &
         $\text{rows}(\texttt{outer})\cdot\Delta c_e (\texttt{inner})$ + $\Delta c_e (\texttt{outer})$  \\
      \texttt{Limited} & $ c_s(\texttt{child})$ &  $ \frac{\text{rows}(\texttt{parent})}{\text{rows}(\texttt{child})}\cdot\Delta c_e(\texttt{child})$ \\
      \texttt{Hash,Sort,Aggregate,Gather} & $\Delta c_s$(\texttt{child}) + $\Delta c_e$(\texttt{child}) &  0 \\
      \texttt{Hash Join, Gather Merge} & $\sum \Delta c_s$(\texttt{child}) &  $\Delta c_{e}$(\texttt{child}) \\
      \bottomrule
    \end{tabularx}
    \vspace{-2em}
\end{table*}

\vspace{-0.5em}
\subsection{Cost Correction and Combination}
\begin{figure}[t]
    \centering
    \includegraphics[width=\linewidth]{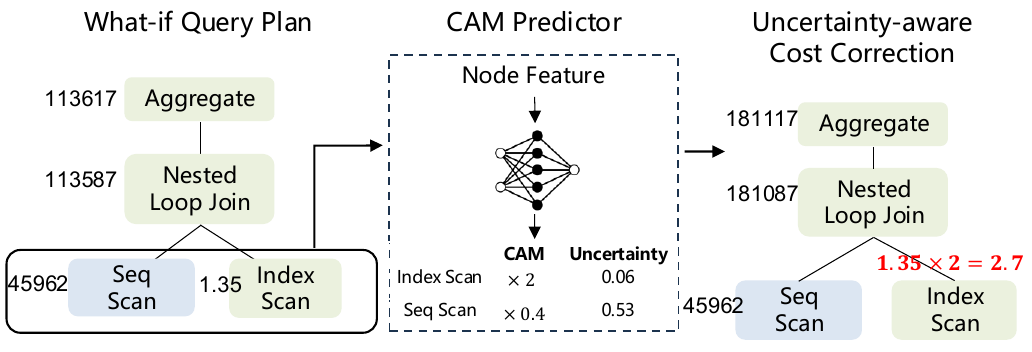}
    \vspace{-1em}
    \caption{An example of uncertainty-aware cost correction} 
    \label{fig:estimation-model}
\end{figure}

\begin{algorithm}
\caption{\texttt{CostCorrection}($P$, $\mathcal{M}$, $\rho$)}
\label{Algo:plan-estimation}
\small
\KwIn{$P$, query plan; $\mathcal{M}$, operator-level CAM predictors; $\rho$, uncertainty threshold.}
\KwOut{$c'(P)$, corrected cost of $P$.}
\For{$o \in P.\text{leaves}$}{
$U(o, \mathcal{M})\leftarrow\alpha \cdot \text{MCD}(o,\mathcal{M}) + (1-\alpha) \cdot \text{Entropy}(o,\mathcal{M})$ \;
    \lIf{$U(o, \mathcal{M}) \leq \rho$}{
        $\texttt{UpdateCost}(P, o, \mathcal{M}(o))$
    }
}
\Return $c'(P)$\;
\end{algorithm}

\begin{algorithm}
\caption{\texttt{UpdateCost}($P$, $o$, $\omega$)}
\label{Algo:update-cost}
\small
\KwIn{$o$, the operator whose cost to be corrected; $\omega$, CAM.}
\lIf{$o$ is leaf}{
$c'_e(o)\leftarrow \omega\cdot c_e(o)$
}
\lElse{
Update $c_s(o)$ and $c_e(o)$ via Table~\ref{table:cost-functions}
}
$\texttt{UpdateCost}(P, o.\text{parent}, \omega)$\;
\end{algorithm}

The uncertainty metric $U(o,\mathcal{M})$ in Equation~\ref{equation:uncertainty} results in a natural operator-level uncertainty-aware cost correction mechanism as illustrated in Algorithm~\ref{Algo:plan-estimation}.
Specifically, we introduce a threshold parameter $\rho$ to control the acceptable degree of model uncertainty.
Cost correction is applied only to \emph{leaf} (e.g., table access and index access) operators with uncertainty below $\rho$, given their prominent impact on query execution time in the context of index tuning~\cite{Wu25}.
If so, we invoke the \texttt{UpdateCost} function detailed by Algorithm~\ref{Algo:update-cost} to update its cost as well as its parent's cost recursively.

For each operator $o$, we update its cost by following an analytic approach~\cite{WuCZTHN13}.
We maintain its \emph{startup cost} $c_s(o)$ and \emph{execution cost} $c_e(o)$, and we only correct the execution cost $c_e(o)$ based on the given CAM.
The total cost of $o$ is the sum of $c_s(o)$ and $c_e(o)$.
Table~\ref{table:cost-functions} presents the formulas supported by PostgreSQL. 
Specifically, we can express the \emph{variation} (i.e., change) of the startup cost (resp. execution cost) of an operator $o$, denoted as $\Delta c_s(o)$ (resp. $\Delta c_e(o)$), in terms of the cost variations of its child operator(s).
For example, for PostgreSQL’s \texttt{Nested-Loop Join} operator, the startup cost variation is the sum of its children’s variations, while the execution cost variation also depends on the number of rows from the outer child.
After computing the variations in the start and execution costs of an operator $o$ via bottom-up propagation, we can simply update $c_s(o)\leftarrow c_s(o)+\Delta c_s(o)$ and $c_e(o)\leftarrow c_e(o)+\Delta c_e(o)$.

Below, we illustrate how to use the cost formulas for propagating leaf cost corrections (or equivalently, cost variations) with a concrete example based on the query plan in Figure~\ref{fig:query-plan-structure}.

\begin{example}
\label{example:correction-process}
Figure~\ref{fig:estimation-model} illustrates the process of uncertainty-aware cost correction on a simplified query plan shown in Figure~\ref{fig:query-plan}.
    The predicted CAMs for leaf operators \texttt{Seq Scan} (SS) and \texttt{Index Scan} (IS) are $0.4$ and $2$, with uncertainty scores $0.53$ and $0.06$. 
    With threshold $0.1$, only \texttt{IS} is corrected:
    $$c_e(\texttt{IS}) = 1.35,\quad
c’_e(\texttt{IS}) = 1.35 \times 2 = 2.70,\quad
\Delta c_e(\texttt{IS}) = 1.35.$$
The corrected cost estimation of \texttt{IS} is then propagated to the parent operator \texttt{Nested-Loop Join} (NLJ) according to the cost formulas in Table~\ref{table:cost-functions}:
\begin{eqnarray*}
c_e'(\texttt{NLJ}) & = & c_{e}(\texttt{NLJ}) + \text{rows}(\texttt{SS})\cdot\Delta c_e(\texttt{IS}) + \Delta c_e(\texttt{SS})\\
&= & 113587 + 50000\times 1.35 + 0 = 181087,
\end{eqnarray*}
\end{example}


One question is how to specify the uncertainty threshold $\rho$.
In our current implementation of \sys, we choose a uniform uncertainty threshold $\rho=0.1$ based on the observed distribution of data and model uncertainty.
It remains an interesting direction for future work to explore more adaptive ways of varying the uncertain threshold $\rho$ as the process of online index tuning proceeds.

\label{sect:classifier:second-stage}

\vspace{-0.5em}
\section{Uncertainty-Aware Index Selection}
\label{sect:enumeration}

Existing online index tuners typically employ RL-based search strategies to balance \emph{exploration} and \emph{exploitation}, which either directly rely on execution feedback or use learned models for reward estimation.  
While direct usage of observed execution feedback provides accurate estimation, it lacks generalization to unseen queries. Learned models improve generalization, but existing approaches~\cite{perera2021dba, wu2022budget, kossmann2022swirl} do not consider model uncertainty as far as we know.

In this section, we show how the uncertainty quantification mechanism in Section~\ref{sec:uncertainty-quantify} can be used to improve RL-based index selection.
We first propose a new uncertainty-aware index value function that rebalances exploration with exploitation.
We then design a variant of the classic $\epsilon$-greedy action/index selection strategy by integrating the uncertainty-aware index value function.

\vspace{-0.5em}
\subsection{Uncertainty-aware Index Value Function}
\label{sect:selection-policy}

We propose an online index value function that dynamically balances between the \emph{immediate reward} (i.e., improvement in workload execution time by deploying an index) and the \emph{long-term reward} (i.e., improvement in model prediction accuracy with the new workload execution time feedback).

\paragraph*{Immediate Reward}
We define the immediate reward of deploying an index $x$ as its \emph{execution benefit} $\text{EB}(x,W)$, i.e., the improvement on workload execution time:
\begin{equation*}
\small
\resizebox{\hsize}{!}{$
    \text{EB}(x,W) =  \frac{\sum\nolimits_{q\in W}c(q,\emptyset)-\sum\nolimits_{q\in W}c(q,x)}{\sum\nolimits_{q \in W} c(q,\emptyset)}=1-\frac{\sum\nolimits_{q\in W}c(q,x)}{\sum\nolimits_{q \in W} c(q,\emptyset)}.  
$}
\end{equation*}

\paragraph*{Long-term Reward}
We define the long-term reward of deploying an index $x$ as its \emph{exploratory value} $\text{EV}(x, W, \mathcal{M})$, i.e., the potential improvement in model prediction.
This is inspired by \emph{active learning}: the more \emph{uncertain} the model $\mathcal{M}$ is about its predictions on (what-if) query plans containing the index $x$ as access paths, the more \emph{valuable} the workload execution time feedback is if we create the index $x$.
In this spirit, we define
$\text{EV}(x, W, \mathcal{M}) = \sum\nolimits_{o\in\mathcal{O}_x} U(o, \mathcal{M}),$
where $U(o, \mathcal{M})$ is the uncertainty metric defined by Equation~\ref{equation:uncertainty}, and $\mathcal{O}_x$ represents the set of all relevant relational operators that access the index $x$.

\paragraph*{Index Value Function}

We now define the \emph{total value} $V(x,W)$ of a candidate index $x$ for the workload $W$ by combining $\text{EB}(x,W)$ and $\text{EV}(x,W,\mathcal{M})$ as follows:
\begin{equation}
\label{equation:total_value}
    V(x,W)= \text{EB}(x,W)\cdot\Big(1+\lambda \cdot\text{EV}(x,W,\mathcal{M})\Big).
\end{equation}
The $\lambda\in(0, 1)$ here is an \emph{exploration weight} that further controls the degree of exploration as online tuning proceeds.
We multiply $\text{EB}(x, W)$ and $\text{EV}(x, W)$ instead of adding them, due to their different semantics.
Intuitively, one can think of $\text{EV}(x, W, \mathcal{M})$ as an uncertainty-based weight placed on $\text{EB}(x,W)$, and the total value is simply the uncertainty-adjusted index benefit.

\vspace{-0.5em}
\subsection{Index Selection Strategy}
We develop a new variant of the classic $\epsilon$-greedy search strategy~\cite{sutton2018reinforcement} for index/action selection by integrating the index value function $V(x,W)$.
In standard $\epsilon$-greedy algorithm, the best action is chosen with probability $1-\epsilon$, and other actions are selected uniformly randomly with probability $\epsilon$, making no distinction between suboptimal choices. 
\emph{Boltzmann exploration}~\cite{peret2004line} addresses this by choosing an action based on its softmax probability w.r.t. the estimated action value. However, it requires tuning a temperature parameter $\tau$.
To retain the distinction capability of Boltzmann exploration without the need of tuning $\tau$, following~\cite{wu2022budget} we redefine the index selection probability 
$\Pr(x) = \frac{V(x, W)}{\sum_{x'\in\mathcal{X}} V(x', W)}.$
That is, we pick an index with probability proportional to its estimated index value.

\paragraph*{Dynamic Adjustment of Exploratory Value}
We further adjust the exploratory value using the exploration weight $\lambda$.
Initially, with little feedback on any index, more exploration is encouraged. After we observe more and more workload execution telemetry as tuning proceeds, exploration needs to be gradually discouraged.
Therefore, we set $\lambda_t = \lambda_{0}\cdot\gamma^t$ when tuning the mini-workload $W_t$, where $\lambda_0$ is the initial exploration weight and $0<\gamma<1$ is a decay factor dictating how fast exploration diminishes.
To respond proactively to workload drift, we further introduce a dynamic reset mechanism controlled by a reset factor $\beta = 1- \frac{N_{\text{unseen}}}{N}$, which represents the proportion of unseen queries in the present mini-workload $W_t$. 
We incorporate this reset mechanism into the exploration weight as $\lambda = \lambda_0\cdot\gamma ^ {\beta t}$.




\paragraph*{Configuration Enumeration}

Algorithm~\ref{Algo:enumeration} presents the details of the index configuration enumeration algorithm based on the above variant of $\epsilon$-greedy index selection strategy.
When a candidate index $x_k$ is selected, we check whether $x_k$ can be pruned according to the following rules: (1) the table on top of which $x_k$ is recommended has reached the maximum number of indexes allowed; (2) $x_k$ can be superseded by a ``covering index'' that has already been selected; or (3) a similar index with the same prefix of (but more) key columns has already been selected.
We include $x_k$ into the final recommended configuration $X_t$ if $x_k$ cannot be pruned.


\begin{algorithm}[t]
\caption{Index Configuration Enumeration}
\label{Algo:enumeration}
\small
\KwIn{$W_t$, the current mini-workload at time $t$; $\mathcal{X}_t$, the candidate indexes for $W_t$; $K$, the maximum number of indexes to be selected;
$\mathcal{M}$, the operator-level CAM predictors; $\lambda_0 = 0.5$, the initial exploration weight; $\gamma$, the decay rate.}
\KwOut{$X_t$, the recommend index configuration for $W_t$.}
$\lambda\leftarrow\lambda_0 \cdot \gamma ^ {\beta t}$\;
\ForEach{$x\in\mathcal{X}_t$} {
    $V(x, W_t)\leftarrow \text{EB}(x,W_t)\cdot\Big(1+\lambda \cdot\text{EV}(x,W_t,\mathcal{M})\Big)$\;
}   
\ForEach{$x\in\mathcal{X}_t$} {
    $\Pr(x) = \frac{V(x, W_t)}{\sum_{z\in\mathcal{X}_t} V(z, W_t)}$\;
}
$X_t\leftarrow\emptyset$\;
\For{$1\leq k\leq K$}{
Select $x_k$ from $\mathcal{X}_t$ w.r.t. $\Pr(x_k)$\;
\lIf{$x_k$ cannot be pruned}
{
    $\mathcal{X}_t\leftarrow \mathcal{X}_t\cup\{x_k\}$
}
}
\Return $X_t$\;
\end{algorithm}

\vspace{-0.5em}
\section{Experimental Evaluation}
\label{sec:evaluation}
We evaluate the performance of \sys and compare it with existing state-of-the-art (SOTA) online index tuning solutions.


\vspace{-0.5em}
\subsection{Experimental Setup}
\label{sect:experiment:set-up}

\paragraph*{Benchmark Datasets}
We use three publicly available benchmark datasets with different scales and characteristics: \textbf{TPC-H}, \textbf{TPC-DS}, and \textbf{JOB} (a.k.a. the ``join order benchmark''~\cite{job-queries}).
Table~\ref{benchmarks} summarizes the properties of these datasets. Following previous studies~\cite{kossmann2020magic, kossmann2022swirl}, we exclude the queries 4, 6, 10, 11, 35, 41, and 95 from \textbf{TPC-DS} because their execution costs are orders of magnitude higher than the other queries.
Including these expensive queries can make the index selection problem less challenging, because indexes that can accelerate any of such queries are clearly favorable.


\paragraph*{Experiment settings}
We conduct all experiments on a workstation with two 24 Core Xeon(R) Gold 5318Y CPU at 2.10GHz and 1TB main memory, running Ubuntu 20.04.6 LTS with PostgreSQL 12.1.
We clear the database buffer pool and the file system cache before running each mini-workload.


\paragraph*{Baselines}
We compare \sys with three online RL-based index tuners/advisors and one offline index tuner/advisor specifically designed for dynamic workloads: 
\begin{itemize}[leftmargin=*]
    \item \textbf{AutoIndex}~\cite{zhou2022autoindex}, which is an online index advisor that uses Monte Carlo tree search (MCTS) for index selection and a deep regression model for index benefit estimation;
    
    \item 
    \textbf{HMAB}~\cite{perera2022hmab}, which applies hierarchical multi-armed bandits for online index selection and learns index benefit from query execution feedback; \textbf{HMAB} further extends and improves over DBA Bandits~\cite{perera2021dba}, which pioneered the idea of modeling online index tuning as contextual bandits.
    \item \textbf{Indexer++}~\cite{sharma2022indexer++}, which is an online index advisor that uses deep Q-learning (DQN) for index selection and prioritized experience sweeping for adaption to dynamic workloads;
    \item \textbf{SWIRL}~\cite{kossmann2022swirl}, which is an index advisor that uses deep reinforcement learning (DRL) to train an index selection policy offline and applies a sophisticated workload embedding model for generalization over unseen queries.
\end{itemize}

\paragraph*{Performance Metrics}
We use \emph{improvement of workload execution time} as our primary performance metric, defined as 
$\frac{\sum_{t=1}^T\Big(C_{\text{exe}}(W_t, \emptyset) - C_{\text{exe}}(W_t,X_t)\Big)}{\sum_{t=1}^T C_{\text{exe}}(W_t, \emptyset)} \times 100\%,$
where $C_{\text{exe}}(W_t, \emptyset)$ is the execution time of the mini-workload $W_t$ without indexes and $C_{\text{exe}}(W_t,X_t)$ is the execution time of $W_t$ with the recommended index configuration $X_t$. 

{\setlength{\tabcolsep}{2pt}
\begin{table}
    \centering
    \caption{Properties of the benchmark databases}
    \label{benchmarks}
    \vspace{-0.5em}
    \begin{tabularx}{\linewidth}{cccccX}
      \toprule
      Benchmark&Dataset&\#Tables&\#Attributes&\#Templates&Size (GB) \\
      \midrule
       \textbf{TPC-H} & Synthetic uniform&8&61&22&10 \\
       \textbf{TPC-DS} & Synthetic skew&24&429&99&10 \\
       \textbf{JOB} & Real-world&21&108&33&6 \\
      \bottomrule
    \end{tabularx}
\end{table}
}

\vspace{-0.5em}
\subsection{Evaluation of Index Benefit Estimation}
\label{sect:exp:index-benefit-estimation}
\rwfour{
We evaluate \sys's \emph{operator-level} index benefit estimation model to assess its accuracy, generalization capability, and impact on index tuning.
We compare \sys against two baselines: the traditional what-if cost estimator and \textbf{LIB}~\cite{shi2022learned}, a state-of-the-art \emph{learned query-level} estimator.
}
\begin{table}%
    \caption{Estimation accuracy of index benefit estimators}
    \vspace{-0.5em}
    \label{exp:estimation-error}
    \centering
    
    \begin{tabularx}{0.8\linewidth}{ccccccccccc}
      \toprule
      \multirow{2}{*}{Dataset}  &  \multirow{2}{*}{What-if} & \multicolumn{2}{c}{\textbf{LIB}}  & \multicolumn{2}{c}{\sys}  \\
      & & $40\%$ & $80\%$ & $40\%$ & $80\%$ \\ 
      \midrule
      
     \textbf{TPC-H} & 0.189 & 0.171 & 0.124 & 0.137 & 0.115 \\
    \textbf{TPC-DS} & 0.062 & 0.118 & 0.046 & 0.056 & 0.039 \\
    \textbf{JOB} & 0.757 & 0.762 & 0.531 & 0.544 & 0.386 \\

      \bottomrule
    \end{tabularx}
\end{table}

\paragraph*{Estimation Accuracy}
We use ``Mean Absolute Error (MAE)'' to measure the accuracy of an index benefit estimator, which is defined as 
$\frac{1}{n} \sum\nolimits_{i=1} ^{n} \Big(|b_c(x_i,q) - b_t(x_i,q)|\Big),$
where $\{x_i\}_{i=1}^n$ represents the set of all candidate indexes.
A smaller MAE means a more accurate estimator.

For the CAM predictors employed by \sys and the \textbf{LIB} baseline, we randomly select 40\% and 80\% of all query templates for model training, and we test the trained models using all query templates.
This is different from a standard ``train-test split'' model validation procedure, where trained models are tested using only holdout data.
\rwthree{We chose to use this setup instead of the standard one, since it is closer to the use case when deploying learned index benefit estimators for real online index tuning.}
Specifically, given that workload drifts are unpredictable, a more reasonable goal for online index benefit estimators is to generalize over all observed query templates.
The setup that we chose therefore emulates the evolving generalization behavior of online index benefit estimators as more and more query templates are observed.

Table~\ref{exp:estimation-error} summarizes the MAE results.
\rwfour{
\sys consistently outperforms both baselines, especially in the data-scarce scenario (with 40\% training data) and for complex workloads such as \textbf{JOB}.
The performance gap highlights advantages of \sys's operator-level models over the query-level models used by \textbf{LIB}, as discussed in Section~\ref{sec:operator-level:cam-pred}.}

{\setlength{\tabcolsep}{2pt}
\begin{table*}[htbp]
\caption{Static and dynamic workloads used in online index tuning evaluation}
\label{table:workload-pattern}
\vspace{-0.5em}
\centering
    {
    \footnotesize
    \begin{tabularx}{0.9\linewidth}{lXlll}
      \toprule
  Workload  & Workload Drift &  $T$ & Templates per $W_t$ & Queries per template  \\
  \midrule
Static  & No & 20 & all templates &  \makecell[l]{ 20 (\textbf{TPC-H}) \\ 10 (\textbf{TPC-DS}, \textbf{JOB})}  \\

Continuous Variation & Change of $20\%$ templates / $W_t$ &  24 & 10 (\textbf{TPC-H}),20 (\textbf{TPC-DS}, \textbf{JOB}) &  20  \\
Periodic Variation & \makecell[l]{Change of $20\%$ templates every \\ 4 (\textbf{TPC-H}) or 5  (\textbf{TPC-DS}, \textbf{JOB}) $W_t$}   & \makecell[l]{24 (\textbf{TPC-H}) \\ 30 (\textbf{TPC-DS}, \textbf{JOB})} & 10 (\textbf{TPC-H}),20 (\textbf{TPC-DS}, \textbf{JOB}) & 20    \\
Cyclic Variation & \makecell[l]{Change of $20\%$ templates / $W_t$,\\ recurring every 15 $W_t$} & 30 &  16 (\textbf{TPC-H}), 20 (\textbf{TPC-DS}, \textbf{JOB}) & 20 \\

      \bottomrule
    \end{tabularx}
    }
\vspace{-2em}
\end{table*}
}

\begin{figure}[t]
    \centering
    \includegraphics[width=0.55\linewidth]{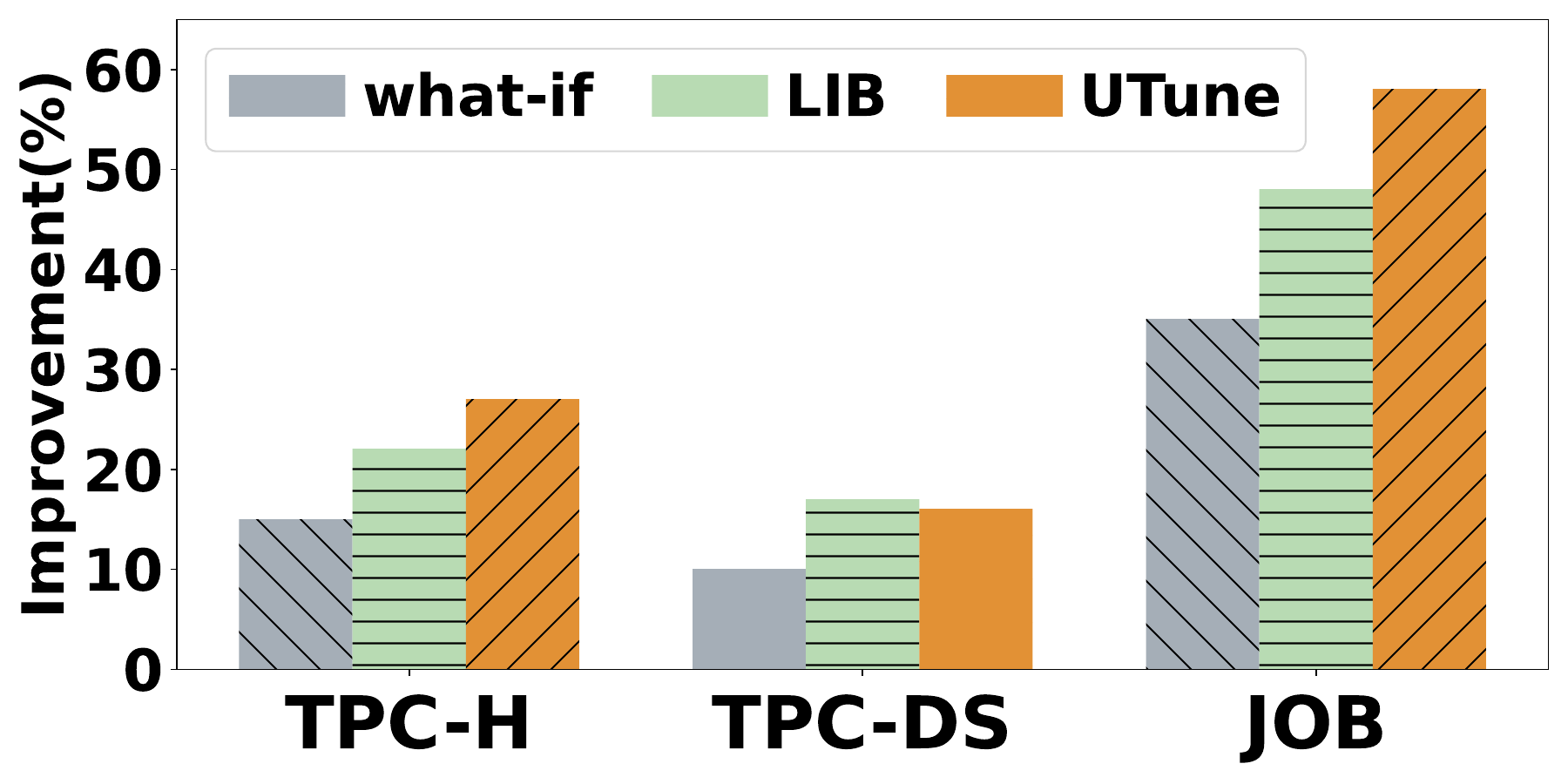}
    \vspace{-0.5em}
    \caption{Improvement of workload execution time when integrating greedy search with index benefit estimators
    }
    \vspace{-0.5em}
    \label{fig:utune-improvement}
\end{figure}

\paragraph*{Impact on Index Tuning}
We are interested in the impact of improved index benefit estimates on end-to-end index tuning.
To this end, we use the classic greedy index selection strategy~\cite{chaud1998autoadmin} and integrate it with the index benefit estimators.
We set the maximum number of indexes allowed to 8.
Figure~\ref{fig:utune-improvement} presents the index tuning results in terms of the improvement on workload execution time using recommended indexes.
Both \textbf{LIB} and \sys outperform what-if cost in selecting effective indexes with knowledge learned from actual query execution feedback data. \sys performs even better than \textbf{LIB} on \textbf{JOB} because of its success in detecting indexes that cause severe query regression. \sys is equally competitive as \textbf{LIB} in correcting what-if cost estimation errors but features a simpler structure and better generalization capability.
Therefore, it can work more effectively for online index tuning with limited amount of query execution feedback data.

\vspace{-0.5em}
\subsection{\rwfour{Validation of Design Choices of CAM Models}}
\label{sect:exp:model-validation}
\begin{figure}[t]
    \centering
     
    \begin{subfigure}[b]{0.48\linewidth}
        \centering
        \includegraphics[width=\textwidth]{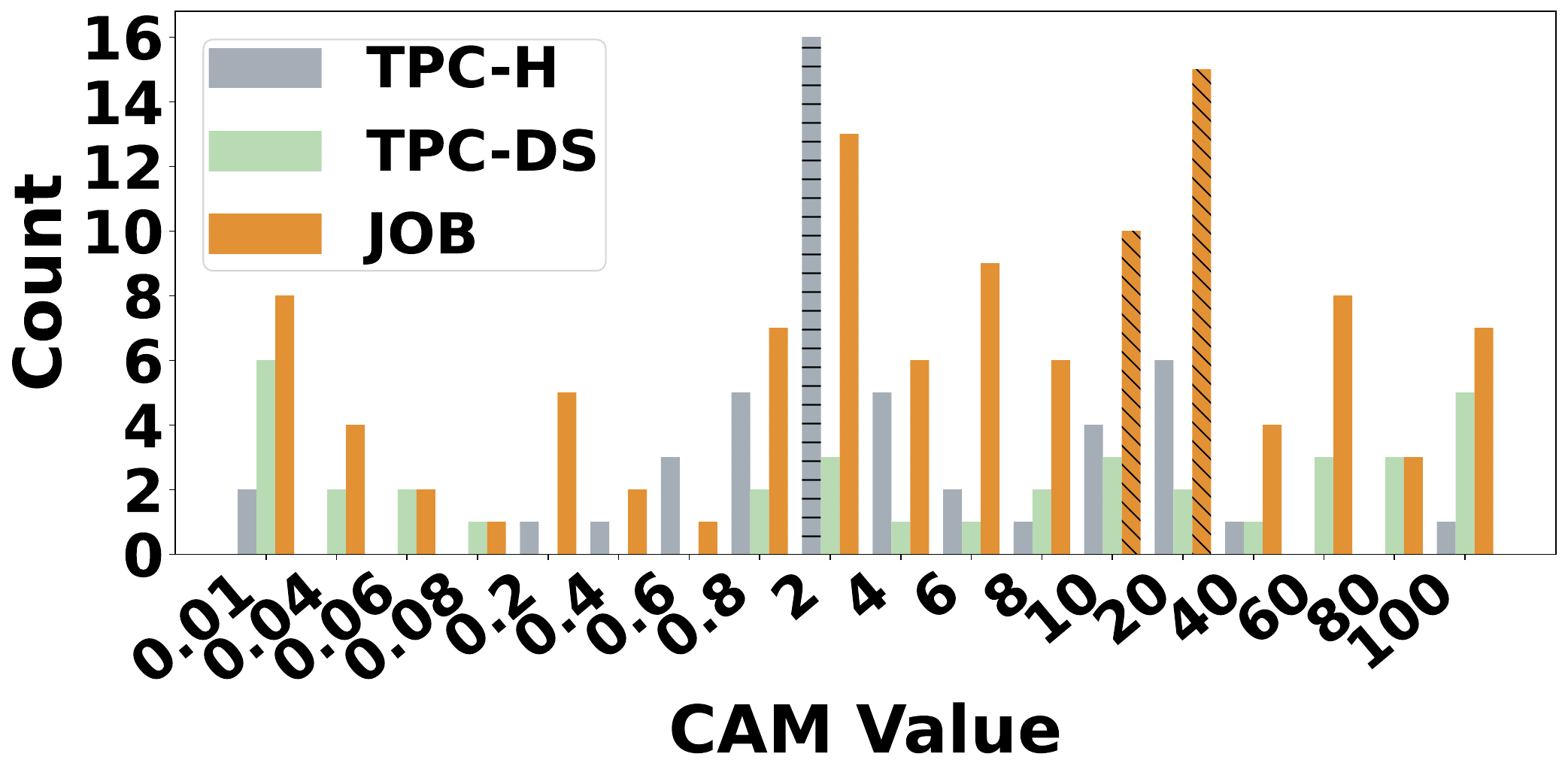}
        \caption{\rwfour{Distribution of $\omega$}}
   \label{fig:exp:design-validation:omega-distribution}
    \end{subfigure}
    \begin{subfigure}[b]{0.48\linewidth}
        \centering
        \includegraphics[width=\textwidth]{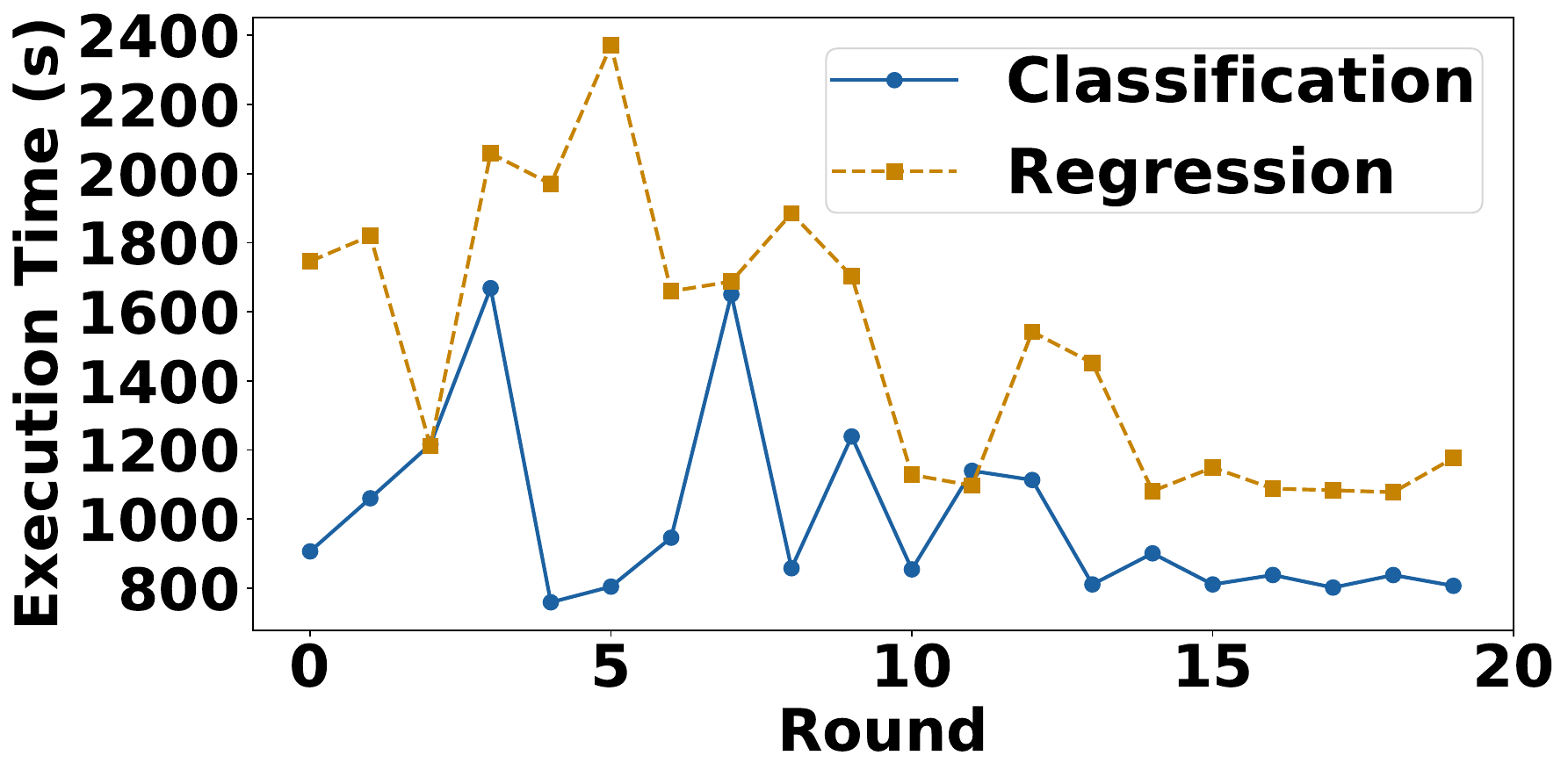}
        \caption{\rwfour{Classifier vs. regressor}}
              \label{fig:exp:design-validation:model-type}
    \end{subfigure}

\vspace{-0.5em}
    \caption{\rwfour{Validation of CAM model design choices.}}
\label{fig:exp:design-validation}
\end{figure}



\rwfour{
Figure~\ref{fig:exp:design-validation:omega-distribution} presents the distributions of the CAM values for the workloads used in our evaluation, confirming that the effective range of $\Omega$ spans a broad spectrum (Section~\ref{sect:estimation:CAM}).
}


\rwfour{
To validate our design choice of formulating CAM prediction as a classification task (Section~\ref{sect:estimation:CAM}), we implemented a regression-based variant of \sys.
To train the regression model, we perform a binary search to find the CAM value (between 0.1 and 100) that minimizes the error between the estimated and actual index benefits, within a specified tolerance.
Figure~\ref{fig:exp:design-validation:model-type} compares the execution time of \textbf{JOB} with the two models. 
The classification model converges to a better configuration (with lower execution time) compared to the regression model because it effectively handles the volatility of \textbf{JOB} with its non-uniform candidate buckets $\Omega$. 
It also identifies the best configuration much faster (around round 4) than the regression model (around round 11) with less exploration costs in the initial rounds. 
}



\vspace{-0.5em}
\subsection{Workloads for Online Index Selection}

We evaluate the performance of \sys and compare it with SOTA index advisors, using static and dynamic workloads.
Table~\ref{table:workload-pattern} summarizes the workloads used in our evaluation. 

\paragraph*{Static Workload}
The query templates do not vary across mini-workloads, and each mini-workload contains all query templates.
Static workload is useful for evaluating the capability of \sys to utilize execution feedback in a stable operational environment.


\paragraph{Dynamic Workload}
We introduce three types of dynamic workload to evaluate \sys in more realistic online index tuning scenarios:
\begin{itemize}[leftmargin=*]
    \item \textbf{(Continuous Variation)} A certain proportion (e.g., 20\%) of query templates will vary in each consecutive mini-workload, which we refer to as a \emph{workload drift} below.
    We use continuous variation to assess an index advisor’s ability to adapt to constantly evolving workload drifts. 
    \item \textbf{(Periodic Variation)} Workload drift occurs every 4 or 5 mini-workloads. We use periodic variation to evaluate how quickly an index advisor can catch up with workload drift.
    \item \textbf{(Cyclic Variation)}  
    A \emph{cycle} comprises a repeating sequence of 15 drifting mini-workloads.
    We use cyclic variation to evaluate an index advisor's ability of reusing its cumulative knowledge learned from past execution history.
\end{itemize}

\begin{figure}[t]
    \centering
     \begin{subfigure}[b]{\linewidth}
        \centering
        \includegraphics[width=\textwidth]{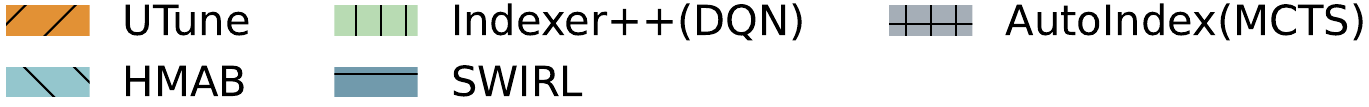}
    \end{subfigure}
  \vspace{-10pt}
    
    \begin{subfigure}[b]{0.3\linewidth}
        \centering
        \includegraphics[width=\textwidth]{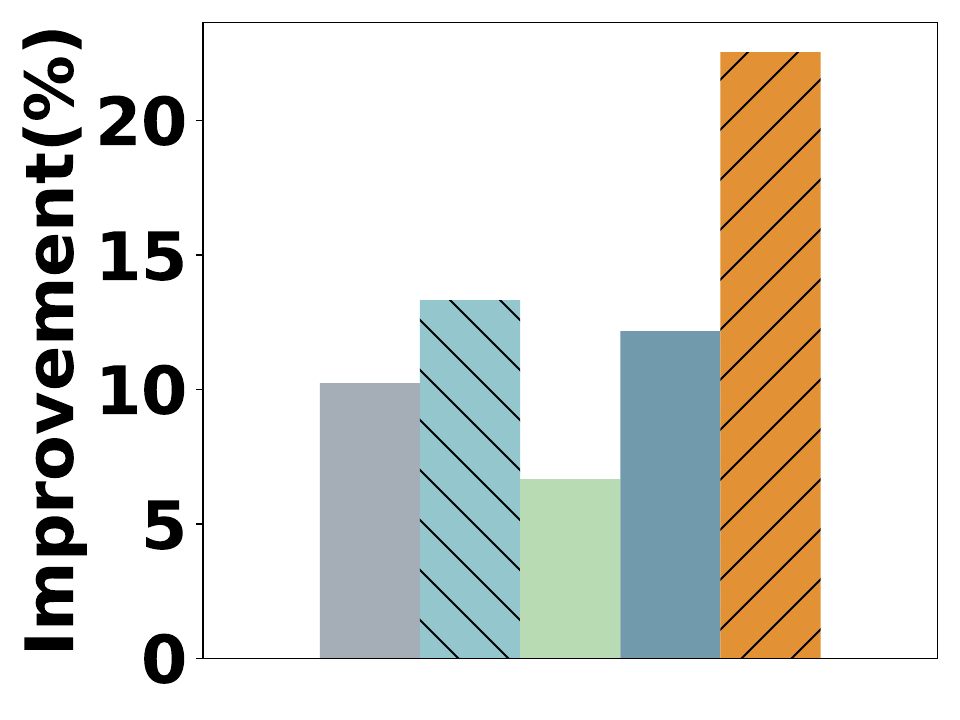}
        \caption{TPC-H}
   \label{fig:exp:static:overall:tpch}
    \end{subfigure}
    \begin{subfigure}[b]{0.3\linewidth}
        \centering
        \includegraphics[width=\textwidth]{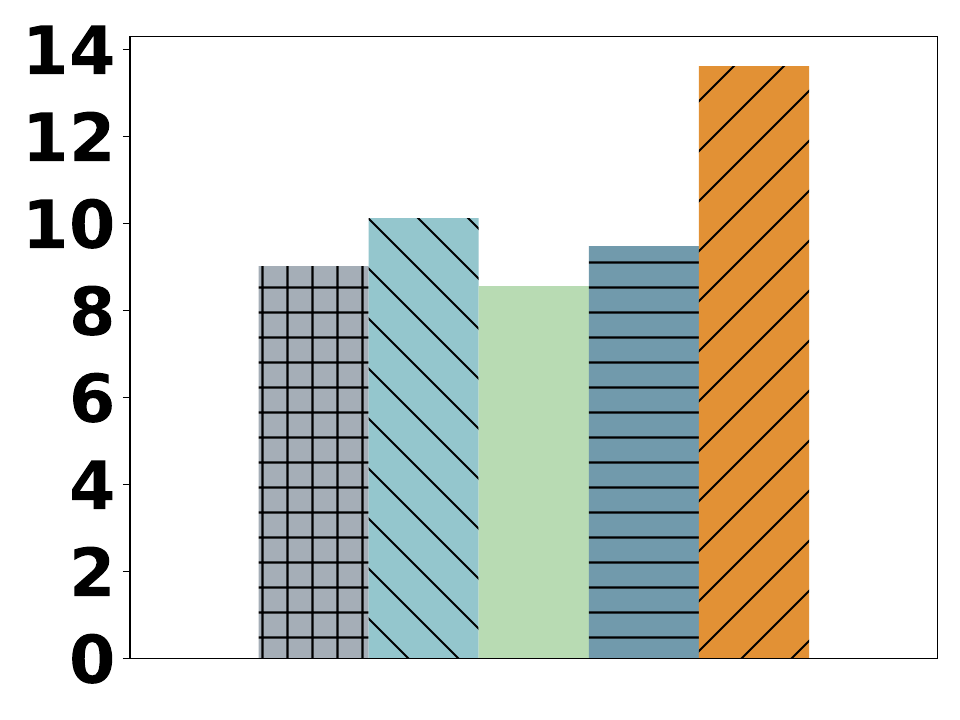}
        \caption{TPC-DS}
              \label{fig:exp:static:overall:tpcds}
    \end{subfigure}
        \begin{subfigure}[b]{0.3\linewidth}
        \centering
        \includegraphics[width=\textwidth]{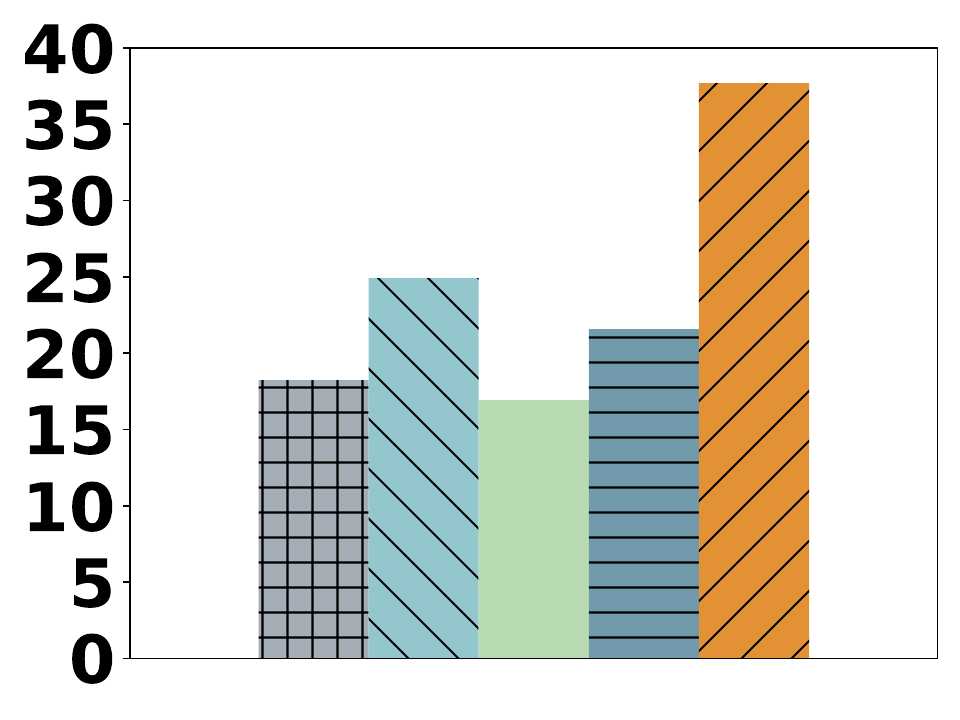}
        \caption{JOB}
              \label{fig:exp:static:overall:tpcds}
    \end{subfigure}
\vspace{-0.5em}
    \caption{Improvement on the execution time of static workload when tuned by different index advisors}
\label{fig:exp:static:overall}
\end{figure}


\vspace{-0.5em}
\subsection{Online Evaluation with Static Workload}




Figure~\ref{fig:exp:static:overall} presents the improvement on the execution time of the static workload when tuned by different online index advisors.
We replay the same workload 20 times. 
We observe that \sys outperforms the best baseline index advisor by $7\%$, $4\%$, and $11\%$ for \textbf{TPC-H}, \textbf{TPC-DS}, and \textbf{JOB}, respectively.
A more detailed analysis can be found in Appendix~\ref{sect:exp:static-workload}.
\begin{figure}
    \centering
    \begin{subfigure}[b]{\linewidth}
        \centering
        \includegraphics[width=\textwidth]{pics/experiments/labels_bars.pdf}
    \end{subfigure}
  \vspace{-2pt}
    \begin{subfigure}[b]{0.3\linewidth}
        \centering
        \includegraphics[width=\textwidth]{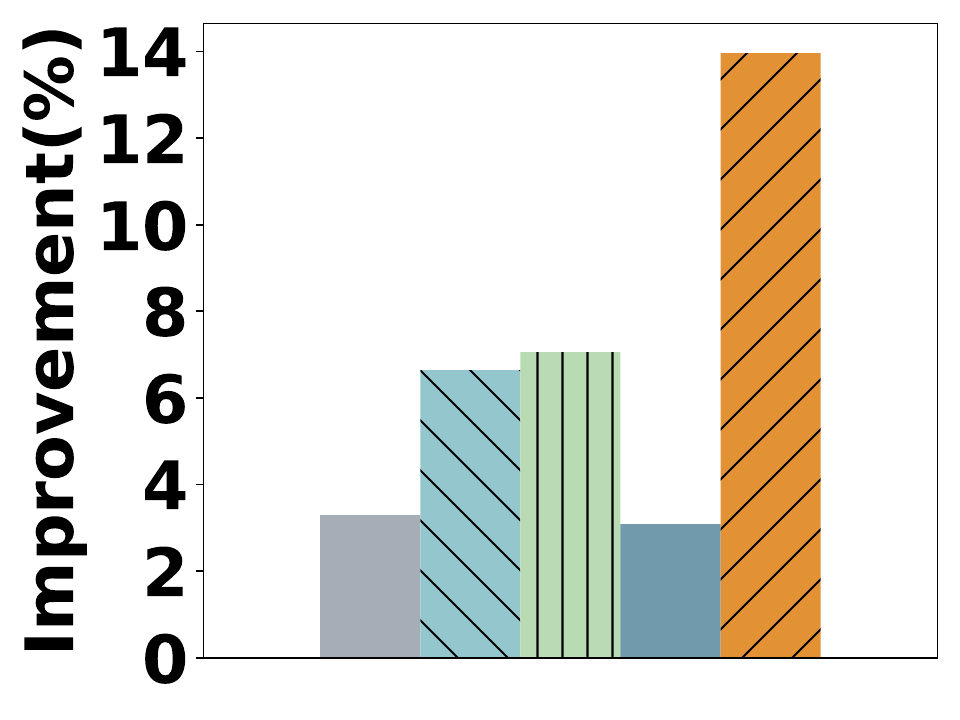}
        \caption{TPC-H}
        \label{fig:exp:dynamic:tpch}
    \end{subfigure}
    \begin{subfigure}[b]{0.3\linewidth}
        \centering
        \includegraphics[width=\textwidth]{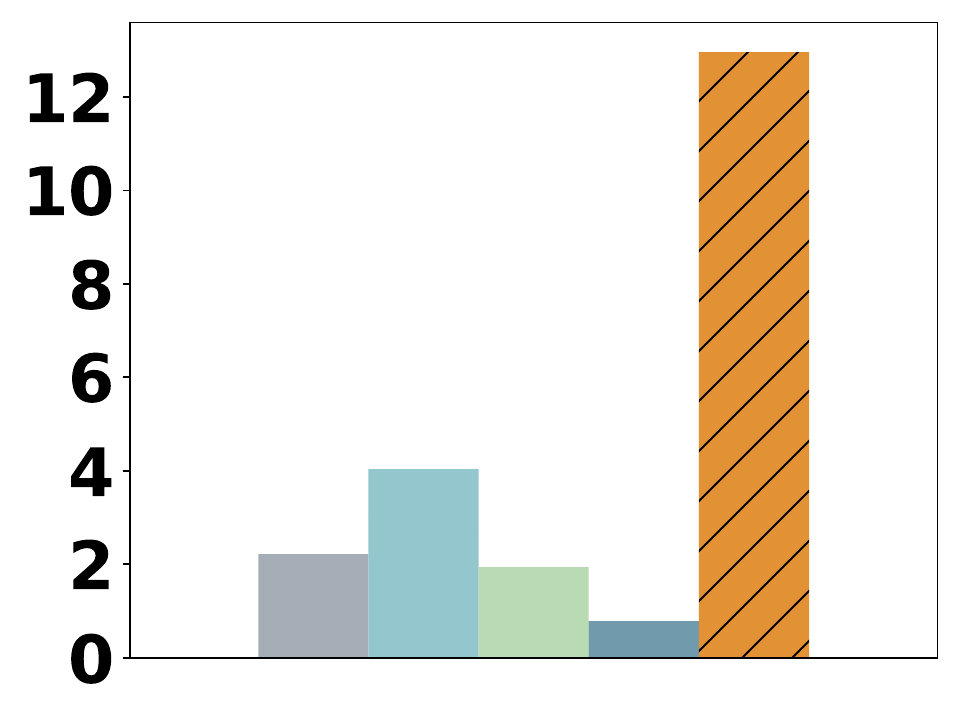}
        \caption{TPC-DS}
        \label{fig:exp:dynamic:tpcds}
    \end{subfigure}
     \begin{subfigure}[b]{0.3\linewidth}
        \centering
        \includegraphics[width=\textwidth]{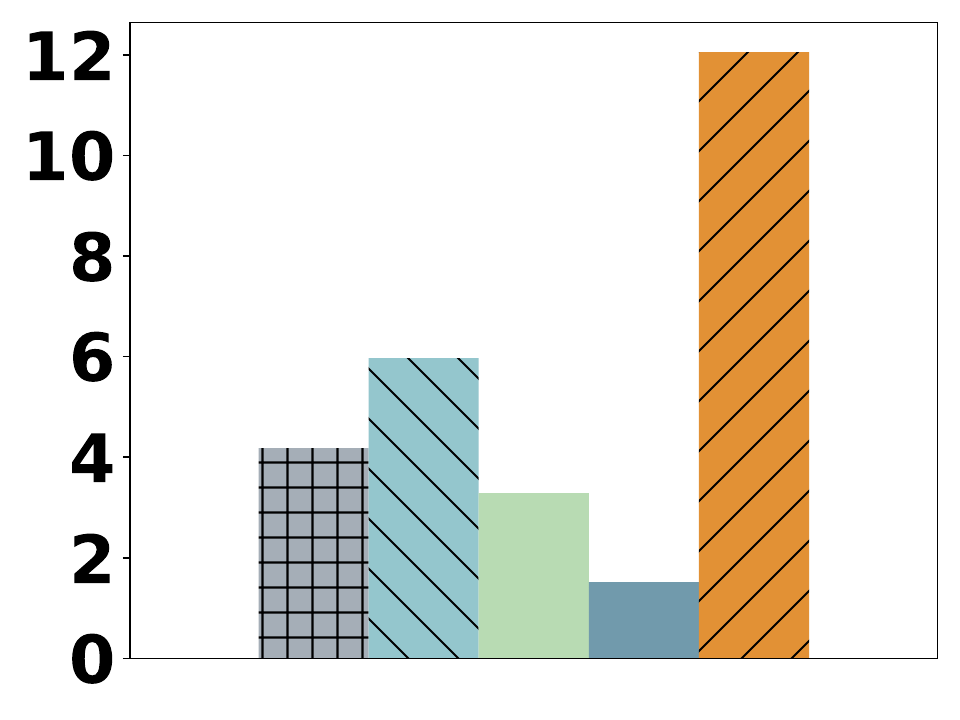}
        \caption{JOB}
        \label{fig:exp:dynamic:job}
    \end{subfigure}
        \vspace{-5pt}
    
    \caption{Improvement of workload execution time when tuned by different index advisors under continuous variation
    }
    \label{fig:exp:overall-dynamic}
\end{figure}

\begin{figure}[tbp]
    \centering
    \begin{subfigure}[b]{\linewidth}
        \centering
        \includegraphics[width=\textwidth]{pics/experiments/labels_bars.pdf}
    \end{subfigure}
    \begin{subfigure}[b]{0.3\linewidth}
        \centering
        \includegraphics[width=\textwidth]{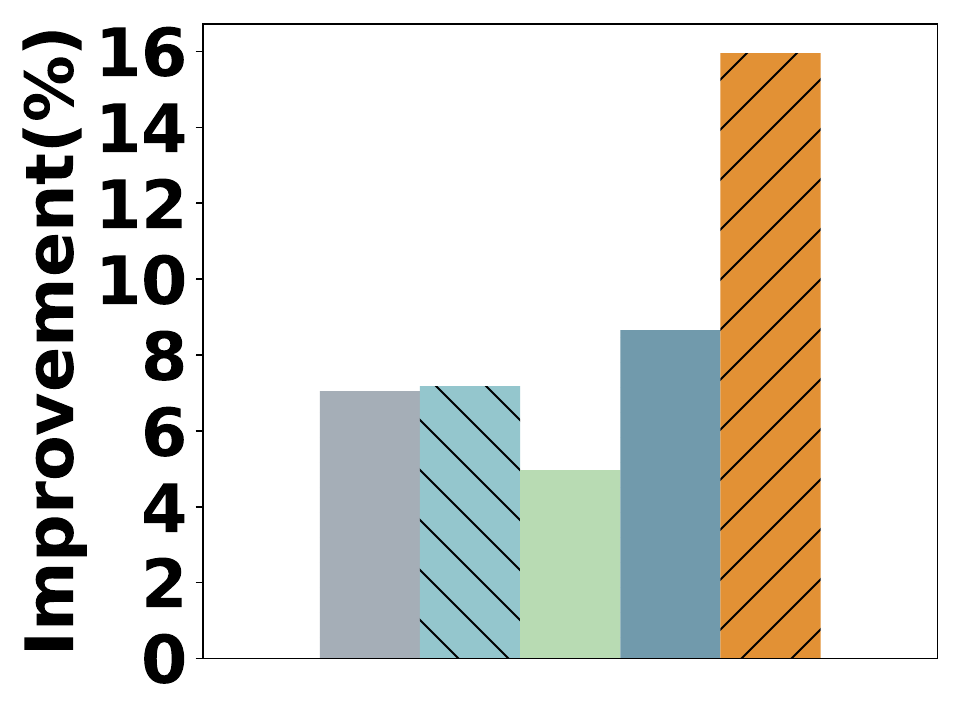}
        \caption{TPC-H}
        \label{fig:exp:overall:session-tpch}
    \end{subfigure}
    \begin{subfigure}[b]{0.3\linewidth}
        \centering
        \includegraphics[width=\textwidth]{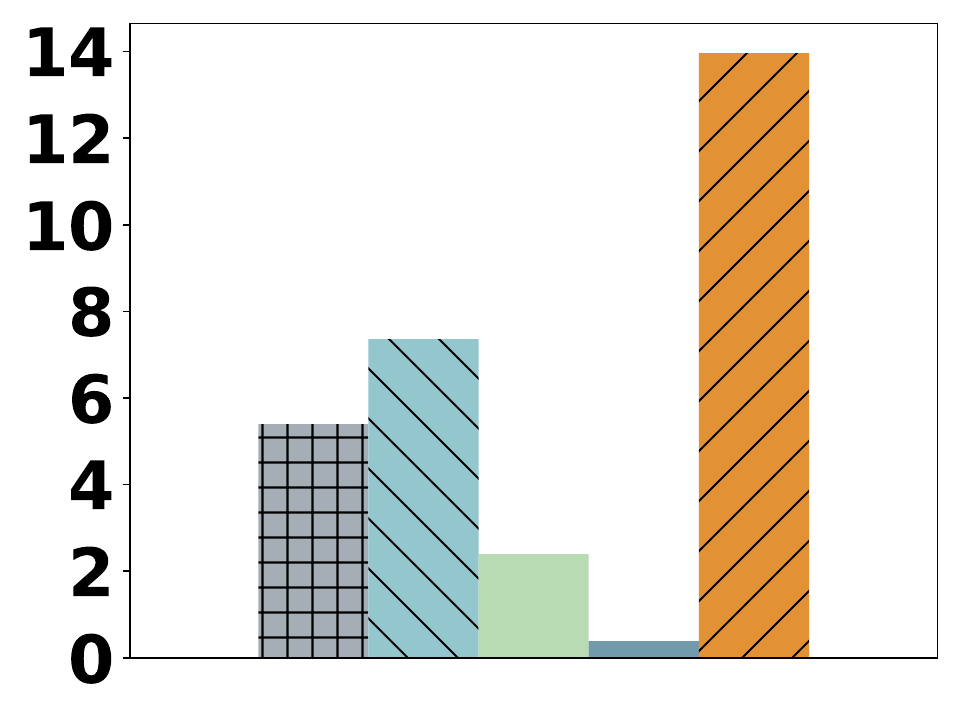}
        \caption{TPC-DS}
        \label{fig:exp:overall:session-tpcds}
    \end{subfigure}
    \begin{subfigure}[b]{0.3\linewidth}
        \centering
        \includegraphics[width=\textwidth]{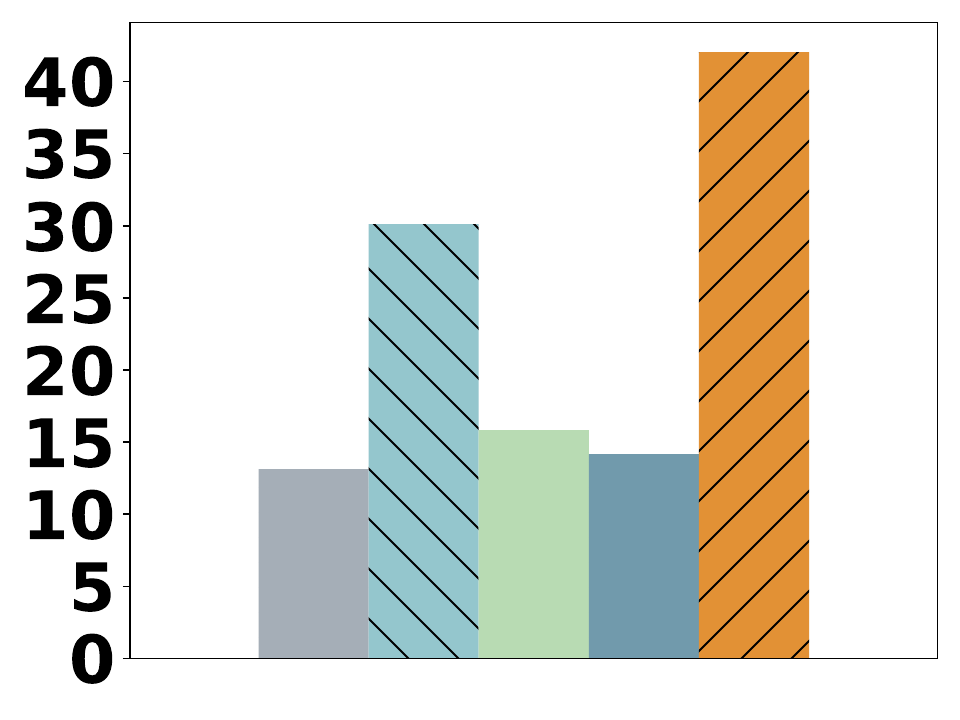}
        \caption{JOB}
        \label{fig:exp:overall:session-job}
    \end{subfigure}
    \vspace{-5pt}
    \caption{Improvement of workload execution time when tuned by different index advisors under periodic variation}
    \vspace{-0.5em}
    \label{fig:exp:overall-session}
    
\end{figure}

\subsection{Online Evaluation with Dynamic Workload}
\subsubsection{Continuous Variation} 

Figure~\ref{fig:exp:overall-dynamic} presents the overall improvement in workload execution time under continuously changing workloads. 
Compared to the \underline{best} baseline, \sys offers $7\%$ more (i.e., from $7\%$ to $14\%$) improvement on \textbf{TPC-H}, $7\%$ more (i.e., from $5\%$ to $12\%$) improvement on \textbf{TPC-DS}, and $6\%$ more (i.e., from $6\%$ to $12\%$) improvement on \textbf{JOB}. 


Continuous variation presents the most challenging case among the dynamic workloads considered in our evaluation.
\textbf{SWIRL} performs the worst because it is designed for offline index tuning. 
Its index selection policy (trained on $W_0$) fails on unseen queries during workload drifts.
Other online baselines such as \textbf{HMAB} leverage query execution feedback passively and cannot react as fast as workload drifts that occur constantly. 
By contrast, \sys exhibits stronger adaptive capability to unseen query templates, attributed to its uncertainty-aware operator-level cost correction framework.
More detailed analysis can be found in Appendix~\ref{sect:exp:continuous-variation}.
\subsubsection{Periodic Variation}

Each mini-workload of \textbf{TPC-H} contains 10 query templates, whereas workload drift occurs every 4 mini-workloads.
That is, the same mini-workload will repeat 4 times before a workload drift occurs.
Similarly, each mini-workload of \textbf{TPC-DS} and \textbf{JOB} contains 20 query templates, and workload drift occurs every 5 mini-workloads.
We chose these settings based on the number of available query templates in each benchmark to allow for sufficient diversity of workload drifts.
Figure~\ref{fig:exp:overall-session} presents the improvement of overall workload execution time under periodic variation.
We observe that \sys outperforms the best baseline index tuner by 7\% on \textbf{TPC-H} (i.e., 16\% vs. 9\%), by 6\% on \textbf{TPC-DS} (i.e., 14\% vs. 8\%), and by 12\% on \textbf{JOB} (i.e., 42\% vs. 30\%).

Periodic variation evaluates how quickly online index advisors converge under a relatively stable workload.
Faster convergence can result in lower index exploration overhead, which is often nontrivial. 
Among the online index advisors, we observe that \sys and \textbf{HMAB} typically converge faster than \textbf{AutoIndex} and \textbf{Indexer++} during the stable period of a mini-workload, but \sys typically converges to a better index configuration, resonating with the observation on static workloads.
We attribute the fast convergence of \sys to its uncertainty-aware selection policy, which prioritizes indexes with low model confidence and thus improves benefit estimates with fewer explorations.
See Appendix~\ref{sect:exp:periodic-variation} for more analysis.

\begin{figure}[tbp]
    \centering
\begin{subfigure}[b]{\linewidth}
        \centering
        \includegraphics[width=\textwidth]{pics/experiments/labels_bars.pdf}
    \end{subfigure}    
    \begin{subfigure}[b]{0.3\linewidth}
        \centering
        \includegraphics[width=\textwidth]{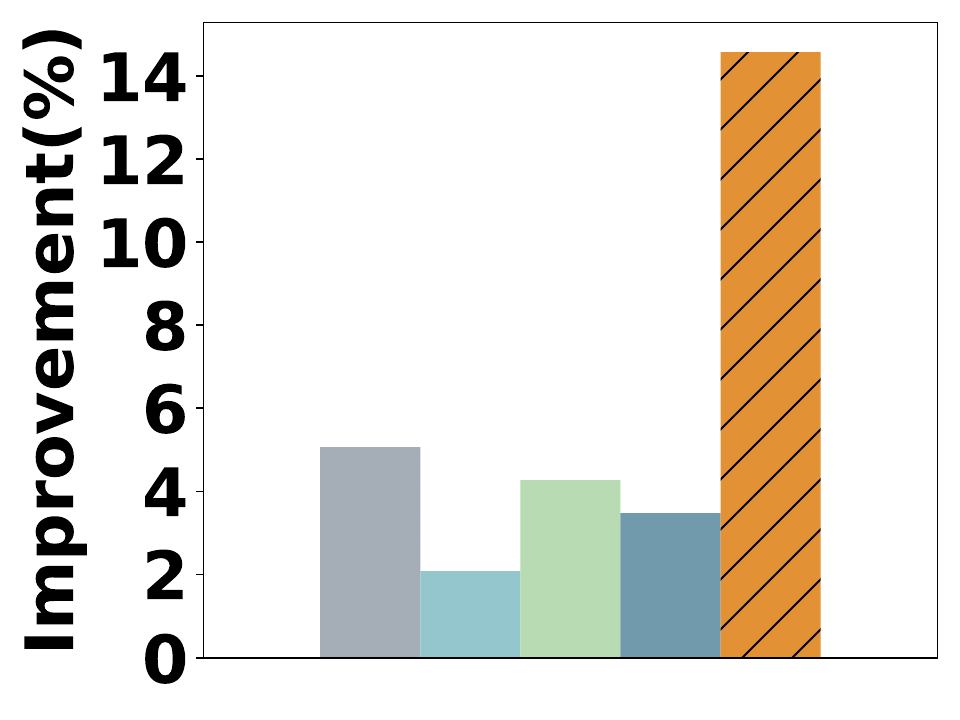}
        \caption{TPC-H}
        \label{fig:exp:overall:session-tpch}
    \end{subfigure}
    \begin{subfigure}[b]{0.3\linewidth}
        \centering
        \includegraphics[width=\textwidth]{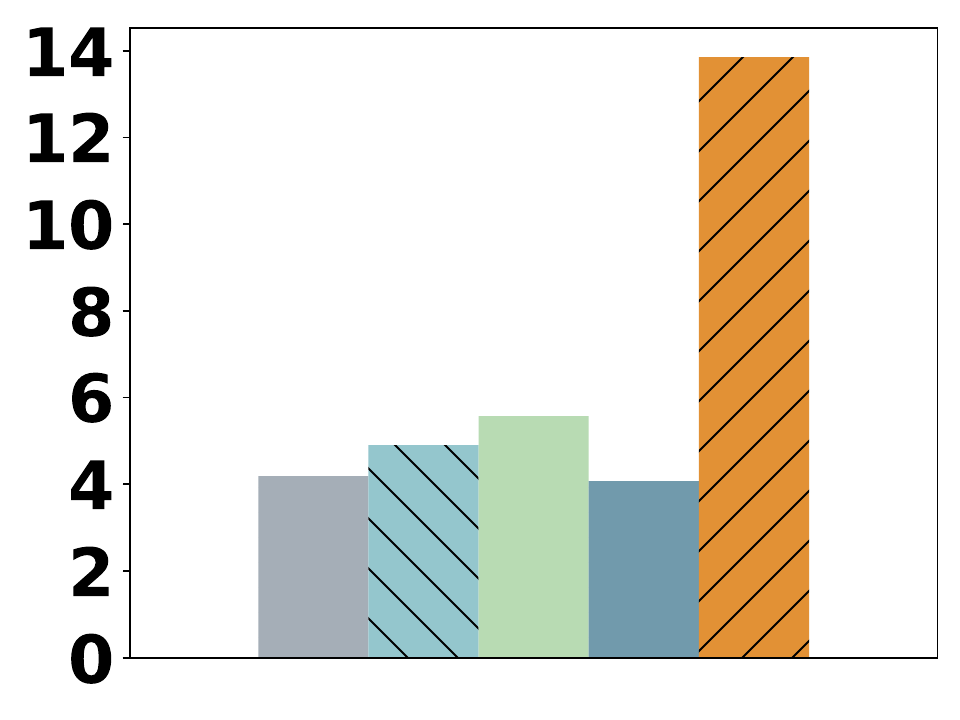}
        \caption{TPC-DS}
        \label{fig:exp:overall:session-tpcds}
    \end{subfigure}
    \begin{subfigure}[b]{0.3\linewidth}
        \centering
        \includegraphics[width=\textwidth]{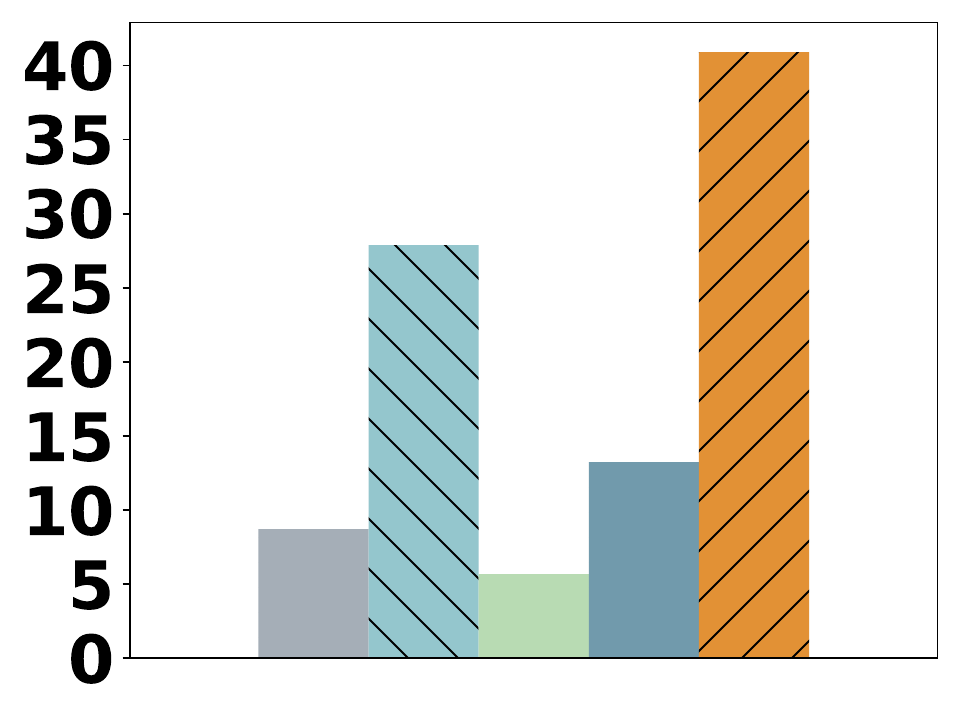}
        \caption{JOB}
        \label{fig:exp:overall:session-job}
    \end{subfigure}
    \vspace{-5pt}
    \caption{Improvement of workload execution time when tuned by different index advisors under cyclic variation}
    \label{fig:exp:overall-cycle}
\end{figure}

\subsubsection{Cyclic Variation}

One common limitation of existing RL-based online index selection policies is ``knowledge forgetting.''
For example, \textbf{HMAB} resets its internal parameters when significant workload drifts are detected, whereas \textbf{Indexer++} adopts \emph{prioritized sweeping} by focusing on the most important updates to the value function.
As a result, these strategies weigh more on their recent search experience than their past search experience.
This is in general not a problem if workload keeps changing, such as in the case of continuous variation.
However, for recurring workload with a long period, such as in the case of cyclic variation, these strategies are not able to utilize their experience learned from the past history.
In contrast, this is not a limitation of \sys, which captures all historical query execution feedback via its operator-level learned index benefit estimators.
Figure~\ref{fig:exp:overall-cycle} presents the improvement of the total workload execution time given by different index advisors for cyclically varying mini-workloads.
\sys improves the best baseline index advisor by 9\% (i.e., 14\% vs. 5\%) on \textbf{TPC-H}, by 8\% (i.e., 14\% vs. 6\%) on \textbf{TPC-DS}, and by 13\% (i.e., 42\% vs. 29\%) on \textbf{JOB}.

\vspace{-0.5em}
\subsection{Impact of Index Tuning Constraints}
\label{sect:exp:index-maintenance}

 In online environments, timely index creation and adaptation are critical.
 \rwfive{
 To quantify the overheads and benefits of indexes, Table~\ref{tbl:index-maintenance} presents a detailed breakdown of index creation overhead (including time and space costs) versus workload execution time under varying index budgets (encompassing both quantity and storage limits).
We omit the results of \textbf{SWIRL} and \textbf{Index++} because they are similar to \textbf{AutoIndex}.
Our evaluation does not include index updates because we focus on \textbf{OLAP} workloads in this paper, following previous work on index tuning~\cite{kossmann2022swirl, wu2022budget}.
However, \sys can be extended to handle OLTP workloads, where update costs are critical, by introducing operator-level cost models for modification operators (e.g., \texttt{INSERT}, \texttt{UPDATE}, and \texttt{DELETE}).}
\begin{table*}[t]
    \caption{\rwfive{Analysis of index maintenance overheads under different constraints}}
    \label{tbl:index-maintenance}
    \vspace{-0.5em}
    \centering
    \scalebox{0.9}{ 
    \begin{tabularx}{\linewidth}{ccXXXXXXXXX}
      \toprule
      
    \multicolumn{2}{c}{Tuning Constraints}
      & \multicolumn{3}{c}{Creation(min)} & \multicolumn{3}{c}{Execution(min)} 
      & \multicolumn{3}{c}{\rwfive{Space(M)}}\\

   &  & UTune & HMAB & MCTS & UTune & HMAB & MCTS & \rwfive{UTune} &  \rwfive{HMAB} & \rwfive{MCTS} \\ 
    \midrule
   \multirow{4}{*}{Number(\#)} & 2 & 0.48 & 0.06 & 0.32 & 38.0 & 39.7 & 40.3 & \rwfive{1,102} & \rwfive{1,807} & \rwfive{1,380} \\
   & 4 & 0.78 & 0.46 & 0.37 & 35.3 & 39.3 & 38.3 & \rwfive{1,898} & \rwfive{1,712} & \rwfive{2,297} \\
   & 8 & 0.86 & 0.4 & 0.44 & 33.7 & 37.3 & 38.7 & \rwfive{2,911} & \rwfive{3,572} & \rwfive{2,580} \\
   & 12 & 1.05 & 0.57 & 0.61 & 33.3 & 37.6 & 40.0 & \rwfive{3,241} & \rwfive{3,758} & \rwfive{3,414} \\
   \multirow{3}{*}{Storage(M)} & 2000 & 0.26 & 0.45 & 0.15 & 36.3 & 40.3 & 41.0 & \rwfive{1,413} & \rwfive{1,513} & \rwfive{1,703}\\
   & 4000 & 0.70 & 0.63 & 0.30 & 35.7 & 38.0 & 39.3 & \rwfive{2,426} & \rwfive{2,415} & \rwfive{1,874}\\
   & 6000 & 0.83 & 0.70 & 0.37 & 34.0 & 38.7 & 39.0 & \rwfive{3,870} & \rwfive{3,729} & \rwfive{3,098} \\

      \bottomrule
    \end{tabularx}
    }
\vspace{-1em}
\end{table*}

\rwfive{The space consumption differs marginally across different algorithms, as it is primarily bounded by the index tuning budget.}
\sys consistently achieves lower execution time than baselines, especially under a tuning budget. Although it incurs higher index creation time as the budget grows (due to the actual index creation during exploration), the corresponding reduction in execution time compensates more for this overhead, making the exploration cost acceptable in practice. 
In contrast, \textbf{HMAB} has lower index creation time but much higher execution time, as its linear reward function over-penalizes candidates on certain tables and therefore leads to suboptimal convergence.
\textbf{AutoIndex} suffers the most significant performance decline as the tuning budget increases (e.g., from 8 to 12 indexes allowed). This is due to the expanded search space, which dramatically increases the exploration overhead of its MCTS-based selection policy, leading to premature and inadequately explored index choices.

\vspace{-0.5em}
\subsection{\rwfive{Analysis of Long-running TPC-DS Queries}}
\label{sect:exp:excluded-queries}
\begin{figure}
    \centering
    \includegraphics[width=0.7\linewidth]{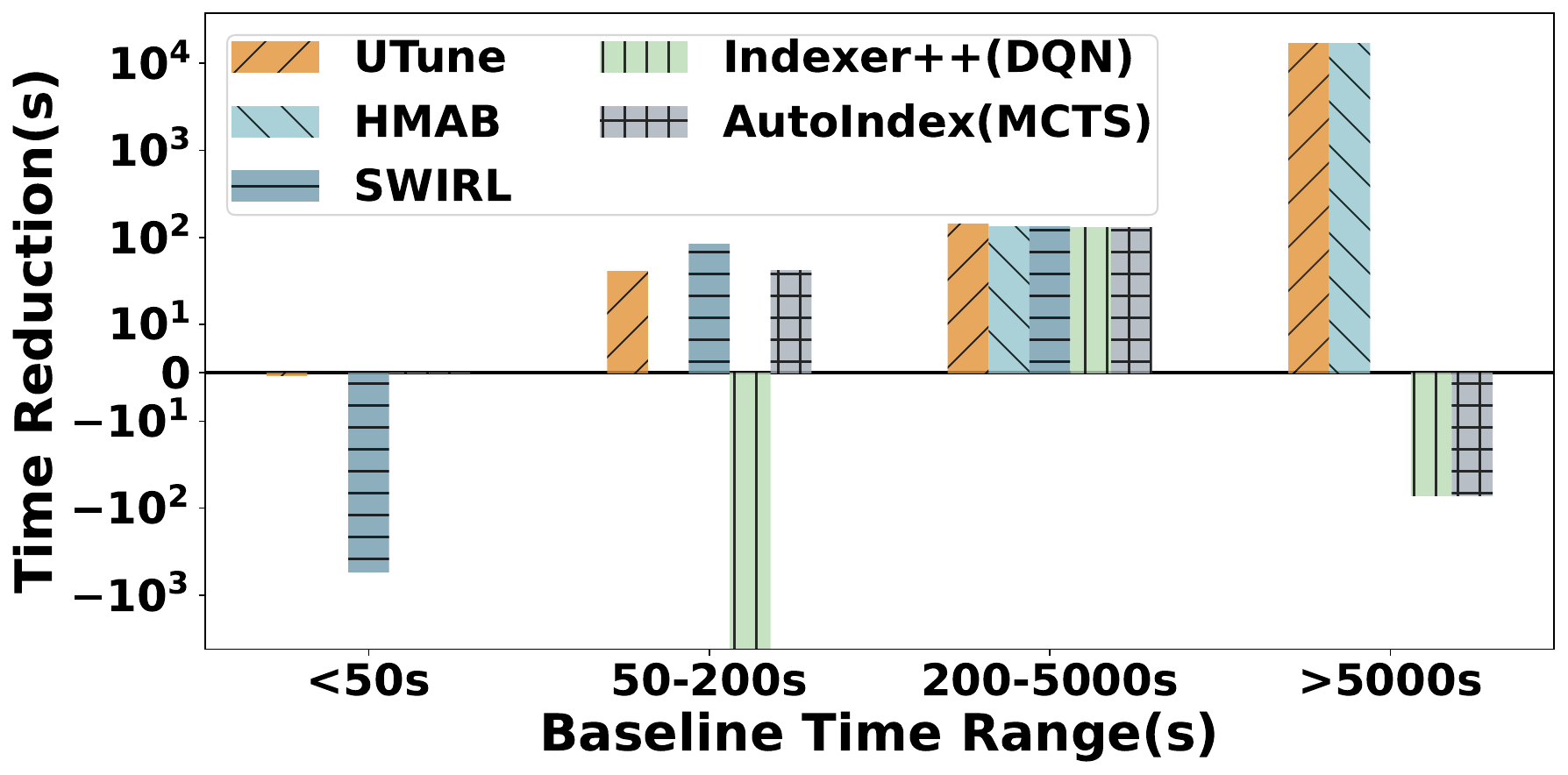}
    \vspace{-0.5em}
    \caption{\rwfive{Excluded long-running \textbf{TPC-DS} queries}}
 \label{fig:exp:excluded-tpcds-queries}
\end{figure}
\rwfive{
We now analyze the performance of \sys for the previously excluded long-running \textbf{TPC-DS} queries.
The average execution time of these queries is over 1,000 seconds, while the average execution time of the other \textbf{TPC-DS} queries is around 6 seconds.
We further classify these queries into four groups based on their execution time.
As shown in Figure~\ref{fig:exp:excluded-tpcds-queries}, \sys achieves orders-of-magnitude performance gains compared to what-if based index advisors (e.g., \textbf{SWIRL}, \textbf{AutoIndex}, \textbf{Index++}).
This significant advantage is driven by the inaccurate what-if cost estimates for these queries.
On the other hand, both feedback-based index advisors, \sys and \textbf{HMAB}, achieve comparable performance.
}

\vspace{-0.5em}
\subsection{\rwfour{Hyper-parameter Sensitivity Analysis}}
\label{sect:exp:parameter-sensitivity}
\begin{figure*}[tbp]
    \centering  
    \begin{subfigure}[b]{0.5\linewidth}
        \centering
\hfill        \includegraphics[width=\textwidth]{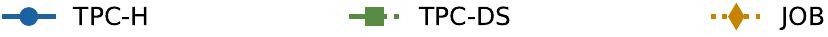}
    \end{subfigure}
    
    \begin{subfigure}[b]{0.24\linewidth}
        \centering
        \includegraphics[width=\textwidth]{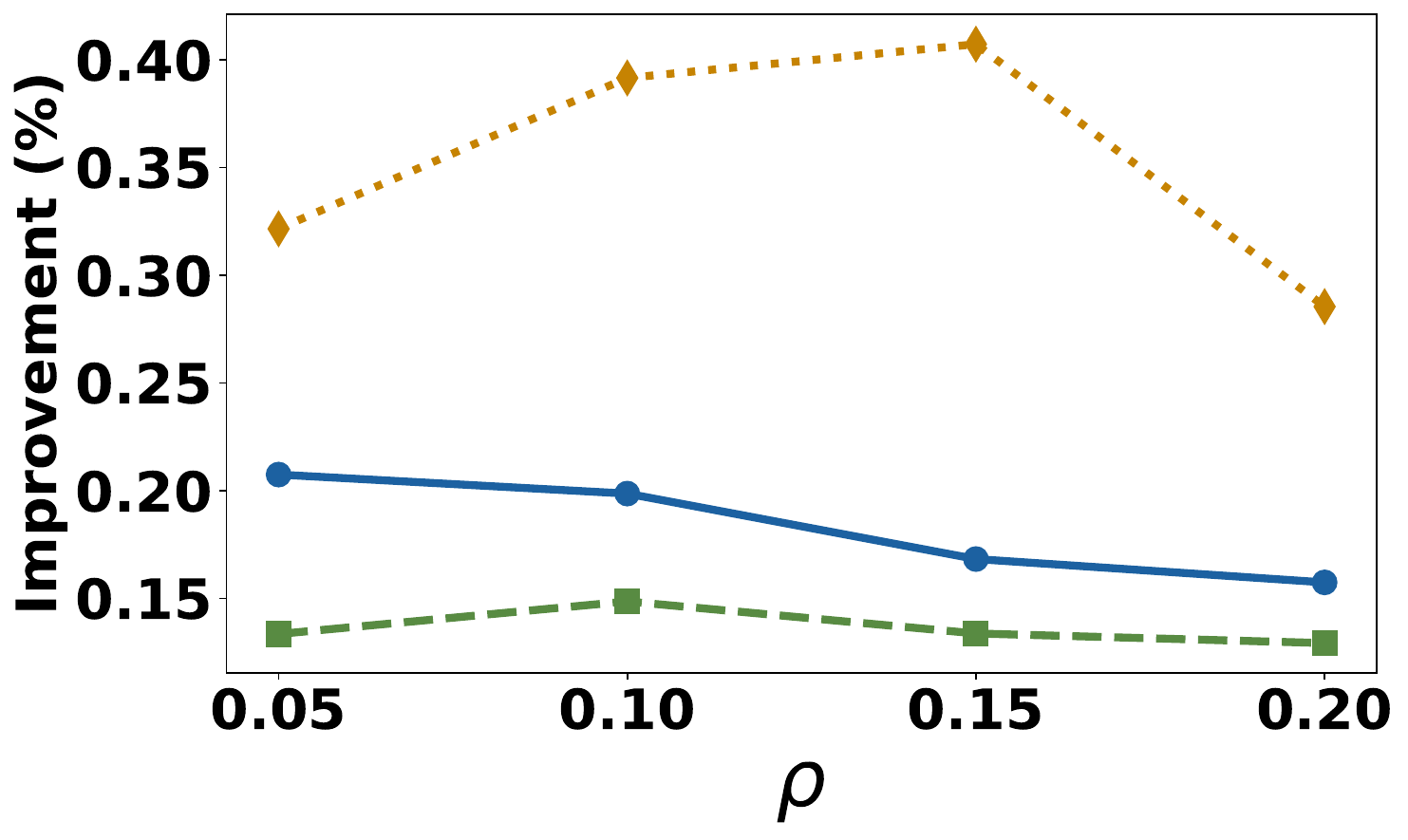}
      \vspace{-1.5em}  \caption{\rwfour{Uncertainty threshold $\rho$}}
        \label{fig:exp:hyper-param-sensitivity:rho}
    \end{subfigure}
    \begin{subfigure}[b]{0.24\linewidth}
        \centering
        \includegraphics[width=\textwidth]{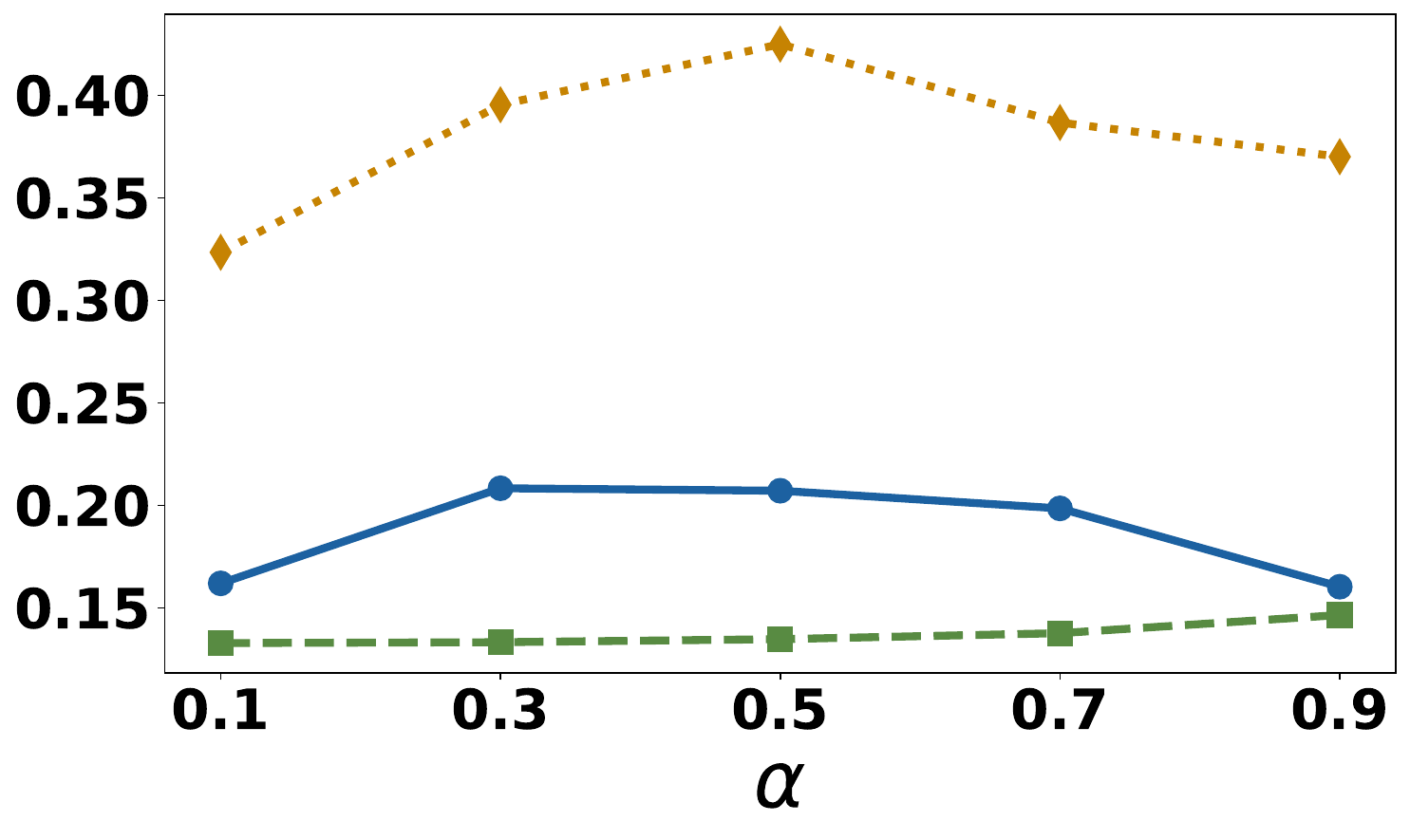}
        \caption{\rwfour{Uncertainty weight $\alpha$}}
        \label{fig:exp:hyper-param-sensitivity:alpha}
    \end{subfigure}
    \begin{subfigure}[b]{0.24\linewidth}
        \centering
        \includegraphics[width=\textwidth]{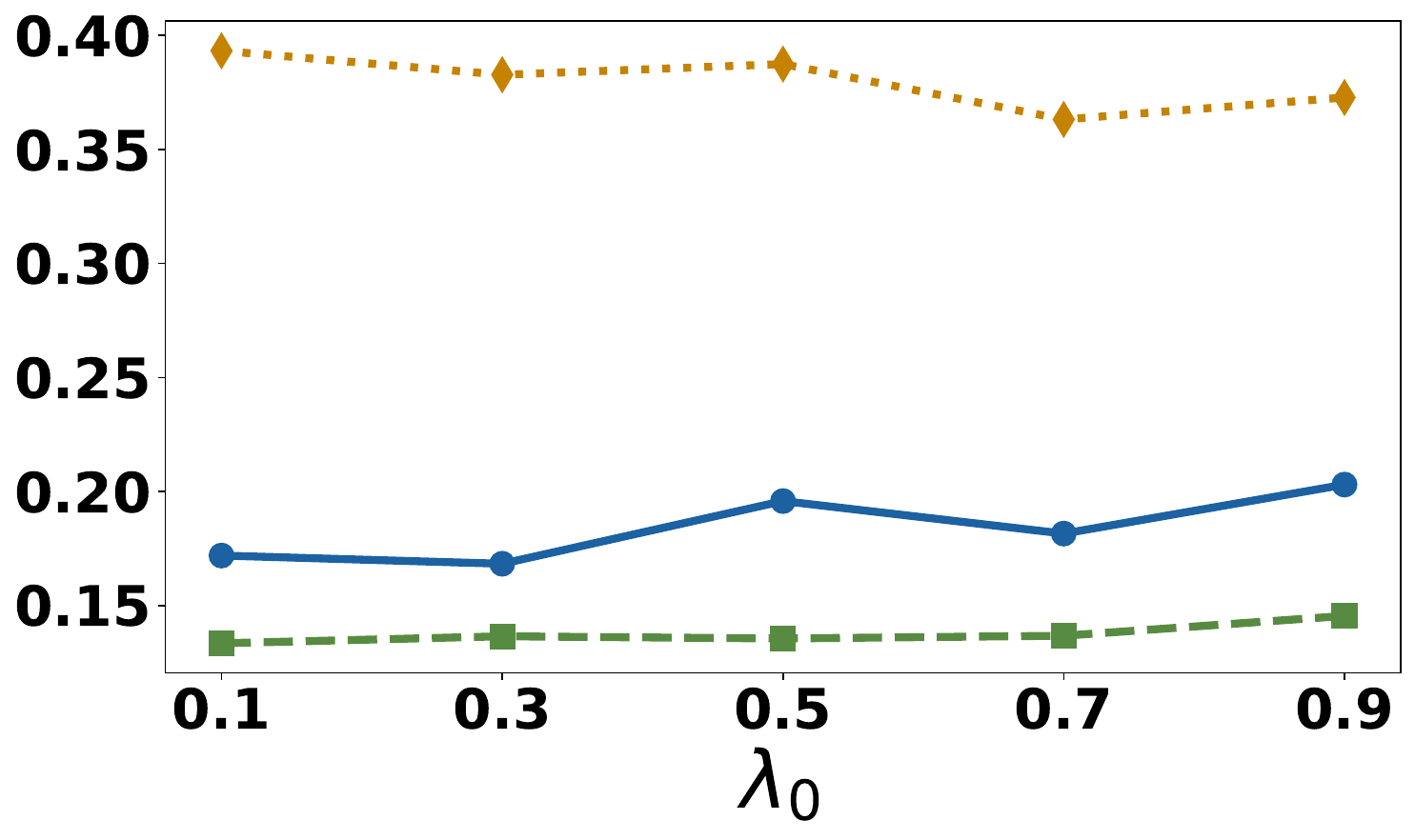}
        \caption{\rwfour{Initial exploration weight $\lambda_0$}}
        \label{fig:exp:hyper-param-sensitivity:lambda-0}
    \end{subfigure}
     \begin{subfigure}[b]{0.24\linewidth}
        \centering
        \includegraphics[width=\textwidth]{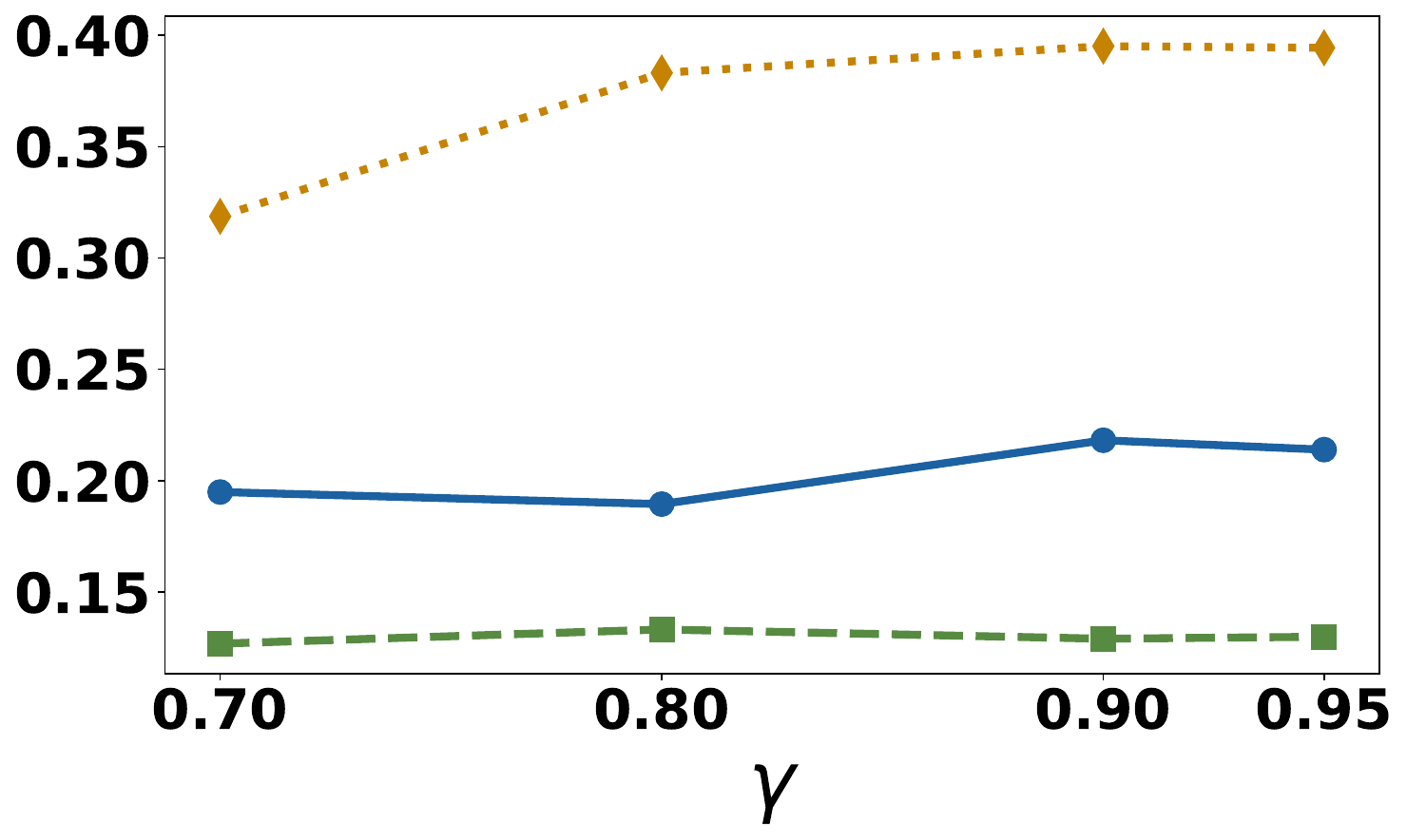}
        \caption{\rwfour{Decay factor $\gamma$}}
        \label{fig:exp:hyper-param-sensitivity:gamma}
    \end{subfigure}
    \vspace{-4pt}
    \caption{\rwfour{Sensitivity of \sys's performance with respect to varying hyper-parameter values}}
    \label{fig:exp:hyper-param-sensitivity}
    \vspace{-2em}
\end{figure*}
\rwone{We analyze the impact of several critical hyper-parameters on the behavior of \sys. Figure~\ref{fig:exp:hyper-param-sensitivity} summarizes the results.}

\paragraph*{\rwone{Index Estimation}}
\rwone{
The uncertainty threshold $\rho$ determines which CAM predictions are trustworthy enough to correct the what-if cost estimates. 
As Figure~\ref{fig:exp:hyper-param-sensitivity:rho} shows, setting $\rho$ around 0.1 provides a stable baseline across different benchmarks, filtering out highly volatile predictions while allowing necessary adjustments.}
\rwone{
Complementing $\rho$, the uncertainty weight $\alpha$ regulates the internal composition of the uncertainty metric, balancing between model and data uncertainty. 
As Figure~\ref{fig:exp:hyper-param-sensitivity:alpha} shows, $\alpha \in [0.3, 0.7]$ works well for most scenarios.}

\paragraph*{\rwone{Index Selection}}
\rwone{
The hyper-parameters $\lambda_0$ and $\gamma$ govern the trade-off between exploration and exploitation during the index selection process.
The initial exploration weight $\lambda_0$ determines the magnitude of the exploration incentive during the cold-start phase, while the decay rate $\gamma$ controls how quickly this incentive diminishes as feedback being collected.
\sys exhibits low sensitivity to these two parameters, as shown in Figures~\ref{fig:exp:hyper-param-sensitivity:lambda-0} and~\ref{fig:exp:hyper-param-sensitivity:gamma}.
In general, setting $\lambda_0 = 0.5$ and $\gamma=0.9$ works well across all workloads.}

\vspace{-0.5em}
\subsection{\rwfive{Case Study}}
\label{sect:exp:case-study}
\rwfive{As a case study to better understand the efficacy of \sys, we present a detailed example from \textbf{JOB}, which involves two candidate indexes: (1) $I_{mi}$,  \texttt{movie\_info(movie\_id)} and (2) $I_{cc}$, \texttt{complete\_cast(status\_id, movie\_id)}.}

\rwfive{\textbf{SWIRL} and \textbf{AutoIndex} rely on what-if cost estimates. They fail on this example in two ways:
\begin{itemize}[leftmargin=*]
    \item (\textbf{Overestimation}) $I_{mi}$ is a popular index on different query templates and is selected frequently because of the predicted high global benefit. However, it can cause severe performance regression~\cite{WuDXNC25} on query \textbf{Q29} (5s $\to$ 29s) due to a \texttt{Nested-Loop Join} with overestimated benefit;
    \item (\textbf{Underestimation}) $I_{cc}$, a covering index for \textbf{Q23}, is usually discarded due to its estimated low benefit ($1\%$), missing a significant actual speedup (23s $\to$ 0.8s).
\end{itemize}
\sys overcomes the pitfalls with the cost correction mechanism. By learning from execution feedback, the CAM models penalize $I_{mi}$ for the observed regression and boosts $I_{cc}$ for its true benefit.
Although \textbf{HMAB} also utilizes execution feedback, it fails to recommend $I_{cc}$ due to its misguided exploration with its simplistic linear reward modeling that tends to prioritize complex and high-cost indexes. 
}

\vspace{-0.5em}
\section{\rwthree{Discussion on Generalizability}}
\label{sec:discussion}

\rwthree{
Although currently built on top of PostgreSQL, \sys can be extended to other DBMSs, distributed settings, and Lakehouse systems.
\sys is based on correcting the query optimizer's cost model by analyzing the execution feedback. 
Therefore, it can be extended to other DBMSs as long as they support \emph{cost-based} query optimization and query \emph{execution feedback}.
The extension involves updating the operator-level cost models in Section~\ref{section:benefit-estimation:update} and the cost correction formulas in Table~\ref{table:cost-functions} 
with respect to the target system. 
}

\rwthree{
\sys can also be extended to distributed settings (such as CockroachDB) and Lakehouse systems (Databricks Lakehouse) that support query cost estimations and execution feedback. 
Although lakehouses generally do not support B-trees, which are commonly used by DBMSs, they can still leverage \sys as long as they support indexing.
For example, Databricks Lakehouse utilizes Z-ordering indexing, allowing \sys to be adapted accordingly.
}

\vspace{-0.5em}
\section{Related Work}

\paragraph*{Classic Index Tuning}
The problem of index tuning has been studied for decades in the \emph{offline} setting, where a static workload and certain constraints (e.g., limitation on storage space) are given and the goal is to find a set of indexes under the constraints that minimizes the workload execution time.
Most commercial and open-source offline index tuning software~\cite{chaudhuri1997efficient,valentin2000db2} employs a cost-based architecture that replies on the what-if API~\cite{chaud1998autoadmin} provided by the query optimizer.
There has been extensive work on making effective what-if query optimizer calls~\cite{wu2022budget,brucato2024wred,Wii,PapadomanolakisDA07,siddiqui2022isum,siddiqui2022distill,WangWNC25} to improve the efficiency of index tuning, but their accuracy remains limited by the underlying cost model of query optimizer due to well-known challenges such as cardinality estimation and analytic cost modeling.

\paragraph*{Online Index Tuning}
Unlike offline index tuning, \emph{online} index tuning~\cite{BrunoC07} operates under workload drifts, which is arguably a more realistic setup in real-world index tuning applications~\cite{das2019automatically,YadavVZ23}.
It is typically formulated as a Markov decision process (MDP) with RL-based solutions~\cite{basu2015cost,
sharma2018case,sharma2021mantis,perera2021dba, zhou2022autoindex, perera2023no,sharma2022indexer++,sadri2020online,wang2024leverage}.
While RL-based index tuning methods have shown promise in adapting to dynamic workloads, they often struggle with the \emph{``cold-start''} problem, such that the lack of reliable execution feedback in early tuning rounds leads to poor initial recommendations and slow convergence. 
As we mentioned in the introduction, recent work on modeling index tuning as contextual bandits~\cite{oetomo2024warm,oetomo2023cutting,zhang2019warm} attempted to mitigate this challenge by incorporating historical knowledge and transfer learning, but failed to capture more complex index benefit functions in real-world workloads due to the simplified assumption of a linear index benefit function.
\paragraph*{Learned Index Benefit Estimators}
Accurate estimation of query execution cost given an index configuration is therefore central to both offline and online index tuning.
What-if cost estimates are relatively cheap to obtain but inaccurate, while true execution feedback is accurate but costly.
This inspires a recent line of work on learned index benefit estimators~\cite{shi2022learned, siddiqui2024ml,ding2019ai,zhao2022queryformer,gao2022smartindex,kossmann2022swirl}, which operate in the ``middle ground'' using query execution data to train learned cost models.
However, existing learned index estimators are mainly designed for offline index tuning and struggle in online settings due to the requirement of large amounts of training data and limited generalization capability over unseen queries.
\paragraph*{Uncertainty Quantification}
One key idea \sys brings into online index tuning is \emph{uncertainty quantification}, which has various applications in related areas such as query execution time estimation and robust query optimization~\cite{ChuHS99,BabcockC05,WuWHN14}.
Uncertainty is essential for learned index benefit estimators trained in an online fashion, especially during early stages of online index tuning where very limited query execution feedback data are available.
Our way of quantifying uncertainty by considering both aleatoric uncertainty (i.e., data uncertainty) and epistemic uncertainty (i.e., model uncertainty) is motivated by the recent work~\cite{yu2024can}.
However, unlike their autoencoder-based approach designed for offline tuning, which faces similar generalization challenges over unseen queries as other offline learned index benefit estimators, \sys leverages operator-level CAM predictors, which generalize better with limited online training data and incur lower computational overhead.
\section{Conclusion}

We have presented \sys, an uncertainty-aware online index tuner that employs operator-level learned index benefit estimators to address two major challenges in online index tuning, namely, (1) limited query execution feedback data for model training and (2) limited generalization capability over unseen queries due to constant workload drifts.
We have designed a low-overhead uncertainty quantification mechanism that incorporates both data uncertainty and model uncertainty, and we have integrated it into both the learned index benefit estimators and the index configuration enumeration component of \sys.
Our experimental evaluation demonstrates that factoring uncertainty information into the index selection strategy of \sys 
can significantly improve the workload execution time and reduce the index exploration overhead in online index tuning, compared to state-of-the-art RL-based index tuners.

\clearpage
\section*{AI-Generated Content Acknowledgement}
The authors used OpenAI’s ChatGPT solely for grammar checking and minor language polishing of the manuscript. The AI system was not used to generate research ideas, technical content, analyses, figures, or experimental results. All scientific contributions, methodologies, and conclusions presented in this paper are entirely the work of the authors.

\bibliographystyle{IEEEtran}
\bibliography{ref}

@inproceedings{oetomo2024warm,
  title={Warm-Starting Contextual Bandits Under Latent Reward Scaling},
  author={Oetomo, Bastian and Perera, R Malinga and Borovica-Gajic, Renata and Rubinstein, Benjamin IP},
  booktitle={2024 IEEE International Conference on Data Mining (ICDM)},
  pages={360--369},
  year={2024},
  organization={IEEE}
}

@article{marcus2019Neo,
author = {Marcus, Ryan and Negi, Parimarjan and Mao, Hongzi and Zhang, Chi and Alizadeh, Mohammad and Kraska, Tim and Papaemmanouil, Olga and Tatbul, Nesime},
title = {Neo: a learned query optimizer},
year = {2019},
issue_date = {July 2019},
publisher = {VLDB Endowment},
volume = {12},
number = {11},
issn = {2150-8097},
url = {https://doi.org/10.14778/3342263.3342644},
doi = {10.14778/3342263.3342644},
journal = {Proc. VLDB Endow.},
month = jul,
pages = {1705–1718},
numpages = {14}
}

@article{oetomo2023cutting,
  title={Cutting to the chase with warm-start contextual bandits},
  author={Oetomo, Bastian and Perera, R Malinga and Borovica-Gajic, Renata and Rubinstein,Benjamin},
  journal={Knowledge and Information Systems},
  volume={65},
  number={9},
  pages={3533--3565},
  year={2023},
  publisher={Springer}
}

@article{zhang2019warm,
  title={Warm-starting contextual bandits: Robustly combining supervised and bandit feedback},
  author={Zhang, Chicheng and Agarwal, Alekh and Daum{\'e} III, Hal and Langford, John and Negahban, Sahand N},
  journal={arXiv preprint arXiv:1901.00301},
  year={2019}
}

@inproceedings{perera2021dba,
  title={DBA bandits: Self-driving index tuning under ad-hoc, analytical workloads with safety guarantees},
  author={Perera, R Malinga and Oetomo, Bastian and Rubinstein, Benjamin IP and Borovica-Gajic, Renata},
  booktitle={2021 IEEE 37th International Conference on Data Engineering (ICDE)},
  pages={600--611},
  year={2021},
  organization={IEEE}
}

@article{perera2022hmab,
  title={HMAB: self-driving hierarchy of bandits for integrated physical database design tuning},
  author={Perera, R Malinga and Oetomo, Bastian and Rubinstein, Benjamin IP and Borovica-Gajic, Renata},
  journal={Proceedings of the VLDB Endowment},
  volume={16},
  number={2},
  pages={216--229},
  year={2022},
  publisher={VLDB Endowment}
}

@article{kossmann2020magic,
  title={Magic mirror in my hand, which is the best in the land? an experimental evaluation of index selection algorithms},
  author={Kossmann, Jan and Halfpap, Stefan and Jankrift, Marcel and Schlosser, Rainer},
  journal={Proceedings of the VLDB Endowment},
  volume={13},
  number={12},
  pages={2382--2395},
  year={2020},
  publisher={VLDB Endowment}
}

@inproceedings{gal2016dropout,
  title={Dropout as a bayesian approximation: Representing model uncertainty in deep learning},
  author={Gal, Yarin and Ghahramani, Zoubin},
  booktitle={{ICML}},
  pages={1050--1059},
  year={2016}
}

@inproceedings{zhou2022autoindex,
  title={Autoindex: An incremental index management system for dynamic workloads},
  author={Zhou, Xuanhe and Liu, Luyang and Li, Wenbo and Jin, Lianyuan and Li, Shifu and Wang, Tianqing and Feng, Jianhua},
  booktitle={2022 IEEE 38th International Conference on Data Engineering (ICDE)},
  pages={2196--2208},
  year={2022},
  organization={IEEE}
}

@article{zhou2024breaking,
  title={Breaking It Down: An In-Depth Study of Index Advisors},
  author={Zhou, Wei and Lin, Chen and Zhou, Xuanhe and Li, Guoliang},
  journal={Proceedings of the VLDB Endowment},
  volume={17},
  number={10},
  pages={2405--2418},
  year={2024},
  publisher={VLDB Endowment}
}

@inproceedings{siddiqui2022isum,
  title={Isum: Efficiently compressing large and complex workloads for scalable index tuning},
  author={Siddiqui, Tarique and Jo, Saehan and Wu, Wentao and Wang, Chi and Narasayya, Vivek and Chaudhuri, Surajit},
  booktitle={Proceedings of the 2022 International Conference on Management of Data},
  pages={660--673},
  year={2022}
}

@inproceedings{wu2022budget,
  title={Budget-aware index tuning with reinforcement learning},
  author={Wu, Wentao and Wang, Chi and Siddiqui, Tarique and Wang, Junxiong and Narasayya, Vivek and Chaudhuri, Surajit and Bernstein, Philip A},
  booktitle={Proceedings of the 2022 International Conference on Management of Data},
  pages={1528--1541},
  year={2022}
}

@article{shi2022learned,
  title={Learned index benefits: Machine learning based index performance estimation},
  author={Shi, Jiachen and Cong, Gao and Li, Xiao-Li},
  journal={Proceedings of the VLDB Endowment},
  volume={15},
  number={13},
  pages={3950--3962},
  year={2022},
  publisher={VLDB Endowment}
}

@article{yu2024refactoring,
  title={Refactoring Index Tuning Process with Benefit Estimation},
  author={Yu, Tao and Zou, Zhaonian and Sun, Weihua and Yan, Yu},
  journal={Proceedings of the VLDB Endowment},
  volume={17},
  number={7},
  pages={1528--1541},
  year={2024},
  publisher={VLDB Endowment}
}

@inproceedings{ding2019ai,
  title={Ai meets ai: Leveraging query executions to improve index recommendations},
  author={Ding, Bailu and Das, Sudipto and Marcus, Ryan and Wu, Wentao and Chaudhuri, Surajit and Narasayya, Vivek R},
  booktitle={{SIGMOD}},
  pages={1241--1258},
  year={2019}
}

@inproceedings{sharma2021mantis,
  title={MANTIS: multiple type and attribute index selection using deep reinforcement learning},
  author={Sharma, Vishal and Dyreson, Curtis and Flann, Nicholas},
  booktitle={Proceedings of the 25th International Database Engineering \& Applications Symposium},
  pages={56--64},
  year={2021}
}

@article{perera2023no,
  title={No DBA? No regret! Multi-armed bandits for index tuning of analytical and HTAP workloads with provable guarantees},
  author={Perera, R Malinga and Oetomo, Bastian and Rubinstein, Benjamin IP and Borovica-Gajic, Renata},
  journal={IEEE Transactions on Knowledge and Data Engineering},
  year={2023},
  publisher={IEEE}
}

@article{siddiqui2024ml,
  title={ML-Powered Index Tuning: An Overview of Recent Progress and Open Challenges},
  author={Siddiqui, Tarique and Wu, Wentao},
  journal={ACM SIGMOD Record},
  volume={52},
  number={4},
  pages={19--30},
  year={2024},
  publisher={ACM New York, NY, USA}
}

@inproceedings{sharma2022indexer++,
  title={Indexer++ workload-aware online index tuning with transformers and reinforcement learning},
  author={Sharma, Vishal and Dyreson, Curtis},
  booktitle={Proceedings of the 37th ACM/SIGAPP Symposium on Applied Computing},
  pages={372--380},
  year={2022}
}

@inproceedings{das2019automatically,
  title={Automatically indexing millions of databases in microsoft azure sql database},
  author={Das, Sudipto and Grbic, Miroslav and Ilic, Igor and Jovandic, Isidora and Jovanovic, Andrija and Narasayya, Vivek R and Radulovic, Miodrag and Stikic, Maja and Xu, Gaoxiang and Chaudhuri, Surajit},
  booktitle={{SIGMOD}},
  pages={666--679},
  year={2019}
}

@INPROCEEDINGS{sadri2020online,
  author={Sadri, Zahra and Gruenwald, Le and Leal, Eleazar},
  booktitle={{ICDEW}}, 
  title={Online Index Selection Using Deep Reinforcement Learning for a Cluster Database}, 
  year={2020},
  volume={},
  number={},
  pages={158-161}
}

@article{chaud1998autoadmin,
  title={AutoAdmin “what-if” index analysis utility},
  author={Chaudhuri, Surajit and Narasayya, Vivek},
  journal={ACM SIGMOD Record},
  volume={27},
  number={2},
  pages={367--378},
  year={1998},
  publisher={ACM New York, NY, USA}
}

@inproceedings{gao2022smartindex,
  title={SmartIndex: An Index Advisor with Learned Cost Estimator},
  author={Gao, Jianling and Zhao, Nan and Wang, Ning and Hao, Shuang},
  booktitle={{CIKM}},
  pages={4853--4856},
  year={2022}
}

@article{zhao2022queryformer,
  title={Queryformer: A tree transformer model for query plan representation},
  author={Zhao, Yue and Cong, Gao and Shi, Jiachen and Miao, Chunyan},
  journal={Proceedings of the VLDB Endowment},
  volume={15},
  number={8},
  pages={1658--1670},
  year={2022},
  publisher={VLDB Endowment}
}

@article{deep2020comprehensive, 
author = {Deep, Shaleen and Gruenheid, Anja and Koutris, Paraschos and Naughton, Jeffrey and Viglas, Stratis}, 
title = {Comprehensive and efficient workload compression}, 
year = {2020},  
publisher = {{PVLDB}}, 
volume = {14}, 
number = {3}
}

@article{siddiqui2022distill,
  title={DISTILL: low-overhead data-driven techniques for filtering and costing indexes for scalable index tuning},
  author={Siddiqui, Tarique and Wu, Wentao and Narasayya, Vivek and Chaudhuri, Surajit},
  journal={Proceedings of the VLDB Endowment},
  volume={15},
  number={10},
  pages={2019--2031},
  year={2022},
  publisher={VLDB Endowment}
}

@article{peret2004line,
  title={On-line search for solving Markov decision processes via heuristic sampling},
  author={P{\'e}ret, Laurent and Garcia, Fr{\'e}d{\'e}rick},
  journal={learning},
  volume={16},
  pages={2},
  year={2004}
}

@inproceedings{grohe2020word2vec,
  title={word2vec, node2vec, graph2vec, x2vec: Towards a theory of vector embeddings of structured data},
  author={Grohe, Martin},
  booktitle={{PODS}},
  pages={1--16},
  year={2020}
}

@article{brucato2024wred,
  title={Wred: Workload Reduction for Scalable Index Tuning},
  author={Brucato, Matteo and Siddiqui, Tarique and Wu, Wentao and Narasayya, Vivek and Chaudhuri, Surajit},
  journal={Proceedings of the ACM on Management of Data},
  volume={2},
  number={1},
  pages={1--26},
  year={2024},
  publisher={ACM New York, NY, USA}
}

@inproceedings{kossmann2022swirl,
  author       = {Jan Kossmann and
                  Alexander Kastius and
                  Rainer Schlosser},
  title        = {{SWIRL:} Selection of Workload-aware Indexes using Reinforcement Learning},
  booktitle    = {{EDBT}},
  pages        = {2:155--2:168},
  year         = {2022},
}

@inproceedings{chaudhuri1997efficient,
  title={An efficient, cost-driven index selection tool for Microsoft SQL server},
  author={Chaudhuri, Surajit and Narasayya, Vivek R},
  booktitle={VLDB},
  volume={97},
  pages={146--155},
  year={1997}
}

@article{wang2024leverage,   
 author       = {Zijia Wang and
                  Haoran Liu and
                  Chen Lin and
                  Zhifeng Bao and
                  Guoliang Li and
                  Tianqing Wang},
  title        = {Leveraging Dynamic and Heterogeneous Workload Knowledge to Boost the
                  Performance of Index Advisors},
  journal      = {Proc. {VLDB} Endow.},
  volume       = {17},
  number       = {7},
  pages        = {1642--1654},
  year         = {2024},

}

@inproceedings{valentin2000db2,
  title={DB2 advisor: An optimizer smart enough to recommend its own indexes},
  author={Valentin, Gary and Zuliani, Michael and Zilio, Daniel C and Lohman, Guy and Skelley, Alan},
  booktitle={{ICDE}},
  pages={101--110},
  year={2000}
}

@inproceedings{basu2015cost,
  title={Cost-model oblivious database tuning with reinforcement learning},
  author={Basu, Debabrota and Lin, Qian and Chen, Weidong and Vo, Hoang Tam and Yuan, Zihong and Senellart, Pierre and Bressan, St{\'e}phane},
  booktitle={{DEXA}},
  pages={253--268},
  year={2015}
}

@article{sharma2018case,
  title={The case for automatic database administration using deep reinforcement learning},
  author={Sharma, Ankur and Schuhknecht, Felix Martin and Dittrich, Jens},
  journal={arXiv preprint arXiv:1801.05643},
  year={2018}
}

@inproceedings{WuCZTHN13,
  author    = {Wentao Wu and
               Yun Chi and
               Shenghuo Zhu and
               Jun'ichi Tatemura and
               Hakan Hacig{\"u}m{\"u}s and
               Jeffrey F. Naughton},
  title     = {Predicting query execution time: Are optimizer cost models
               really unusable?},
  booktitle = {ICDE},
  year      = {2013},
  pages     = {1081-1092},
}

@inproceedings{AkdereCRUZ12-brown-icde,
  author    = {Mert Akdere and
               Ugur \c{C}etintemel and
               Matteo Riondato and
               Eli Upfal and
               Stanley B. Zdonik},
  title     = {Learning-based Query Performance Modeling and Prediction},
  booktitle = {ICDE},
  year      = {2012},
  pages     = {390-401}
}

@article{LiKNC12,
  author       = {Jiexing Li and
                  Arnd Christian K{\"{o}}nig and
                  Vivek R. Narasayya and
                  Surajit Chaudhuri},
  title        = {Robust Estimation of Resource Consumption for {SQL} Queries using
                  Statistical Techniques},
  journal      = {Proc. {VLDB} Endow.},
  volume       = {5},
  number       = {11},
  pages        = {1555--1566},
  year         = {2012}
}

@article{Wu25,
  author       = {Wentao Wu},
  title        = {Hybrid Cost Modeling for Reducing Query Performance Regression in
                  Index Tuning},
  journal      = {{IEEE} Trans. Knowl. Data Eng.},
  volume       = {37},
  number       = {1},
  pages        = {379--391},
  year         = {2025}
}

@inproceedings{YadavVZ23,
  author       = {Ritwik Yadav and
                  Satyanarayana R. Valluri and
                  Mohamed Za{\"{\i}}t},
  title        = {{AIM:} {A} practical approach to automated index management for {SQL}
                  databases},
  booktitle    = {{ICDE}},
  pages        = {3349--3362},
  year         = {2023}
}

@inproceedings{ma2018query,
  title={Query-based workload forecasting for self-driving database management systems},
  author={Ma, Lin and Van Aken, Dana and Hefny, Ahmed and Mezerhane, Gustavo and Pavlo, Andrew and Gordon, Geoffrey J},
  booktitle={Proceedings of the 2018 International Conference on Management of Data},
  pages={631--645},
  year={2018}
}

@inproceedings{BrunoC07,
  author       = {Nicolas Bruno and
                  Surajit Chaudhuri},
  title        = {An Online Approach to Physical Design Tuning},
  booktitle    = {{ICDE}},
  pages        = {826--835},
  year         = {2007}
}

@misc{job-queries,
title = {Join Order Benchmark},
author = {Viktor Leis},
howpublished = {\url{https://github.com/gregrahn/join-order-benchmark}}
}

@article{Wii,
  author       = {Xiaoying Wang and
                  Wentao Wu and
                  Chi Wang and
                  Vivek R. Narasayya and
                  Surajit Chaudhuri},
  title        = {Wii: Dynamic Budget Reallocation In Index Tuning},
  journal      = {Proc. {ACM} Manag. Data},
  volume       = {2},
  number       = {3},
  pages        = {182},
  year         = {2024}
}

@inproceedings{PapadomanolakisDA07,
  title={Efficient use of the query optimizer for automated physical design},
  author={Papadomanolakis, Stratos and Dash, Debabrata and Ailamaki, Anastasia},
  booktitle={{VLDB}},
  pages={1093--1104},
  year={2007}
}

@article{yu2024can,
  title={Can Uncertainty Quantification Enable Better Learning-based Index Tuning?},
  author={Yu, Tao and Zou, Zhaonian and Xiong, Hao},
  journal={arXiv preprint arXiv:2410.17748},
  year={2024}
}

@book{sutton2018reinforcement,
  title={Reinforcement learning: An introduction},
  author={Sutton, Richard S and Barto, Andrew G},
  year={2018},
  publisher={MIT press}
}

@article{WuWHN14,
  author       = {Wentao Wu and
                  Xi Wu and
                  Hakan Hacig{\"{u}}m{\"{u}}s and
                  Jeffrey F. Naughton},
  title        = {Uncertainty Aware Query Execution Time Prediction},
  journal      = {Proc. {VLDB} Endow.},
  volume       = {7},
  number       = {14},
  pages        = {1857--1868},
  year         = {2014}
}

@inproceedings{ChuHS99,
  author       = {Francis C. Chu and
                  Joseph Y. Halpern and
                  Praveen Seshadri},
  editor       = {Victor Vianu and
                  Christos H. Papadimitriou},
  title        = {Least Expected Cost Query Optimization: An Exercise in Utility},
  booktitle    = {{PODS}},
  pages        = {138--147},
  year         = {1999}
}

@inproceedings{BabcockC05,
  author       = {Brian Babcock and
                  Surajit Chaudhuri},
  title        = {Towards a Robust Query Optimizer: {A} Principled and Practical Approach},
  booktitle    = {{SIGMOD}},
  pages        = {119--130},
  year         = {2005}
}

@article{WuZZL24,
  author       = {Yang Wu and
                  Xuanhe Zhou and
                  Yong Zhang and
                  Guoliang Li},
  title        = {Automatic Index Tuning: {A} Survey},
  journal      = {{IEEE} Trans. Knowl. Data Eng.},
  volume       = {36},
  number       = {12},
  pages        = {7657--7676},
  year         = {2024}
}

@article{WangWNC25,
  author       = {Xiaoying Wang and
                  Wentao Wu and
                  Vivek R. Narasayya and
                  Surajit Chaudhuri},
  title        = {Esc: An Early-Stopping Checker for Budget-aware Index Tuning},
  journal      = {Proc. {VLDB} Endow.},
  volume       = {18},
  number       = {5},
  pages        = {1278--1290},
  year         = {2025}
}

@article{WuDXNC25,
  author       = {Wentao Wu and
                  Anshuman Dutt and
                  Gaoxiang Xu and
                  Vivek R. Narasayya and
                  Surajit Chaudhuri},
  title        = {Understanding and Detecting Query Performance Regression in Practical
                  Index Tuning: [Experiments {\&} Analysis]},
  journal      = {Proc. {ACM} Manag. Data},
  volume       = {3},
  number       = {6},
  pages        = {1--26},
  year         = {2025}
}

\clearpage
\appendices

\begin{figure*}
    \centering
    \begin{subfigure}[b]{0.18\linewidth}
        \centering
        \includegraphics[width=\textwidth]{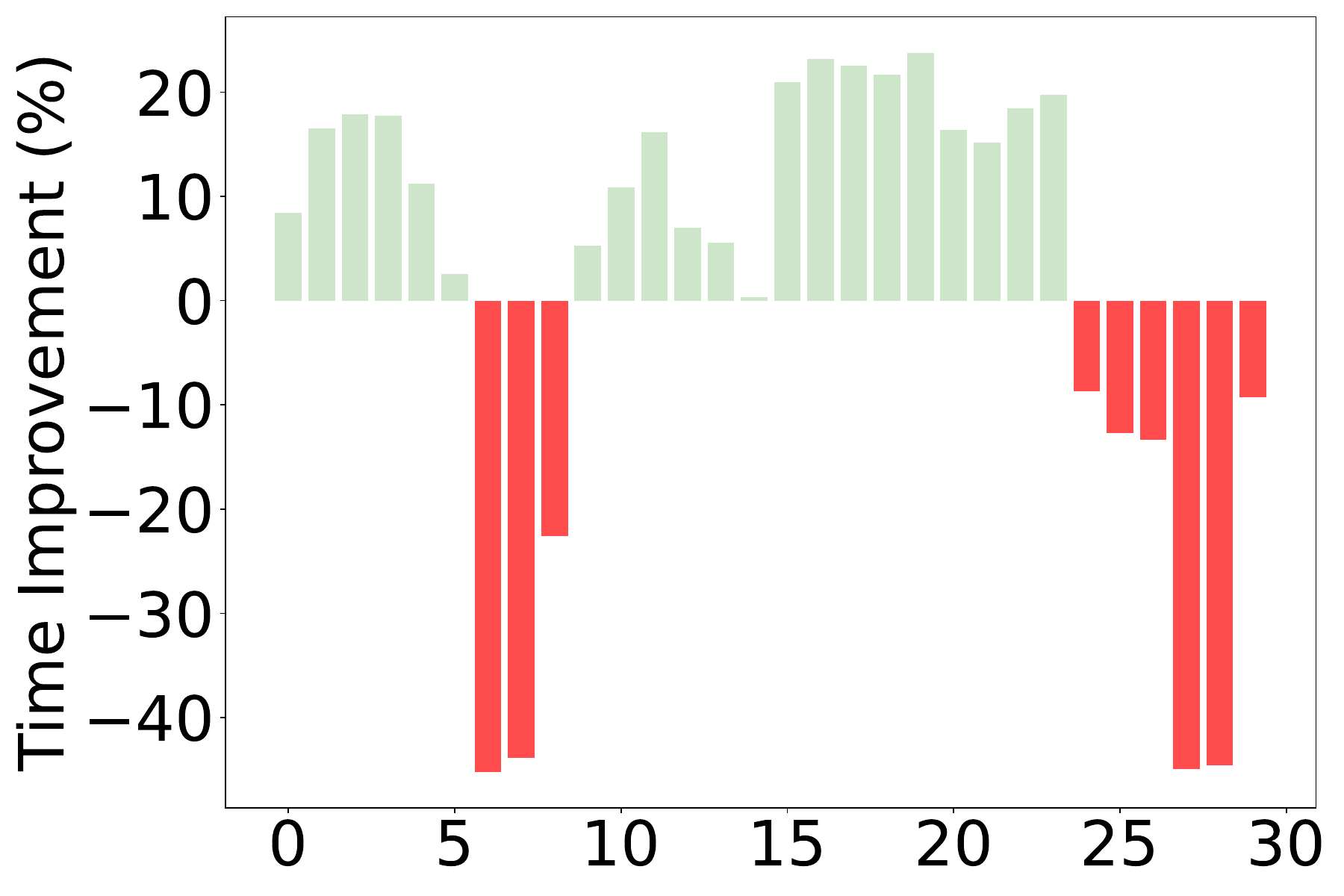}
        \caption{AutoIndex (MCTS)}
        \label{fig:exp:dynamic:tpch}
    \end{subfigure}
    \begin{subfigure}[b]{0.18\linewidth}
        \centering
        \includegraphics[width=\textwidth]{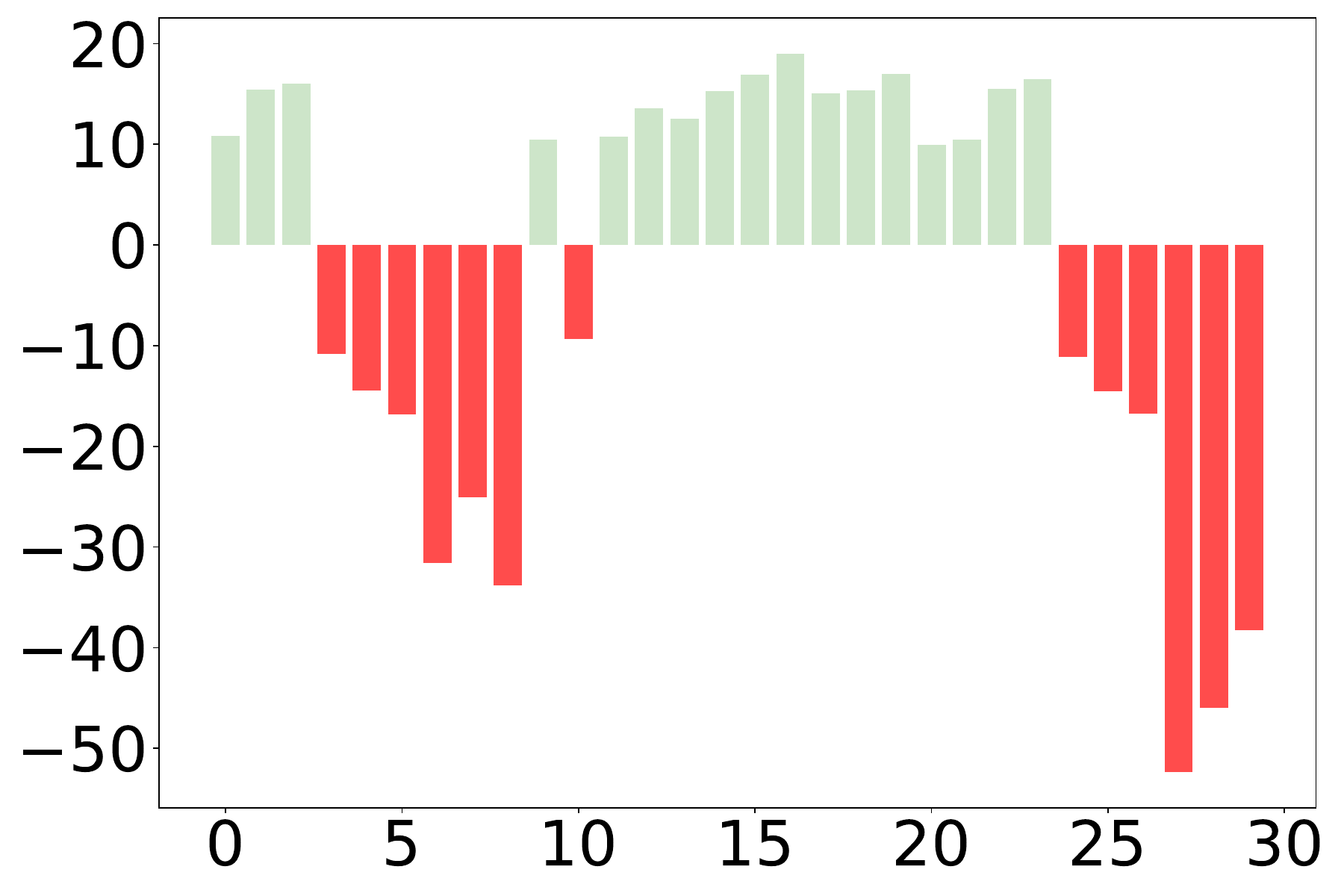}
        \caption{SWIRL}
        \label{fig:exp:dynamic:tpcds}
    \end{subfigure}
        \begin{subfigure}[b]{0.18\linewidth}
        \centering
        \includegraphics[width=\textwidth]{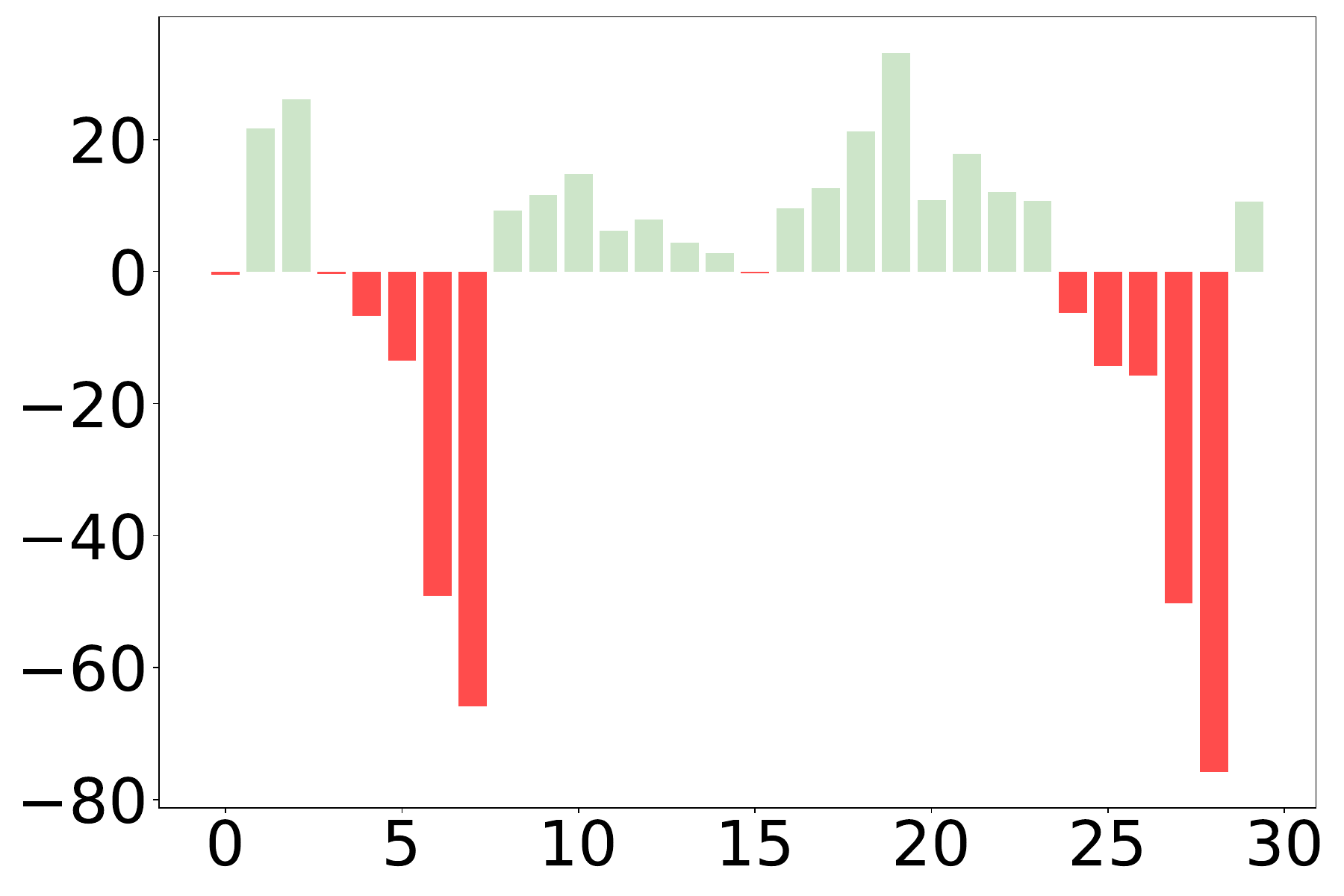}
        \caption{Indexer++(DQN)}
        \label{fig:exp:dynamic:job}
    \end{subfigure}
        \begin{subfigure}[b]{0.18\linewidth}
        \centering
        \includegraphics[width=\textwidth]{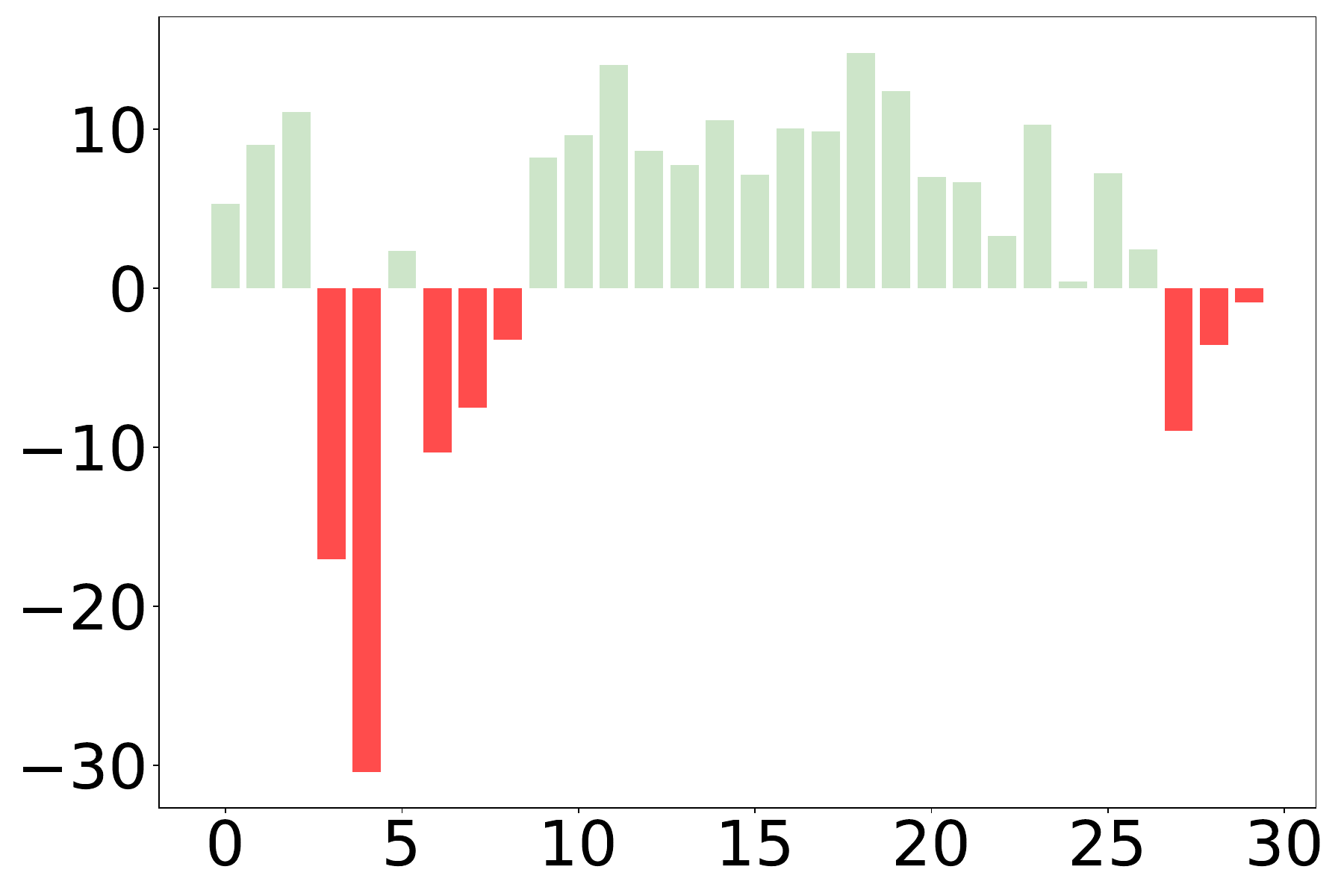}
        \caption{HMAB}
        \label{fig:exp:dynamic:job}
    \end{subfigure}
     \begin{subfigure}[b]{0.18\linewidth}
        \centering
        \includegraphics[width=\textwidth]{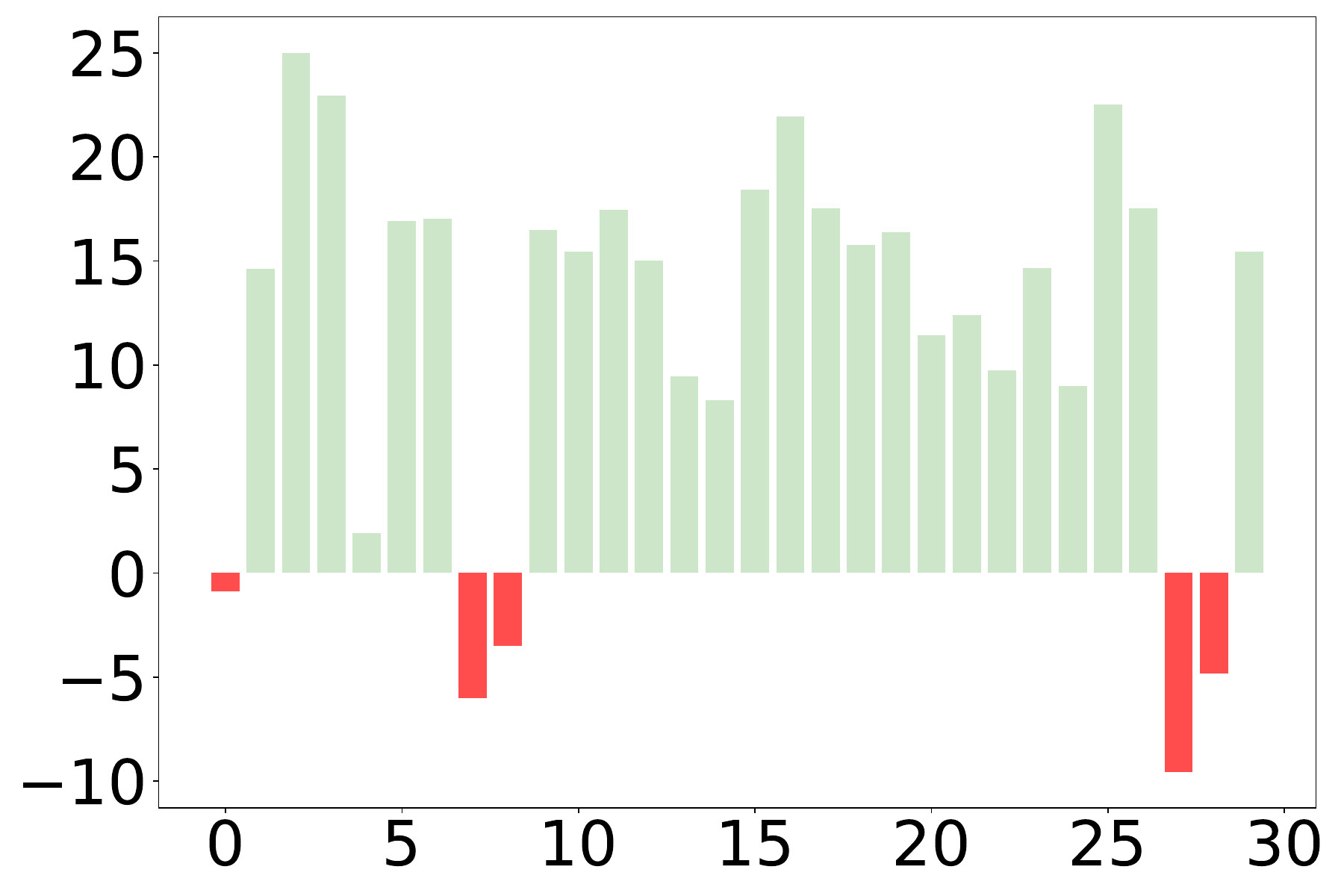}
        \caption{\sys}
        \label{fig:exp:dynamic:job}
    \end{subfigure}
    \vspace{-0.5em}
    \caption{Improvement of execution time for each mini-workload over \textbf{TPC-DS} with continuous variation
    }
    \vspace{-1.5em}
    \label{fig:exp:dynamic}
\end{figure*}

\begin{figure*}
    \centering
    \begin{subfigure}[b]{0.18\textwidth}
        \includegraphics[width=\textwidth]{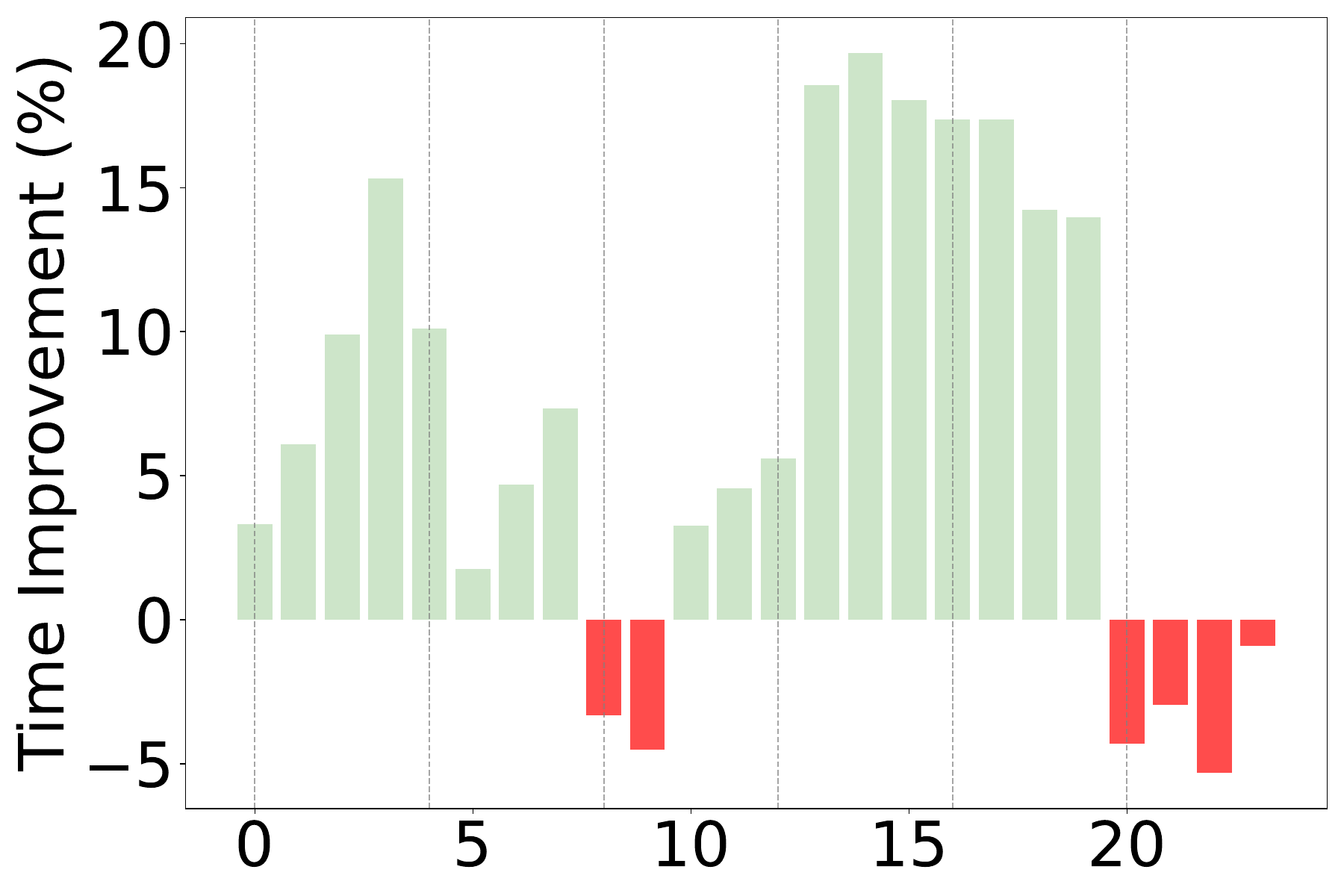}
        \caption{AutoIndex (MCTS)}
        \label{fig:exp:session:time-by-roundMCTS}
    \end{subfigure}
    \begin{subfigure}[b]{0.18\textwidth}
        \includegraphics[width=\textwidth]{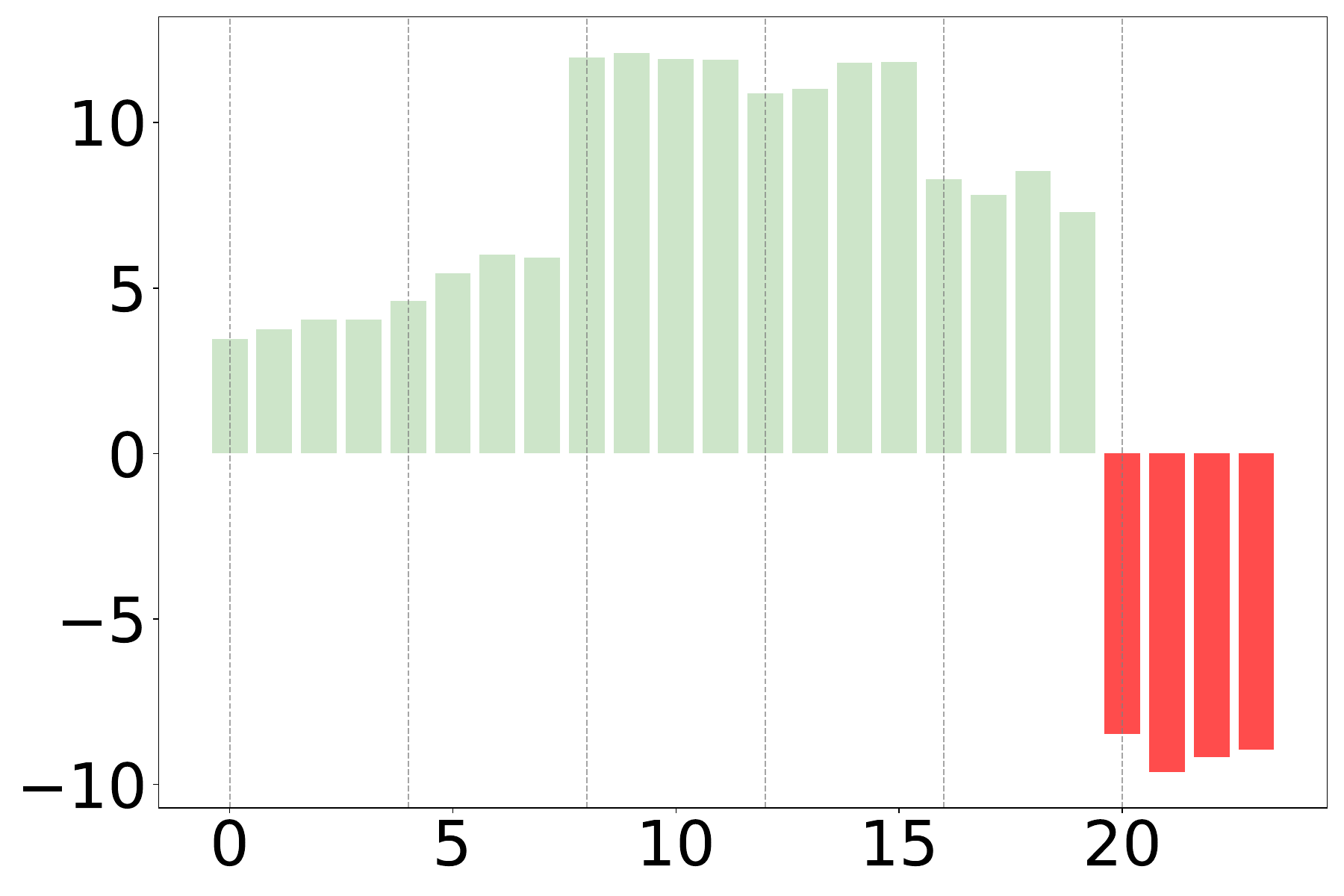}
        \caption{SWIRL}
        
        \label{fig:exp:session:time-by-round:SWIRL}
    \end{subfigure}
    \begin{subfigure}[b]{0.18\textwidth}
        \includegraphics[width=\textwidth]{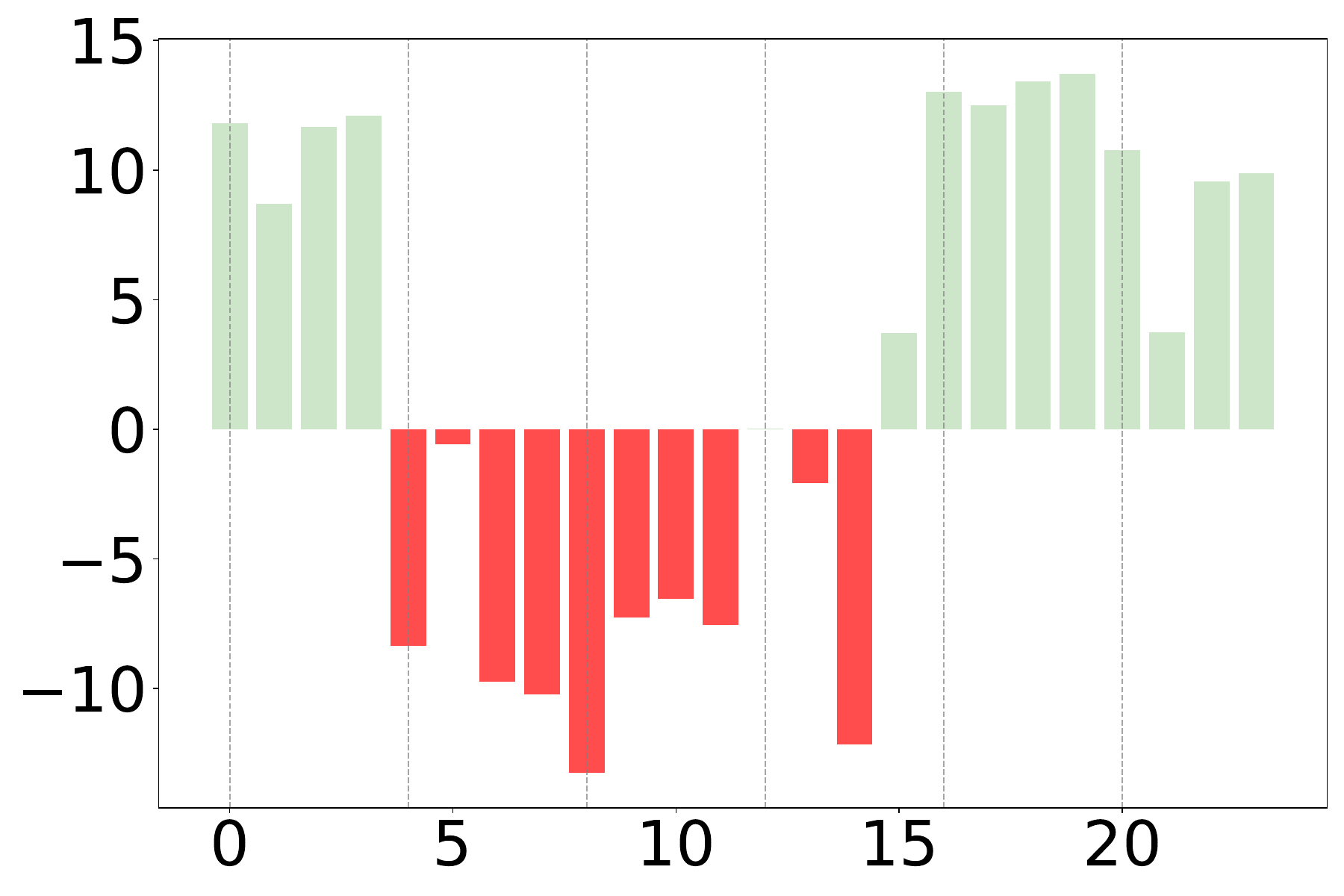}
        \caption{Indexer++ (DQN)}
        \label{fig:exp:session:time-by-round:DQN}
    \end{subfigure}
        \begin{subfigure}[b]{0.18\linewidth}
        \centering
        \includegraphics[width=\textwidth]{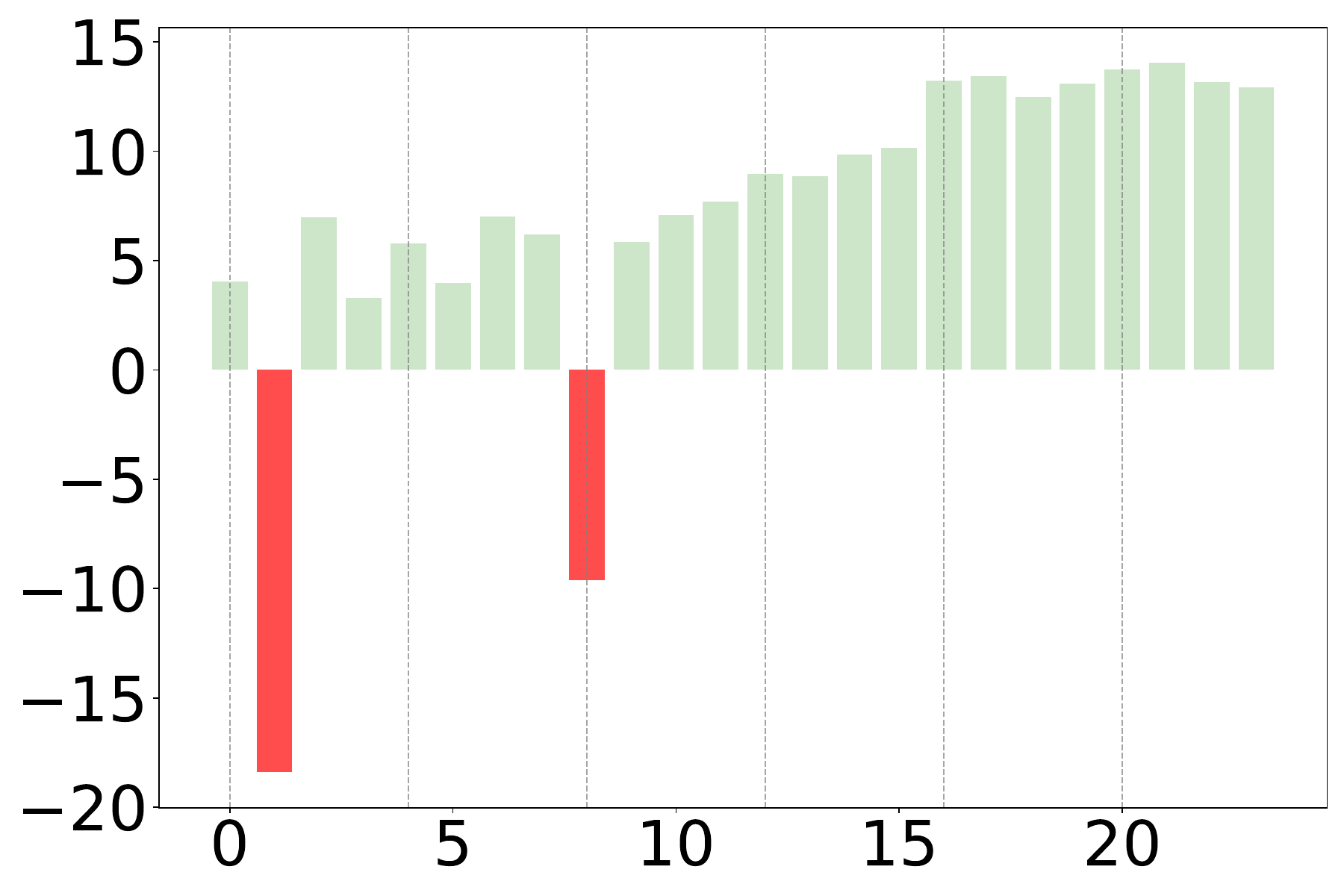}
        \caption{HMAB}
        \label{fig:exp:session:time-by-round:DBA}
    \end{subfigure}
     \begin{subfigure}[b]{0.18\linewidth}
        \centering
        \includegraphics[width=\textwidth]{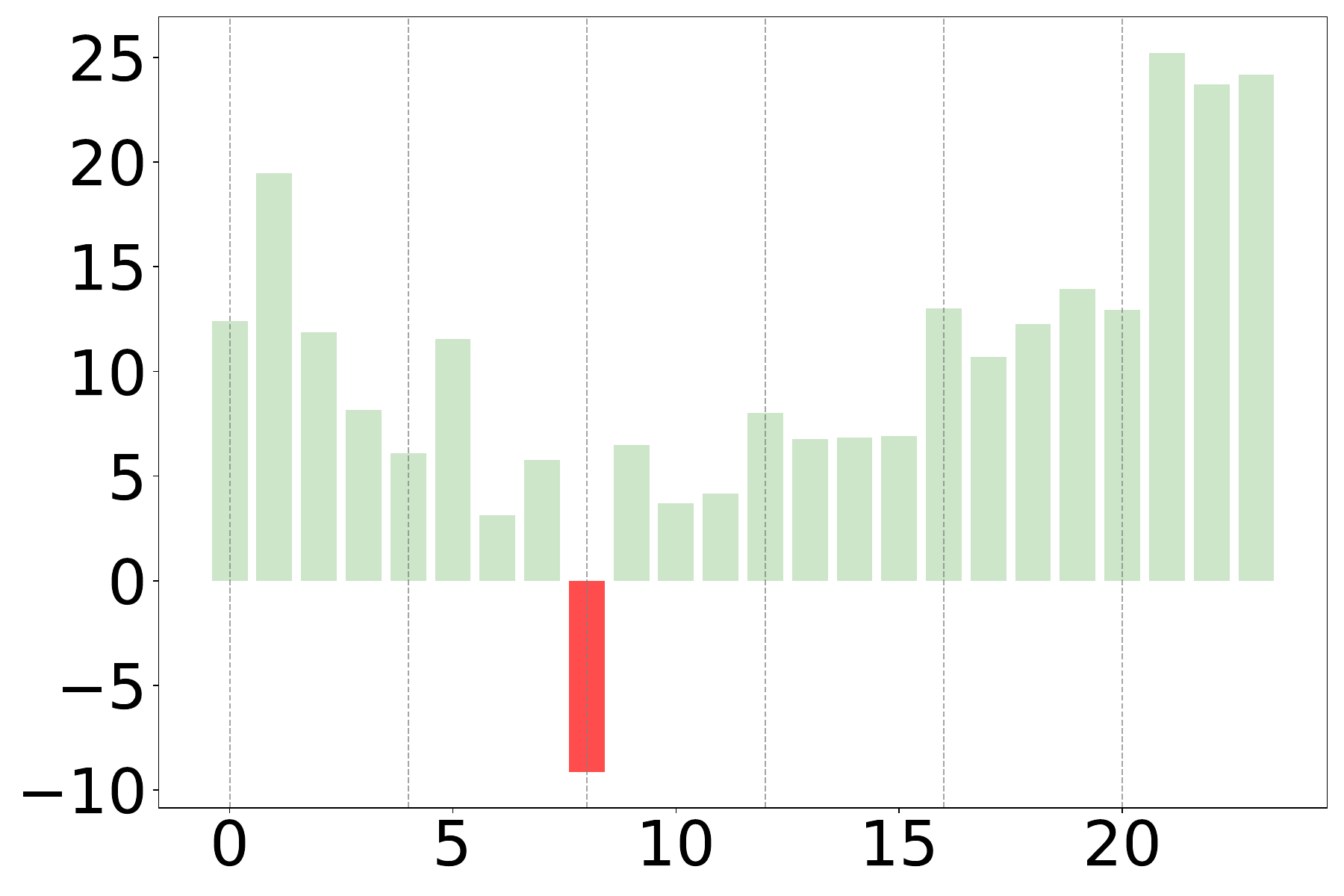}
        \caption{\sys}
        \label{fig:exp:session:time-by-round:UTune}
    \end{subfigure}
        \vspace{-5pt}
    \caption{Improvement of execution time for each mini-workload over \textbf{TPC-H} with periodic variation}
    \vspace{-1em}
    \label{fig:exp:session:time-by-round}
\end{figure*}

\section{Continuous Variation for \textbf{TPC-DS}}
\label{sect:exp:continuous-variation}

Figure~\ref{fig:exp:dynamic} further presents the execution time improvement (green bar) and regression (red bar) for each mini-workload over \textbf{TPC-DS}. \sys exhibits much less regression than other baselines, highlighting its robustness to workload drifts.
This stems from \sys's uncertainty-aware index benefit estimation framework, which avoids attractive but highly uncertain index choices in the first place.
Among the baseline index advisors, \textbf{HMAB} has the least degree of regression, as it regularly re-evaluates its index choices under shifting workloads, gradually discarding indexes with overestimated benefits and thereby reducing the chance of regression.

\section{Periodic Variation for \textbf{TPC-H}}
\label{sect:exp:periodic-variation}
Figure~\ref{fig:exp:session:time-by-round} further presents the execution time improvement/regression for each mini-workload over \textbf{TPC-H}.
Again, we observe that the chance of regression is much lower for \sys compared to the other baseline index advisors.
Although all index advisors experience periodic performance drops every 4 mini-workloads due to recurring workload drifts, \sys recovers more quickly, highlighting its robustness against workload drifts.
Notably, at $t=8$, \textbf{SWIRL} maintains performance while other index advisors suffered degradation due to a new query template \textbf{TPC-H} Q14. This is because \textbf{SWIRL}'s workload-aware selection policy reacts immediately to drifts, unlike the ``trial-and-error'' strategies of other methods that take multiple rounds of online feedback. However, \textbf{SWIRL}'s adaption to unseen query templates is not guaranteed. For example, at $t=20$, \textbf{SWIRL} performs the worst during another drift.

\section{\rwone{Impact of workload drift}}
\label{sect:exp:workload-drift}
\begin{figure}
\centering
\begin{subfigure}[b]{\linewidth}
     \centering   
     \includegraphics[width=\textwidth]{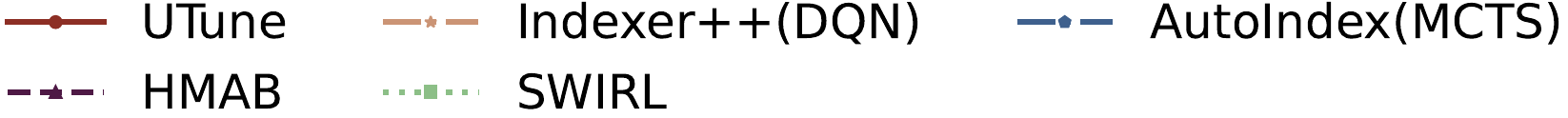}
\end{subfigure}
\begin{subfigure}[t]
{\linewidth}
    \centering      
    \includegraphics[width=0.45\textwidth]{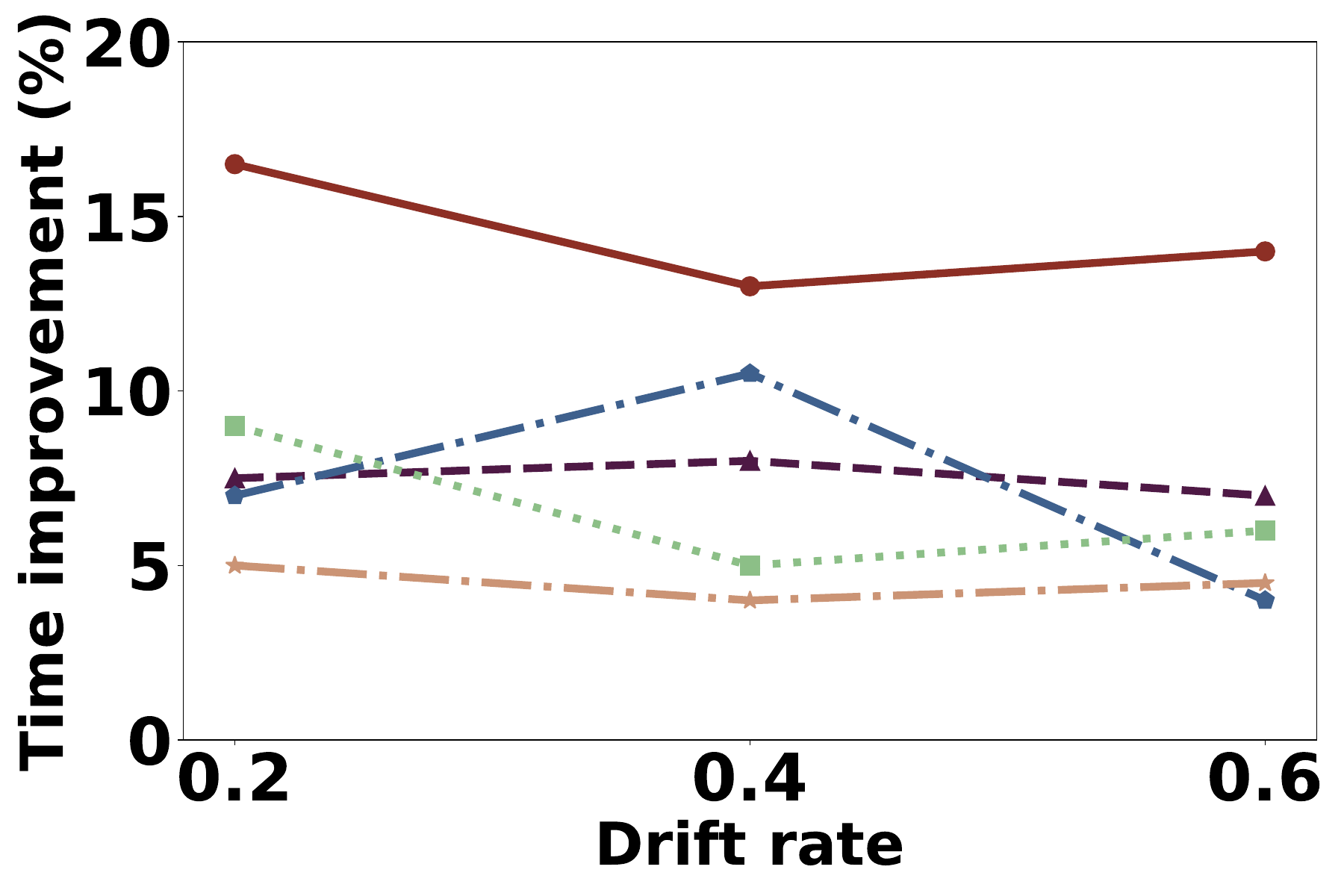}  
\end{subfigure}
\hfill
\vspace{-1.5em}
\caption{Impact on workload execution time improvement by varying the workload drift rate}
\label{exp:drift-rate}
\end{figure}

\rwone{In our previous evaluations, we have fixed the \emph{workload drift rate} to 20\%, i.e., 20\% of the query templates in the current mini-workload are replaced during workload drift.
Varying the drift rate of the workload may also have an impact on the performance of \sys, and we further examine it here.
Figure~\ref{exp:drift-rate} presents the improvement of total workload execution time when increasing the workload drift rate from 20\% to 60\% for the periodically varying workload on \textbf{TPC-H}.
As expected, the performance of all tested index advisors drops with increased degree of workload drift.
However, \sys is much less impacted compared to the other baseline index advisors.
Among the baselines, \textbf{HMAB} maintains consistent improvement, though much lower than that of \sys.
The improvement from \textbf{AutoIndex} first increases and then drops sharply, indicating performance instability with respect to varying workload drifts.
The improvements from \textbf{Indexer++} and \textbf{SWIRL} follow the same trend as that of \sys, though with significant deficits.}

\begin{figure}[t]
\vspace{-0.5em}
        \centering
    \begin{subfigure}[b]{\linewidth}
        \centering
        \includegraphics[width=0.9\textwidth]{pics/experiments/labels_line.pdf}
    \end{subfigure}
    \begin{subfigure}[b]{\linewidth}
        \centering
        \includegraphics[width=0.5\textwidth]{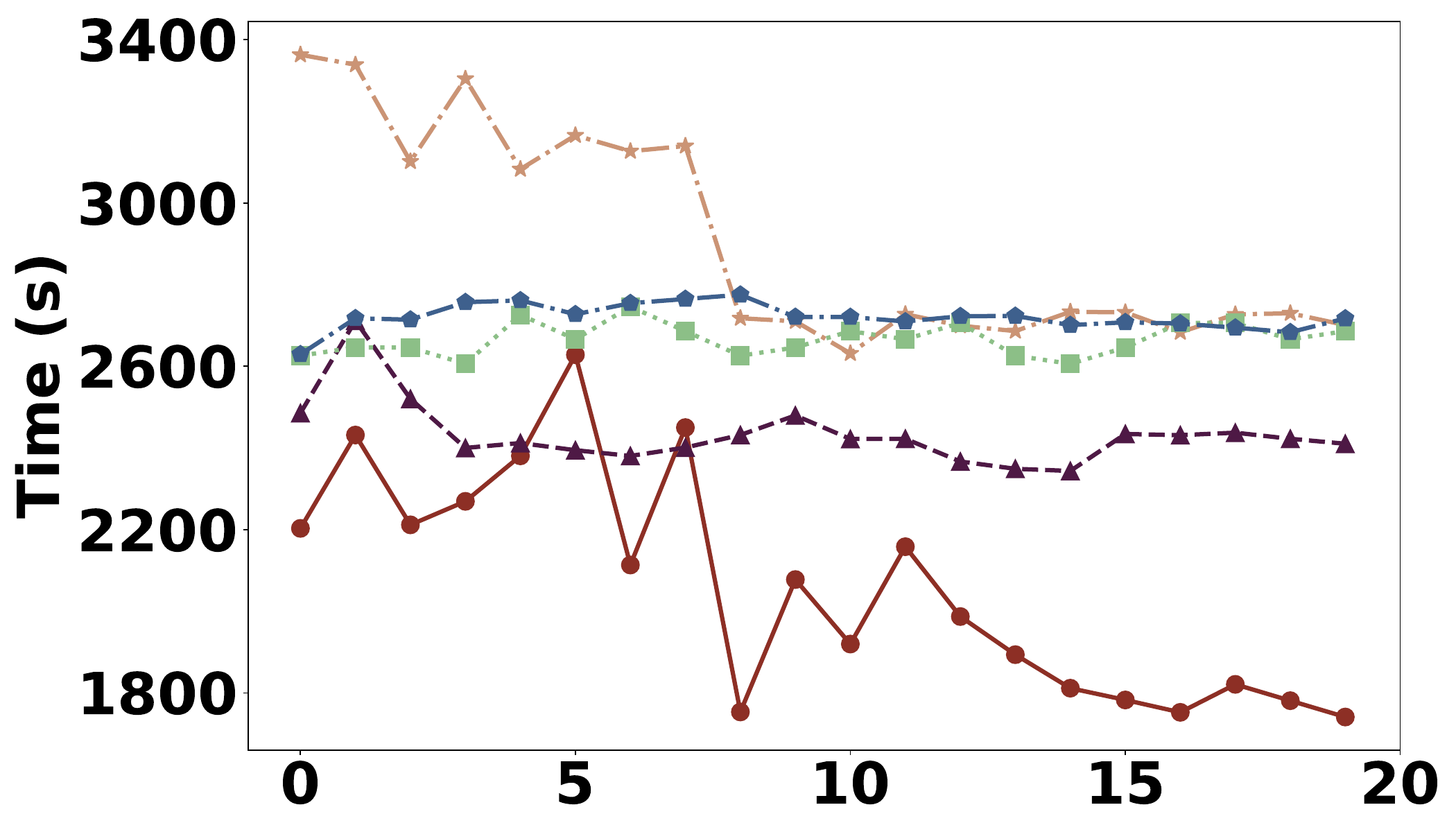}
    \end{subfigure}
    \vspace{-1.5em}
    \caption{Execution time of each mini-workload with the static \textbf{TPC-H} workload}
    \label{fig:exp:static}
\end{figure}

\section{Detailed Analysis of Static Workload}
\label{sect:exp:static-workload}
Compared to index advisors that rely on what-if calls, such as \textbf{SWIRL}, \textbf{AutoIndex}, and \textbf{Indexer++}, we attribute the improvement of \sys to its ability of utilizing actual query execution time feedback.
Specifically, significant what-if cost estimation errors on \textbf{JOB} hinder the performance of baseline index advisors while the uncertainty-aware approach taken by \sys allows for effective what-if cost correction.
Although in theory, one could replace what-if calls in \textbf{SWIRL}, \textbf{AutoIndex}, and \textbf{Indexer++} with real execution feedback, naively doing so would cause excessive exploration overhead of these RL-based index advisors and therefore forfeiting their usability for online index tuning.
It remains interesting to investigate the feasibility of replacing the what-if calls by learned index benefit estimators, such as the one used by \sys, for these RL-based index advisors, which we leave for future research.

We draw a closer comparison between \sys and \textbf{HMAB}, both using execution time as a reward. 
\textbf{HMAB} formulates index selection as \emph{contextual bandits} and assumes a linear relationship between index features and their rewards~\cite{perera2022hmab}. 
This causes negative feedback on one index to be unfairly propagated to other structurally similar candidates in the same table.
Additionally, its ``upper confidence bound(UCB)'' selection strategy encourages exploring complex indexes with uncertain benefits, often yielding poor initial index recommendations.
In contrast, instead of assuming linear reward function, \sys follows a more standard paradigm with the goal of learning the right reward function from observed execution data, based on operator-level CAM predictors and uncertainty-aware cost correction. 
In addition, by incorporating uncertainty into index selection, \sys avoids the ``overfitting'' problem evidenced by the convergence behavior of \textbf{HMAB}. 

To illustrate this, Figure~\ref{fig:exp:static} presents the improvement of workload execution time observed in each replay of the static workload on \textbf{TPC-H}.
We observe that \sys can gradually lower the workload execution time with more replays, whereas \textbf{HMAB} quickly converges to some suboptimal configuration without making further progress.

\section{\rwone{Example of Index Interaction}}
\label{sect:exp:index-interaction}
\rwone{Following the case study presented in Section~\ref{sect:exp:case-study}, we dive deeper into the compound benefit of $\{I_{mi},I_{cc}\}$.
While $I_{mi}$ causes regression in isolation for \textbf{Q29}, the combination of $\{I_{mi},I_{cc}\}$ is optimal, since $I_{cc}$ accelerates the lookups required on the inner side of the \texttt{Nested-Loop Join} and therefore fixes the regression caused by $I_{mi}$:} 
\begin{enumerate}
    \item \rwone{In the initial rounds, $I_{mi}$ is selected because of its high what-if estimated benefit, while $I_{cc}$ is selected because of its high \emph{epistemic uncertainty} (for exploration). }
    \item \rwone{
    In the following rounds, $I_{cc}$ is dropped as exploration shifts, but $I_{mi}$ remains selected due to its beneficial feedback from many query templates. Executing $I_{mi}$ in isolation causes a severe regression on \textbf{Q29}. Crucially, this creates a conflict: 
    the high benefit of $I_{mi}$ observed previously contradicts the current poor performance. 
    The performance volatility manifests a spike in the \emph{aleatoric uncertainty} of the operator-level models.}
    \item \rwone{Instead of discarding $I_{mi}$ due to the regression, the uncertainty-aware strategy interprets this high \emph{aleatoric uncertainty} as a signal to prompt \sys for further exploration of $I_{mi}$ under different index configurations.}
    \item \rwone{Eventually, the combination $\{I_{mi}, I_{cc}\}$ is sampled again. The presence of $I_{cc}$ fixes the regression, stabilizing the performance of $I_{mi}$. The selection shifts to a stable state with high total value (in Equation~\ref{equation:total_value}), leading \sys to converge on this beneficial configuration.}
    \end{enumerate}

\section{\rwfive{Details of Candidate Generation}}
\label{sect:exp:candidate-gen}
\rwfive{Following classical rule-based approaches~\cite{perera2022hmab, zhou2024breaking}, \sys generates index candidates that target distinct dimensions of query execution: 
(1) \emph{payload filtering}, where we generate indexes on \texttt{Select} predicates to optimize table access methods;
(2) \emph{join optimization}, where we generate indexes on \texttt{Join} keys that allow the query optimizer to alter join order or join operation (e.g., \texttt{Hash Join} $\to$ \texttt{Nested-Loop Join});
(3) \emph{data ordering}, where we also generate candidates for \texttt{Order By} and \texttt{Group By} columns, which allows the query optimizer to eliminate expensive \texttt{Sort} operators.
}

\end{document}